\documentclass[a4paper,fleqn,usenatbib]{mnras}
\usepackage{amsmath} 
\usepackage{amssymb}
\usepackage{newtxtext,newtxmath}
\usepackage{colordvi}

\usepackage[T1]{fontenc}
\usepackage{ae,aecompl}

\usepackage{graphicx}	
\usepackage[caption=false]{subfig}
\usepackage{natbib}
\usepackage{multirow}

\title[Self-consistent analysis of the stellar populations of massive starbursts at $z\sim1$]
{Ultraviolet-to-far-infra-red self-consistent analysis of the stellar
  populations of massive starburst galaxies at intermediate redshifts}

\author[N. Espino-Briones et al.]{
N\'estor Espino-Briones$^{1,2}$\thanks{E-mail: nespino@ucm.es, permanent: nespitron@gmail.com}, 
Pablo G. P\'erez-Gonz\'alez$^{3,1}$, 
Jaime Zamorano$^{1}$ and
\newauthor~Luc\'ia Rodr\'iguez-Mu\~noz$^{4}$
\\
$^{1}$Departamento de F\'isica de la Tierra y Astrof\'isica, Fac. CC. F\'isicas, Universidad Complutense de Madrid, Plaza de las Ciencias 1, Madrid, E-28040, Spain\\
$^{2}$Instituto Nacional de Astrof\'isica, \'Optica y Electr\'onica, Luis E. Erro No. 1, Tonantzintla, C.P. 72840, Puebla, M\'exico\\
$^{3}$Centro de Astrobiolog\'ia, Instituto Nacional de T\'ecnica Aeroespacial, Carretera de Ajalvir km 4, Torrej\'on de Ardoz, Madrid, E-28850, Spain\\
$^{4}$Dipartimento di Fisica e Astronomia ``Galileo Galilei'', Universit\`a degli Studi di Pavoda, Vicolo dell'Osservatorio 3, I-35122 Pavoda, Italy
}

\date{Accepted 2022 March 9. Received 2022 March 4; in original form 2021 August 23}

\pubyear{2022}

\begin{document}
\label{firstpage}
\pagerange{\pageref{firstpage}--\pageref{lastpage}}
\maketitle

\begin{abstract}
We study in detail the properties of the stellar populations of 111 massive 
($\log(M_{\star}/\mathrm{M}_{\sun}) \ge 10)$ dusty (FIR-selected) starburst ($SFR/SFR_\mathrm{MS}>2$) 
galaxies at $0.7<z<1.2$. For that purpose, we use self-consistent methods that analyse the UV-to-FIR
broadband observations in terms of the stellar light and dust re-emission with energy-balance techniques. 
We find that the emission of our starburst galaxies can be interpreted as a recent star formation 
episode superimposed on a more evolved stellar population. On average, the burst age is $\sim~80$ Myr and
 its attenuation $\sim 2.4$ mag. Assuming our starburst galaxies at half their lifetimes, we infer a
 duration of the starburst phase of $\sim160$ Myr. The median stellar mass and \textit{SFR} are 
$\log(M_\star/\mathrm{M}_{\sun})\sim10.6$ and $\sim220~\mathrm{M}_{\sun}~$yr$^{-1}$. Assuming 
this \textit{SFR} and the inferred duration of the starburst phase, the stellar mass added during
this phase corresponds to $\sim40$ per cent the median stellar mass of our sample. The young-population
age determines the position of our galaxies in the $M_{\star}-SFR$ plane. Galaxies located at the 
largest distances of the MS  present shorter young-population ages. The properties of the underlying 
stellar population cannot be constrained accurately with our broadband data. We also discuss the impact of 
including the FIR data and energy-balance techniques in the analysis of the properties of the stellar
populations in starburst galaxies.
\end{abstract}

\begin{keywords}
galaxies -- galaxies: starburst -- galaxies: photometry -- infrared: galaxies -- galaxies: evolution
\end{keywords}



\section{Introduction}


Studying the processes governing the formation and evolution 
of galaxies, deep optical and infra-red (IR) surveys have determined 
that most of the star-forming galaxies up to $z\sim 5$ exhibit a 
correlation between their stellar mass ($M_{\star}$) and star 
formation rate (\textit{SFR}), which is alluded to as main sequence
(MS; e.g., \citealt{noe07,sal07,dro08}). The scatter of the MS at 
fixed stellar mass is modest ($\sim0.3$~dex; e.g., \citealt{noe07,kur16}), 
which is explained as evidence of a steady star-formation mode which
requires smooth gas accretion from the intergalactic medium 
\citep{ren09,gen10}. The outliers above the MS are classified as 
starburst\footnote{In this paper, we use the term `starburst' to refer
to galaxies whose \textit{SFR} is at least twice that on the MS at 
fixed $M_{\star}$ (i.e., $1\sigma$ above the MS), and the term `starburst phase' to allude to the 
time-scale in which such elevated \textit{SFR} is sustained.}
 galaxies. Examples of these objects with very prominent star formation
are typically dust-enshrouded systems like ultra luminous red galaxies 
(ULIRGs, \citealt{san96,elb07}), and high-redshift ($z \goa 2$)
submillimetre galaxies (SMGs; \citealt{tac08,gom18}). Dusty star-forming galaxies 
(dSFGs) dominate the star-forming population in the mass range
$\log(M_{\star}/\mathrm{M}_{\sun})\sim 10.0-10.5$ at $z<1.5$, and they are a factor 
$3-5$ more important than unobscured star-forming galaxies at $\log(M_{\star}/\mathrm{M}_{\sun}) \gtrsim 10.5$
\citep{mar16}. For $\log(M_{\star}/\mathrm{M}_{\sun}) > 10.5$ and at $0<z<2.5$, more than 90 per cent
of the total (ultraviolet+IR) star formation is obscured \citep{whi17}.
Novel observations have disclosed the appearance 
of a significant dusty starburst population up to $z\sim 4-5$ \citep{bis18,don20}. 

The elevated \textit{SFRs} of starbursts are probably triggered by 
extreme processes like mergers (e.g., \citealt{dma08,cib19}) or disc
instabilities (e.g., \citealt{cev10,rom16}). These intense \textit{SFRs} are
expected to persist for short time-scales ($\sim100-400$~Myr; \citealt{tac08,mie17}) 
due to the limited gas supply. 

A distinct interpretation for starbursts
--derived from $z \gtrsim 1$ far-IR (FIR) selected samples-- suggests 
that outliers of the MS correspond to high specific \textit{SFR} galaxies 
housing young stellar populations. These galaxies are captured while still 
assembling most of their $M_{\star}$ in order to arrive to the MS \citep{man16}.

A way to understand the starburst event in the evolution of galaxies is
estimating their physical properties, e.g., $M_{\star}$ and \textit{SFR}. For that
purpose, multi-wavelength observations are typically used, normally biased towards
the rest-frame ultraviolet (UV) and optical ranges. However, for dusty starbursts, the impact of 
dust in the spectral energy distributions (SEDs) of galaxies --absorbing the UV
 to near infra-red (NIR) photons-- is very significant. Consequently, only taking 
into account that this absorbed energy is re-emitted in the FIR window, we can more 
robustly determine the stellar population properties of starburst galaxies.

Stellar population synthesis (SPS) models are widely
used to extract the physical properties of dusty starbursts. Going beyond the determination
of basic properties such as $M_{\star}$ and \textit{SFR}, they can also be used to 
explore star formation histories (SFHs), stellar ages, and dust content of galaxies
from their observed UV-to-FIR SEDs (see review in \citealt{con13}). 

The gross effect of dust attenuation in the emission
of the stellar populations of a galaxy is typically represented with an attenuation
curve (e.g., \citealt{cal00,cha00}). The amount of dust attenuation needs to be heavily 
constrained in order to avoid an age-attenuation degeneracy \citep{pfo12,con13}. FIR 
data can help to constrain the stellar-light attenuation, if it is imposed an energy 
balance between the stellar light absorbed by dust and its re-emission \citep{tak99,bur05}. 
Hence, energy-balance techniques arise as a good choice to model the observed
UV-to-FIR SEDs of dusty starbursts (e.g., \citealt{dac08,nol09}).

The possibility to carry out a detailed analysis of the dust emission from intermediate 
redshift galaxies has been significantly enhanced by missions such as \textit{Spitzer} and 
\textit{Herschel}. The \textit{Herschel Space Observatory} \citep{pil10} observed at FIR wavelengths 
across and beyond the dust-emission peak of starburst galaxies allowing to 
constrain properly their total IR luminosity ($L_\mathrm{TIR}$\footnote{Hereafter, the
integrated luminosity from 8 to $1000~\micron$.}). The Spectral and Photometric Image 
REciever (SPIRE; \citealt{gri10}) mapped $\sim 350$~deg$^2$ at 250, 350 and 500~$\micron$ 
for the \textit{Herschel} Multi-tiered Extragalactic Survey 
(HerMES; \citealt{oli12}). The large area surveys carried out by IR missions are limited in depth restricting
statistical studies to the most luminous and massive sources. Nevertheless, dust-obscured
star formation is paramount to understand the cosmic evolution of the star-forming population. 

The main goal of this paper is understanding the implications of the starburst
event in terms of the stellar population properties of galaxies above the MS.
For that purpose, we aim at determining such properties at intermediate ($z\sim1$) redshift, 
an epoch when dSFGs practically dominate the production of stars in the Universe (e.g.,
\citealt{per05,lef05,cas14}). We use a FIR-selected sample to ensure that we constrain the peak
of the dust emission in an accurate manner, and thus we select bona fide starburst galaxies. 
These $\sim 100$ dusty starbursts are introduced making publicly available their UV-to-FIR photometry.
Combining these multi-wavelength data, we fit observed UV-to-FIR SEDs with self-consistent models 
that consider the obscuration of starlight and dust emission with energy-balance techniques. 
We use two codes which manage this self-consistent analysis to determine the stellar-population 
properties of our dusty starburst sample. Utilising different codes, we assess the
robustness and identify systematics on the derived solutions. Considering the dusty and 
starbursting nature of our objects, the method we use is fitting two stellar populations 
(depicted by a young population on top of an evolved one) models to their observed SED with
and without FIR data. By doing this, we check the impact of the FIR information in breaking 
the age-attenuation degeneracy and in the derivation of stellar parameters. In particular, we
investigate the time-scales of galaxies in the starburst phase, and the amount of stellar mass
assembled in such phase.

The paper is structured as follows: In Section \ref{sec:dat} we detail the multi-wavelength data compiled and
catalogued for this work. Section \ref{sec:fimeca} describes our FIR photometric extraction and band-merging
methodology. In Section \ref{sec:sase}, we detail our procedure to derive photometric redshifts, and 
to select massive dusty starbursts using such photometric and spectroscopic redshift estimates.
Section \ref{sec:scm} explains our auto-consistent SED fitting methods to explore the stellar properties of 
dusty starbursts. In Section \ref{sec:FIRprior}, we assess the importance of the FIR-data usage in the 
determination of such properties. Section \ref{sec:stepro} contextualises our characterization  
of the starburst stellar parameters with other samples. Finally, a summary and the main conclusions of this work
are given in Section \ref{sec:clo}.

Throughout, we adopt $\Omega_\mathrm{M}=0.3$, $\Omega_\Lambda=0.7$, $h=0.7$ cosmology. All 
magnitudes quoted are in AB system. All statistical properties are reported as 
$50\mathrm{th}^{+\mathrm{to~}84\mathrm{th~percentile}}_{-\mathrm{to~}16\mathrm{th~percentile}}$.

\section{Data description and reduction}\label{sec:dat}

This work is based on the analysis of a sample of galaxies selected in
the \textit{Subaru/XMM-Newton} Deep Survey field (SXDS\footnote{The
  SXDS field is also referred as the UDS (ultra deep survey) included
  in the \textit{UKIRT} infrared deep sky survey (UKIDSS;
  \citealt{law07})}; $\alpha =
02^\mathrm{h}18^\mathrm{min}00^\mathrm{s}$, $\delta =
-05\degr00\arcmin00\arcsec$). This field is a broad-area ($\sim1$
deg$^2$) sky region counting with multi-wavelength data gathered for
studying the evolution of galaxies and the nature of extragalactic
X-ray sources \citep{fur08}. We have collected a variety of UV-to-FIR
observations, the main characteristics of these data are described in
Table \ref{tab:sxdsdc} and in the following Subsections. Since in this
work we are interested in IR-bright galaxies, we first describe the
\textit{Spitzer} and \textit{Herschel} data used to select the sample
of FIR emitters, and then we present the rest of ancillary
datasets assembled for our analysis.

\subsection{\textit{Spitzer} data}\label{sec:spda}

\textit{Spitzer} observations of the SXDS/UDS field with both the IRAC
and MIPS instruments were carried out within the legacy program for
UDS (hereafter SpUDS; \citealt{dun07}). The available basic calibrated
data (BCD) products for all wavelengths between 3.6 and
70~$\micron$ were downloaded from the \textit{Spitzer} heritage
archive, reduced and calibrated.

For the MIPS-24 data, artefact and background removal, flat--fielding
and mosaicking were done using \textsc{mopex} (version 18.3.1;
\citealt{mak05}). We used an output square pixel of 1.2~arcsec. 
The average exposure time is
$\sim3300$~s per pixel. Source detection and photometry for the
MIPS-24 mosaic were carried out with an iterative procedure, which
uses a point spread function (PSF) technique plus aperture correction
(see, e.g., \citealt{per05}). We identified sources with
\textsc{sextractor} \citep{ber96} using the coverage map as a
weighting image, performing a series of detection passes to remove
bright sources and allow the detection of fainter emitters. 
Photometry was measured following a PSF-fitting method and using the \textit{phot} and
\textit{allstar} tasks from the \textsc{daophot} package in the Image
Reduction and Analysis Facility (\textsc{iraf\footnote{\textsc{iraf}
    is distributed by the national optical astronomy observatory,
    which is operated by the association of universities for research
    in astronomy (AURA), Inc., under cooperative agreement with the
    national science foundation.}}). Given the full width half maximum
(FWHM) of the MIPS-24 PSF ($\sim6$~arcsec), some sources are
overlapped. Although \textit{allstar} deals with this crowdedness
effectuating concurrent fits to multiple objects, six passes were
required to recover the faintest sources.  We extracted PSF photometry
from all the sources of these passes together. 
The errors of the MIPS-24 photometry were
derived from the sky uncertainty. The largest sky rms determination
between \textit{phot} and \textit{allstar} is compared with 3 other
sky rms estimates (as described in appendix A of \citealt{per08}). The
final error is set to the largest value of the 5 estimations. Our
final MIPS-24 catalogue has a median $5\sigma$ threshold of
$\sim70~\mu$Jy.

For the MIPS-70 data, dark horizontal and vertical stripes were
removed by filtering with the \textit{cleanup70.tcsh} script of the
germanium reprocessing tools (\textsc{gert}) and the mosaic was built
with \textsc{mopex}. The presence of bright sources in the mosaic
produces a boost of the filtering effect, causing a partial
subtraction of source flux. We got rid of such problem following the
procedure described in \citet[][see their sections 3.1--3.3]{fra09}.
Our final mosaic has a square-pixel size of 4.0~arcsec, and 
an average exposure time of $\sim1430$~s per
pixel. MIPS-70 photometry was also carried out with a PSF technique plus
aperture correction, but using a prior-based method which takes
advantage of the depth of the IRAC and MIPS-24 observations (see
Section \ref{sec:fircat}). The errors of the MIPS-70 photometry were
estimated from the sky uncertainty in a similar way to that described
for the MIPS-24 photometry. Our final MIPS-70 catalogue has median
$5\sigma$ threshold of $\sim9$~mJy.

To check the robustness of our MIPS catalogues, we computed
differential source number counts, and we compared them with MIPS
counts presented in the literature.  For the MIPS-24 data, our results
and those of \citet{pap04} present differences smaller than 10 per cent in
median for the flux interval $0.07-5$~mJy. For the MIPS-70 data, our
results are consistent within 5 per cent with those derived by \citet{fra06}
for GOODS-N at flux densities between $\sim6-10$~mJy (after performing
completeness corrections). Our values are also compatible within 20 per cent
with the results estimated by \citet{fra09} for COSMOS within the flux
range $\sim10-100$~mJy.

SpUDS IRAC observations were carried out in map mode for its four
channels, 3.6, 4.5, 5.8, and 8.0~$\micron$. The average exposure time
per pixel is $\sim8700$~s for the 4 channels.
 All the data were reduced with the \textit{Spitzer}
pipeline which produces BCD, and were mosaicked using the technique of
\citet{hua04}, which comprises pointing honing, distortion correction,
and mosaicking using a pixel scale half of the original ($\sim0.6$~arcsec) 
with a drizzling strategy. Source extraction and photometry
were performed as explained in the appendix A of \citet{per08}. We
detected sources with \textsc{sextractor} in the 3.6 and 4.5-$\micron$
images separately, and then we combined both lists of sources.
Aperture photometry was determined for each channel IRAC image, fixing
the IRAC $3.6+4.5$ positions in the four IRAC bands, and forcing
measurements. We obtained final integrated fluxes measuring in
circular apertures of 2-arcsec radius, and applying aperture
corrections based on empirical PSFs. The limiting magnitudes (defined
as the third quartile of the magnitude distribution of our sample,
hereafter also for mid-infrared (MIR), NIR, UV and optical data) are indicated in
Table \ref{tab:sxdsdc}. The photometric uncertainties in the IRAC
catalogues were set to the largest value obtained with the 3 sky rms
estimates method described in the appendix A of \citet{per08}.

\subsection{\textit{Herschel} data}\label{sec:heda}

The FIR information used in this study includes Photodetector Array 
Camera and Spectrometer (PACS; \citealt{pog10}) and SPIRE data
for the UDS field, which was observed as part of the key program
guaranteed time (KPGT) HerMES. We downloaded all public PACS data
inside the UDS region from the \textit{Herschel} science archive,
including targets with longer exposure time as: UDS-SCUBA (also for
HerMES), and the open-time cycle 2 (OT2) UDS-CANDELS (PI Mark
Dickinson). Regarding SPIRE, we downloaded also data from shallow
observations taken using PACS/SPIRE parallel mode (for \textit{XMM}
VIDEO 1, which is in HerMES too) and data from deep observations taken
by UDS-CANDELS. We built the mosaics using the \textit{Herschel}
interactive processing environment (\textsc{hipe}) version 11, which
includes the common PACS \citep{wie09} and SPIRE \citep{dow10}
photometer pipelines, and some taylor-made methods for reducing data
of both instruments.

We used output pixel sizes of 1.2 and 2.4~arcsec for the PACS 100
and 160-$\micron$ mosaics. Source detection and photometry, for both
PACS bands, were carried out based in a prior-position method (see
Section \ref{sec:fircat}) and PSF (FWHM $\sim7$, 12~arcsec for the
green and red filters) fitting. We applied correction factors to the
PACS fluxes to account for the finite sizes of the PSFs and for losses
due to the high-pass filtering performed by the reduction pipeline
\citep{pop12,bal14}. The errors of the PACS 100 and 160-$\micron$
photometry were estimated from the sky uncertainty in a similar way to
that described for the MIPS-24 photometry (see Section \ref{sec:spda}).
Considering that we have used PACS data with different exposure times,
and that such differences are appreciable in the sky level of the
mosaic, we did catalogues for 3 different areas: UDS-CANDELS,
UDS-SCUBA and UDS-NORMAL. The median $5\sigma$ sensitivities of the
PACS filters in these areas are detailed in Table \ref{tab:sxdsdc}.

The SPIRE observations of the SXDS/UDS field were done in scan map
mode. Timelines are processed for each bolometer in the 3 channels,
using the SPIRE photometer pipeline. The timelines of the full set of
observations were gridded on to sky to create the mosaics for each
bolometer array. These mosaics were made using the destriper algorithm
and the naive map--maker \citep{ros10,smi12}, included in
\textsc{hipe}, with default pixel sizes of 6, 10 and 14~arcsec for the
250, 350 and 500-$\micron$ bands.

The PSF sizes of the SPIRE channels are large (FWHM $\sim18$, 25 and
36~arcsec), and the 3 bands are confusion limited
($5\sigma_\mathrm{conf}\sim29$, 32, 34~mJy; \citealt{ngu10}). 
Source detection and photometry for the 3 SPIRE
bands were also carried out with a prior-position method (see Section \ref{sec:fircat}) 
and PSF fitting. Then, we applied a calibration based on the beam sizes which
were assumed Gaussians, and a correction factor ($\sim10$ per cent) to
consider losses due to pixelation in the PSFs. The errors of the SPIRE
250, 350 and 500-$\micron$ photometry were estimated from the sky
uncertainty (see Section \ref{sec:spda}). Although we used SPIRE data
with different exposure times inside the UDS region, the background of
the SPIRE maps appears uniform. The final catalogues have median
$5\sigma$ thresholds of $\sim13$, 18, and 18~mJy for the SPIRE 250,
350 and 500-$\micron$ channels.

\begin{table*}
\caption[Characteristics of the data compiled for the SXDS/UDS field]{Characteristics of the data compiled for the SXDS/UDS field. (1) Name of the observing band. (2) Effective wavelength of the filter+detector. (3) Third quartile of the magnitude (flux) distribution  (limiting magnitudes for the FUV to IRAC-8.0 bands). (4) Magnitudes (fluxes) of a detection with SNR=5. (5) Median of the PSF FWHM in arcseconds. (6) Area covered by the surveys in deg$^2$. (7) Source density per square arcminute up to the $m_{\rm Q_3}$ for FUV to 8-$\micron$ bands and up to $m_{5\sigma}$ for the 24 to 500-$\micron$ channels. (8) Source from which the data were obtained.

  $^\alpha$ PACS limiting magnitudes for regions with different exposure time: UDS-CANDELS, UDS-SCUBA, and UDS-NORMAL, respectively.  $^\textrm{a}$ \textit{Spitzer} Program \#40021 (PI: Dunlop). $^\textrm{b}$ \textit{Herschel} Program, Proposal ID: \texttt{KPGT\_soliver\_1} (PI: Oliver). $^\textrm{c}$ \textit{Herschel} Program, Proposal ID: \texttt{OT2\_mdickins\_1} (PI: Dickinson)}
\label{tab:sxdsdc}
\begin{center}
\begin{tabular}{lcccccccc}
\hline \hline
& Band & $\lambda_{\textrm{eff}}$ & $m_{\textrm{Q}_3}$ & $m_{5\sigma}$ & FWHM & Area & Surf. Dens. & Source \\
& (1) & (2) & (3) & (4) & (5) & (6) & (7) & (8) \\
\hline
\multirow{10}{*}{\rotatebox[origin=c]{90}{\parbox[c]{4.5cm}{\centering UV, optical and NIR Data}}} & \vspace{3pt} FUV & $153.9$ nm & $24.8$ & $24.4$ & 5.5 & 3.2 & 2.0 & \textit{GALEX} GTO \\
& \vspace{3pt} NUV & $231.6$ nm & $24.4$ & $24.2$ & 5.5 & 3.2 & 4.4 & \textit{GALEX} GTO \\
& \vspace{3pt} \textit{B} & $441.9$ nm & $27.6$ & $29.2$ & 0.9 & 1.2 & 135.3 & \textit{Subaru}/Suprime-Cam \\
& \vspace{3pt} \textit{V} & $545.6$ nm & $27.3$ & $28.9$ & 0.9 & 1.2 & 143.9 & \textit{Subaru}/Suprime-Cam \\
& \vspace{3pt} \textit{R}$_\mathrm{c}$ & $651.8$ nm & $27.1$ & $28.8$ & 0.8 & 1.2 & 130.2 & \textit{Subaru}/Suprime-Cam \\
& \vspace{3pt} \textit{i}$'$ & $766.9$ nm & $27.0$ & $28.7$ & 0.9 & 1.2 & 137.3 & \textit{Subaru}/Suprime-Cam \\
& \vspace{3pt} \textit{z}$'$ & $906.8$ nm & $26.4$ & $28.0$ & 0.9 & 1.2 & 108.9 & \textit{Subaru}/Suprime-Cam \\
& \vspace{3pt} \textit{J} & $1.25~\micron$ & $24.2$ & $25.4$ & 0.8 & 0.8 & 55.3 & \textit{UKIRT}/WFCAM \\
& \vspace{3pt} \textit{H} & $1.64~\micron$ & $23.0$ & $24.1$ & 0.8 & 0.8 & 40.3 & \textit{UKIRT}/WFCAM \\
& \vspace{3pt} \textit{K} & $2.21~\micron$ & $22.9$ & $24.1$ & 0.8 & 0.8 & 56.7 & \textit{UKIRT}/WFCAM \\
\hline
\multirow{15}{*}{\rotatebox[origin=c]{90}{\parbox[c]{6.5cm}{\centering Mid/Far Infrared Data}}}& \vspace{3pt} IRAC-3.6 & $3.56~\micron$ & $24.2$ & $23.1$ & 2.1 & 1.0 & 36.6 & \textit{Spitzer} GTO$^\textrm{a}$ \\
& \vspace{3pt} IRAC-4.5 & $4.50~\micron$ & $24.3$ & $23.2$ & 2.1 & 1.0 & 35.1 & \textit{Spitzer} GTO$^\textrm{a}$ \\
& \vspace{3pt} IRAC-5.8 & $5.74~\micron$ & $23.3$ & $22.0$ & 2.3 & 1.0 & 25.9 & \textit{Spitzer} GTO$^\textrm{a}$ \\
& \vspace{3pt} IRAC-8.0 & $7.93~\micron$ & $23.3$ & $21.8$ & 2.3 & 1.0 & 23.9 & \textit{Spitzer} GTO$^\textrm{a}$ \\
& \vspace{3pt} MIPS-24 & $23.8~\micron$ & $19.4~(63~\mu\textrm{Jy})$ & $19.3~(70~\mu\textrm{Jy})$ & 6 & 1.2 & 6.0 & \textit{Spitzer} GTO$^\textrm{a}$ \\
& \vspace{3pt} MIPS-70 & $72.5~\micron$ & $14.7~(5~\textrm{mJy})$ & $14.0~(9~\textrm{mJy})$ & 18 & 1.2 & 0.1 & \textit{Spitzer} GTO$^\textrm{a}$ \\
& \vspace{3pt} PACS-100$^\alpha$ & $102.4~\micron$ & $\dots$ & $16.3~(1.1~\textrm{mJy})$ & 7 & 0.1 & 4.2 & \textit{Herschel} OT2$^\textrm{c}$ \\
& \vspace{3pt}  &  & $\dots$ & $15.4~(2.4~\textrm{mJy})$ & & 0.2 & 1.3 & \textit{Herschel} KPGT$^\textrm{b}$ \\
& \vspace{3pt}  &  & $\dots$ & $14.9~(4.1~\textrm{mJy})$ & & 0.6 & 0.7 & \textit{Herschel} KPGT$^\textrm{b}$ \\
& \vspace{3pt} PACS-160$^\alpha$ & $165.6~\micron$ & $\dots$ & $15.2~(3.1~\textrm{mJy})$ & 12 & 0.1 & 3.3 & \textit{Herschel} OT2$^\textrm{c}$  \\
& \vspace{3pt}  &  & $\dots$ & $14.5~(5.8~\textrm{mJy})$ & & 0.2 & 1.5 & \textit{Herschel} KPGT$^\textrm{b}$ \\
& \vspace{3pt}  &  & $\dots$ & $14.0~(8.9~\textrm{mJy})$ & & 0.6 & 0.8 & \textit{Herschel} KPGT$^\textrm{b}$ \\
& \vspace{3pt} SPIRE-250 & $253.2~\micron$ & $14.7~(4.8~\textrm{mJy})$ & $13.6~(13~\textrm{mJy})$ & 18 & 2.0 & 0.8 & \textit{Herschel} KPGT$^\textrm{b}$ \\
& \vspace{3pt} SPIRE-350 & $355.9~\micron$ & $14.6~(5.5~\textrm{mJy})$ & $13.3~(18~\textrm{mJy})$ & 26 & 2.0 & 0.2 & \textit{Herschel} KPGT$^\textrm{b}$ \\
& \vspace{3pt} SPIRE-500 & $511.3~\micron$ & $14.6~(5.5~\textrm{mJy})$ & $13.3~(18~\textrm{mJy})$ & 36 & 2.0 & 0.1 & \textit{Herschel} KPGT$^\textrm{b}$ \\
\hline
\end{tabular}
\end{center}
\end{table*}

To test the reliability of our \textit{Herschel} catalogues, we
computed source differential number counts, and we compared them with
literature. For our final sample of starbursts (see Section \ref{sec:sbsel}),
fractions of $\sim 92$ and $\sim 91$ per cent of the flux measurements 
are in the ranges $\sim4-40$ and $\sim 7-60$ mJy for PACS-100 and -160, respectively. 
For these flux ranges, our counts present differences smaller than 25 and 20 per cent with
the previous estimates for different fields of \citet{ber11} and \citet{mag13}.
For the SPIRE channels, the flux densities of our starburst sample range in
$\sim 10-95$, $\sim 15-60$, and $\sim 15-35$ mJy for 250, 350 and 500 $\micron$, respectively.
In these flux ranges, our counts show differences smaller than 15, 20 and 20 per cent 
with the earlier estimations of \citet{oli10} for HerMES, and the ones of \citet{bet12}
for COSMOS and GOODS-N. For the SPIRE bands, we also compared the flux densities estimated
for our starburst sample in our prior-based catalogue (see Section \ref{sec:fircat}) with
those provided by \citet[estimated with priors too]{swi14}. Both methods yield compatible
SPIRE flux densities to $\lesssim 20$ per cent accuracy.

\subsection{UV, optical, NIR and spectroscopic data}\label{sec:ancda}

In this Section, we describe the UV-to-NIR and spectroscopic datasets
of the SXDS/UDS field compiled for counterpart identification of each
FIR emitter, and photometric redshift estimations. Table
\ref{tab:sxdsdc} summarises the main characteristics of all data
collections including wavelengths, the third quartile of the source
magnitude distribution (giving information about the depth of the
catalogs), magnitudes corresponding to the $5\sigma$ threshold above
the sky level, FWHM of the PSFs, areas of the surveys, source number
densities, and the survey from which data were obtained for each band.

The \textit{Galaxy evolution explorer} (\textit{GALEX};
\citealt{mar05}) observed the SXDS field, as a part of the data for
\textit{XMM-Newton} large scale structure survey, in both
\textit{GALEX} filters. We downloaded the reduced images and
photometric catalogues for the area overlapping with the SXDS/UDS
field. These images have an average exposure time of $\sim24$~ks in 
both channels.

The ground-based optical imaging of the SXDS field was taken with
Suprime-Cam \citep{miy02} on the \textit{Subaru} telescope. Five
continuous cross-shaped Suprime-Cam pointings were required to cover
the SXDS field. The \textit{Subaru} imaging data used in our study
comprised the broadband filters \textit{B}, \textit{V},
\textit{R}$_\mathrm{c}$, \textit{i}$'$ and \textit{z}$'$. We
downloaded the five tile images of each one of the 5 filters for SXDS
data release 1. Then, we performed source detection and photometry 
using \textsc{sextractor} with typical methods.

The SXDS field was also observed with the \textit{United Kingdom
  infrared} telescope wide field camera (\textit{UKIRT} WFCAM;
\citealt{cas07}) for the ultra deep survey (UDS; Almani et al. in
preparation) which is the deepest constituent of the \textit{UKIRT}
infrared deep sky survey (UKIDSS; \citealt{law07}). The area of the
UDS field was covered using the nominal observing mode. The UDS data
used in our study encompass imaging in the broadband filters
\textit{J}, \textit{H} and \textit{K}. We downloaded the four tiles of
object and confidence images for these 3 filters from WFCAM science
archive in its data release 8. We carried out source extraction an 
photometry in each filter tile using \textsc{sextractor} with standard 
procedures.

Various surveys have done a spectroscopic follow-up in
the SXDS/UDS field. Multi-object spectra were taken by \citet{gea07}
using the low-dispersion survey spectrograph (LDSS2) on the
\textit{Magellan} telescope. They were studying galaxies belonging to
groups and clusters at $z\sim0.5$. Spectra of galaxies of a cluster
candidate at $z\sim1.4$ were obtained by \citet{van07} using the deep
imaging multi-object spectrograph (DEIMOS) on the \textit{Keck-2}
telescope. Faint quasi-stellar objects at $1.57<z<3.29$ were followed
by \citet{sma08} using the AAOmega spectrograph on the
\textit{Anglo-Australian} telescope. X-ray sources were also followed
using the faint object camera spectrograph (FOCAS) on the
\textit{Subaru} telescope by \citet{aki15}, who also observed the
bright sources in the SXDS region with the 2-degree field (2dF)
spectrograph, obtaining redshifts in the range $0.1 < z<4.0$. The UDSz
program (PI O. Almaini; see also \citealt{mcl13,bra13}) has targeted
$\sim3500$ NIR-selected galaxies using the focal reducer and
low-dispersion spectrograph (FORS2) or the visible multi-object
spectrograph (VIMOS), obtaining 1512 secure redshifts in the range
$0<z<4.8$.

\section{Far infrared and merged catalogues}\label{sec:fimeca}

In this Section, we present the prior-based FIR photometric catalogues
and the band-merging technique used to characterize our
sources through the analysis of UV-to-FIR SEDs.

The FIR photometric extraction follows the method described in
\citet{per10}; see also \citet{raw16} and \citet{rod19}. 
The UV-to-FIR spectro-photometric merged catalogue of the
SXDS/UDS field was built within the \textsc{rainbow} cosmological
database\footnote{\url{https://arcoirix.cab.inta-csic.es/Rainbow_navigator_public/}} scheme
\citep{per08,bar11a}. We briefly describe these methods in the
following 2 Subsections.

\subsection{Prior-based FIR cataloguing}\label{sec:fircat}

Our method to detect sources in the MIPS-70, PACS, and SPIRE data
combines high-confidence direct-detection positions (see Section
\ref{sec:spda}) with prior-based positions.

First, we cross--correlated the MIPS-70 direct-detection catalogue with
the MIPS-24 one in a 3-arcsec radius. We compared the WCS of both 
catalogues, getting an rms accuracy smaller than 1.4~arcsec. 
The PACS, and SPIRE-250 mosaics were aligned to MIPS-24 astrometry
using an overall X-Y offset based on the positions of the brightest
(few tens) \textit{Herschel} point sources, obtaining a WCS rms accuracy
smaller than 0.9, 1.3, and 3~arcsec between the MIPS-24, and
PACS-100/160, and SPIRE-250 mosaics. The SPIRE-350 and -500 images
were aligned to the WCS of the SPIRE-250 and -350 mosaics, getting WCS
alignments with rms smaller than 5 and 7~arcsec between the SPIRE 350
and 250, and the 500 and 350-$\micron$ maps, respectively. Thus, the
overall WCS rms accuracy obtained for each FIR band is smaller or similar
than half of its pixel size.

Source detection and photometry in the MIPS-70 and \textit{Herschel}
bands depend on the prior catalogues obtained using the higher-spatial
resolution IRAC and MIPS-24 mosaics, and on direct detections on each
FIR band. The different PSF sizes of the IRAC and MIPS-24 filters
compared to the MIPS-70 and \textit{Herschel} bands implied that we
may have several prior sources merged in one single source in the
redder bands, i.e., we assumed that sources that can be deblended in 
longer wavelength data must be separated by a minimum distance. We 
did some testing to find the optimal distance for each FIR band. In 
order to cope with this, 2 procedures were implemented. 
First, the IRAC and MIPS-24 prior catalogues were cut to
a $5\sigma$ threshold to avoid spurious identifications in the FIR
catalogues. Then, we purged the input prior list by substituting close
(within distances smaller than the MIPS-70 and \textit{Herschel} PSF
FWHM) groups of prior sources with a single pseudo--source located in
the position of the brightest source of that group. The adoption of a
position for a pseudo--source is not crucial, because small centring 
offsets are permitted within the concurrent PSF fitting method.

For the MIPS-70 map, we grouped prior positions within 3/4 of the
MIPS-70 FWHM diameters of each other. We combined the purged list of
prior sources with the $\geq5\sigma$ direct-detection MIPS-70 list and
applied the PSF fitting technique as described in Section
\ref{sec:spda}.

For the PACS maps, we combined positions obtained by direct detections
from several passes in the respective mosaic, with those of the prior
sources inside each PACS map. We dealt with the different PSF sizes of
the IRAC and MIPS-24 bands, and the PACS channels by grouping objects
within 5 and 7~arcsec diameters of a given candidate for the 100 and
160-$\micron$ filters, keeping only a direct detection or a
pseudo--source. We fitted empirical PSFs to the purged candidate list
of each band as explained in Section~\ref{sec:heda}, tolerating small
concomitant recentring offsets.

For the SPIRE-250 map, the same procedure used for PACS mosaics was
followed, changing the prior catalogue grouping diameter to
12~arcsec. For the SPIRE-350 and 500-$\micron$ mosaics, we used as
prior 250 and 350-$\micron$ source positions, respectively. For these
two redder channels, the grouping diameters were 12 and 17~arcsec, but
we gave preference to prior positions compared to direct-detection
ones.  We employed the PSF fitting technique fed by the purged list of
candidates of the 3 channels as described in Section \ref{sec:heda}.
We did not allow recentring in the \textit{phot+allstar} tasks for
the 350 and 500-$\micron$ bands. Then, the final detections in the 350
and 500-$\micron$ bands are based on SPIRE-250 positions.

\subsection{Spectro-photometric merged catalogues}\label{sec:merca}

Our main goal is studying the relation between the stellar and dust
emission of starburst galaxies, which requires UV-optical-NIR SEDs for
each FIR source. In this work, we have focused on the common $\sim0.7$~deg$^2$ 
surveyed by \textit{Spitzer}, \textit{Herschel},
\textit{Subaru} and \textit{UKIRT} telescopes, selecting galaxies in
the FIR and looking for counterparts in the other datasets. Our
selection includes objects with at least 2 MIPS-70 and/or \textit{Herschel}
fluxes above a $4\sigma$ threshold and a $5\sigma$ MIPS-24 detection, which acts
as a bridge between the wide-beam FIR bands and the higher spatial
resolution UV/optical/NIR filters.

The \textsc{rainbow} code matches the source coordinates from a master
selection catalogue (24~$\micron$ in our case) to a referential
optical band (\textit{R}$_\mathrm{c}$ in our study). This reference
helps to diminish the radius of the subsequent cross--correlations with
the other photometric bands and the spectroscopic catalogue. A
2.5-arcsec radius was used to match the positions of the 24-$\micron$
source list and the \textit{R}$_\mathrm{c}$ objects. Within this
radius, multiple optical identifications were found for several sources
(20 per cent of the total number of MIPS-24 emitters). In such
case, all optical objects within this radius were kept as possible
counterpart for the 24-$\micron$ source. Starting from the 
\textit{R}$_\mathrm{c}$ coordinates, aperture-matched photometry
was carried out in the available UV, optical, NIR, and IRAC bands, 
and we also looked for a spectroscopic redshift using a 0.8-arcsec 
search radius. Photometric
uncertainties from optical to IRAC images are computed, concurrently
to the measurement of flux density, using the 3 estimates method
mentioned in Section \ref{sec:spda} (for a full description of the
\textsc{rainbow} procedures, see \citealt{per08,bar11a}).

Considering the different PSF sizes for the MIPS-70 and
\textit{Herschel} channels, distinct radii were used to cross--match the
referential identifications with the FIR bands (4.0, 2.8, 3.5, 6.0,
9.0 and 12.0~arcsec for MIPS $70~\micron$, PACS 100 and $160~\micron$,
SPIRE 250, 350 and $500~\micron$). For the MIPS and \textit{Herschel}
channels, the integrated flux was assumed to be that determined from the
PSF fitting and aperture correction (see Sections \ref{sec:spda} and
\ref{sec:heda}). In the case of multiple referential identifications,
i.e., multiple sources in the optical/NIR bands matching the 
same FIR detection, we have assigned to these multiple identifications
only one FIR counterpart, their corresponding match from each one of the MIPS 
and \textit{Herschel} channels.

It is interesting to evaluate how many 24-$\micron$ sources are
associated to a given $4\sigma$ source detected at longer wavelengths. 
On average, one MIPS-70, PACS or SPIRE-250 source can be identified 
with $\sim1.3$ MIPS-24 objects. In particular, a significant fraction of $\sim80$ 
per cent of MIPS-70, PACS or SPIRE-250 sources have only one MIPS-24 counterpart,
$\sim10$ per cent have 2, and $\sim10$ per cent have more than 2. For 350 and 500-$\micron$
sources the average number of MIPS-24 possible counterparts rises to $\sim2$.

We might also consider how many optical counterparts are associated to
the $4\sigma$-FIR identifications for the multi-wavelength catalogue.
The fraction of MIPS-70, PACS-100/160 objects with 1 and 2 optical
counterpart(s) is $\sim70$ and 20 per cent, and the fraction of sources with
more than 2 associations is $\sim10$ per cent in both cases. A fraction of $\sim60$ 
per cent of SPIRE-250 objects have only one optical counterpart, $\sim35$ per cent
have 2-3 optical matches, and $\sim5$ per cent have more than 3. A fraction of 
$\sim80$ per cent of SPIRE-350 sources have 1-3 optical counterparts, and $\sim20$ 
per cent have 4-7 optical matches. A fraction of $\sim70$ per cent of 500-$\micron$ 
objects have 1-3 optical counterparts, and $\sim30$ per cent have 4-8 optical matches.

Thus, we have selected a sample of 2999 sources detected in 2
or more FIR bands (apart from the MIPS-24 detection) at the
${4\sigma}$ level (at least) in an area of 0.7~deg$^2$. 
These sources are identified with 3925 optical objects satisfying the limiting 
magnitude criteria of Table \ref{tab:sxdsdc}, and the average number of optical 
counterparts for the FIR sources is $\sim1.3$. We note, however, that 
$\sim80$ per cent of the MIPS-24 galaxies can be identified unambiguously with 
an optical source. Since MIPS-70 and \textit{Herschel} data present worse spatial
resolutions, an uncertainty remains on the assignment of FIR fluxes to
the correct (or most probable) UV/optical/NIR/MIR (and MIPS-24)
counterpart. To cope with such uncertainty, it is compelling to assess how many of 
these $4\sigma$-threshold FIR sources present a PACS detection. As mentioned above, $\sim80$
per cent of the PACS sources exhibit a single MIPS-24 association and only one 
optical counterpart. There are  2536 sources detected in 2 or more FIR filters including 
a PACS channel, which are associated to 3245 optical objects.

\section{Selection of a sample of dusty starbursts at $\lowercase{\mathbfit{z}\mathbf{\sim 1}}$}\label{sec:sase}

In this Section, we describe how we built a sample of
\textit{bona fide} dusty starburst galaxies when the span of the peak 
of the cosmic dusty SFR density ends ($z\sim 1$, \citealt{cas12,bri17}).
To build such a sample, we need reliable (spectroscopic or
photometric) redshifts, stellar masses, and \textit{SFRs} for the IR
emitters described in the previous Section. In order to have reliable
\textit{SFR} estimations for our galaxies, we considered a \textit{main sample}
composed by sources with a MIPS-24 detection and measured
high-confidence fluxes in at least 2 more FIR bands ($\lambda_\mathrm{obs}\geq 70~\micron$)
including a PACS detection, for which accurate total IR luminosities could be estimated. 
We added a \textit{complementary sample} with sources only detected by MIPS
at 24~$\micron$ (some also detected in 1 FIR filter) in order to maximize 
the completeness of our final sample, although admitting the lower reliability 
of their IR-based \textit{SFR} estimations. 
Focusing on the main sample (see Section \ref{sec:sbsel}), and considering
that the space density of ULIRGs decreases 2 orders of magnitude from $z\sim 2$ to 
$z\sim 0.6$ \citep{dud20}, we decided to centre our analysis on the redshift
range $0.7\leq z\leq 1.2$.

In the following Sections, we describe how we estimated photometric
redshifts, stellar masses, and \textit{SFRs} for the IR sources presented in
Section \ref{sec:fimeca}. Then, we select starburst galaxies, defined as 
those sources which are significantly and reliably above the MS
of galaxies in the \textit{SFR} vs. $M_{\star}$ plane. Finally, we give
statistical properties of our final sample of galaxies (adding up the
main and complementary samples).

\subsection{Estimations of photometric redshifts, stellar masses and \textit{SFRs}}
\label{sec:zphot}

We estimated photometric redshifts ($z_\mathrm{phot}$) using the
\textsc{eazy} code \citep{bra08}, and the SEDs described in the previous Section for
the parent sample formed by the 3925 MIPS-24 emitters which are
covered by the IRAC, WFCAM, and Suprime-Cam surveys.
In our SED analysis, we only considered fluxes measured at the $\geq4\sigma$ level.

In order to calibrate our photometric redshifts, we used spectroscopic
redshifts, which were available for about one third of the parent
sample, virtually all ($\sim90$ per cent) at $z<1.3$, and $\sim 30$ per cent
at $0.7 \leq z \leq 1.2$, our range of interest. The photometric redshift
accuracy was quantified using the normalized median absolute deviation
of $\Delta z=z_\mathrm{phot}-z_\mathrm{spec}$ ($\sigma_\mathrm{NMAD}$,
\citealt{ilb06}). The scatter of the whole spectroscopic sample is
$\sigma_\mathrm{NMAD}=0.030$, and the median of the common used
estimator results $|\Delta z|/(1+z_\mathrm{spec})=0.020$. Defining
$\eta$ as the fraction of outliers with $|\Delta
z|/(1+z_\mathrm{spec})>0.2$, we get $\eta=12$ per cent. For sources at 
$0.7 \leq z \leq 1.2$, the values are $\sigma^{z}_\mathrm{NMAD}=0.028$, and 
$\eta^{z}=10$ per cent.

Our $\sigma^{z}_\mathrm{NMAD}$ value is similar to that ($\sigma^\mathrm{B}_\mathrm{
  NMAD}=0.025$) estimated for each subsample of MIPS-24 emitters 
derived from WFC3-F160W-selected objects in the GOODS-S, UDS and EGS 
fields \citep{bar19}, but obtaining a higher $\eta$ value (10 compared 
with 4 per cent). Our statistics for the reliability of photometric 
redshifts are also similar to those found by other authors focused in the
analysis of FIR sources (see, e.g., \citealt{ber11,cas12}).

Using these photometric or spectroscopic (when available) redshifts, 
we estimated stellar masses and \textit{SFRs}. Previous to such estimations, 
we excluded from our work all X-ray
emitters (using the catalogue from \citealt{aki15}), which account for 4 per cent
of the parent sample. Another 4 per cent of sources were identified as
obscured active galactic nuclei (AGN) candidates without X-ray counterpart using the criteria
in \citet{don12}. Thus, the total fraction of AGN candidates is
$\sim8$ per cent, which is similar with that obtained from other
\textit{Herschel}-based studies (e.g., a 7 per cent is found in \citealt{gru13}).

Stellar masses were derived from UV, optical
and NIR data by selecting the best-fitting template of a reference set
constructed using one and two stellar populations (see
\citealt{per08}).

Concerning total \textit{SFRs}, we calculated $SFR_\mathrm{UV+IR}$ for the parent sample of galaxies using
the recipe found in \citet{bel05}, which considers the unobscured UV
luminosity jointly with UV photons absorbed by dust and re-emitted in the
IR (see \citealt{ken98}).  The equation providing \textit{SFRs} from UV and total
IR luminosities --assuming a \citet{sal55} initial mass function
(IMF)-- is:
\begin{equation}\label{eq:sfruia}
SFR_{\rm UV+IR}(\mathrm{M}_{\sun}~{\mathrm{yr}^{-1}})=1.8 \times 10^{-10}[3.3L(0.28)+L_\mathrm{TIR}]/\mathrm{L}_{\sun},
\end{equation}
where $L(0.28)=\nu L_\nu(0.28)$ is the monochromatic luminosity at
$0.28~\micron$ rest frame, obtained by interpolating in the
best-fitting stellar population model used for the $M_\star$ estimation.

The total IR luminosities, $L_\mathrm{TIR}$, are derived from
\textit{Spitzer} and \textit{Herschel} data by computing the median
value of the luminosities provided by the best-fitting templates
extracted from the the libraries of dust emission models given in
\citet{cha01}, \citet{dal02}, and \citet{rie09} (hereafter CE01, DH02,
and R+09). When several IR fluxes are available (MIPS-24 plus
other PACS and/or SPIRE bands), all libraries provide similar results,
with differences lower than 0.15~dex and a typical rms scatter
around 0.05~dex. When only the 24-$\micron$ measurement is available,
the estimations are more uncertain, since a large extrapolation is
needed to translate from a single monochromatic luminosity in the
rest-frame MIR to the $L_\mathrm{TIR}$\, which is dominated by the
dust-peak emission, typically located at 50 to 200$~\micron$. In order to
account for this issue, we used the method described in \citet{ruj13},
especially designed to cope with the lack of FIR data for MIPS-24
emitters, to get total IR luminosities for MIPS-only sources,
$L_\mathrm{TIR}^\mathrm{R+13}$. We compared the
$L_\mathrm{TIR}^\mathrm{R+13}$ values with $L_\mathrm{TIR}$
estimations for \textit{Herschel} sources. This comparison showed 
a $\sim0.2$-dex scatter between both estimates. 
Using the $L_\mathrm{TIR}^\mathrm{R+13}$ and the 3 IR model libraries,
we computed the expected fluxes for the undetected FIR bands. We only
kept in our sample those galaxies whose synthetic fluxes are
compatible with a non-detection in the FIR channels (compared to
$5\sigma$-detection thresholds). For complementary sources,
we assigned the 0.2-dex scatter as uncertainty for each FIR synthetic flux.

\subsection{Selection of dusty massive starburst galaxies and
  construction of the final sample}\label{sec:sbsel}

\begin{figure}
\centering
\includegraphics[width=\columnwidth]{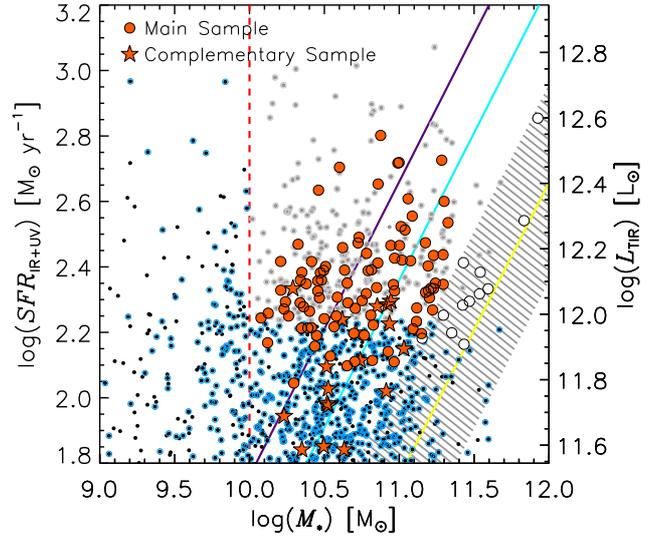}
\caption{Stellar mass, $M_\star$, vs. total \textit{SFR}, $SFR_\mathrm{IR+UV}$,
  plot for the galaxies in our sample with 24-$\micron$ counterpart 
  \textit{(filled black circles)} and with PACS association \textit{(open blue circles)} at
  $0.7 \leq z \leq 1.2$. We display the relation for the main sequence
  at $z\sim1$ of \citet[\textit{solid yellow line}]{elb07} with its
  $68$\% confidence level \textit{(hashed grey area)}. 
  The \textit{filled grey circles} depict values for the 237 dSFGs satisfying the $L_\mathrm{TIR}$
  and $M_{\star}$ criteria. The \textit{dashed red line} marks the $M_{\star}$ lower limit.
  The \textit{filled orange circles} show values for the IR-bright starburst of the main
  sample (with $SFR_\mathrm{IR+UV}$ above the MS+$1\sigma$ scatter).
  The \textit{filled orange stars} show values for dusty starbursts in the complementary
  sample. The \textit{open black circles} show
  IR-bright objects inside the MS$\pm 1\sigma$ area, which are excluded from
  this study.  The \textit{solid cyan} and \textit{solid purple line}
  depict the aforenamed MS plus $2\sigma$ and $3\sigma$ scatter,
  respectively.
  \label{fig:mssfrse}}
\end{figure}

From now on, we restrict our work to galaxies in the redshift range
$0.7 \leq z \leq 1.2$. The choice of the redshift limits is based on 2
points: (1) We are interested in studying IR-bright starbursts 
when the span of the peak of the cosmic dusty \textit{SFR} density ends.
We define such starbursts as dSFGs whose position in the \textit{SFR} vs. 
$M_{\star}$ plane is at least $1\sigma$ above the MS (see below).
(2) We want to maximize the number of sources in our sample, especially 
those for which accurate total IR luminosities can be obtained (so we 
are certain about their dusty starburst nature), which implies detections
in several FIR bands. We are also interested in assessing the duration
of the starburst phase of powerful sources in the galaxy evolution 
scheme, so we cut the sample at $L_\mathrm{TIR} \goa 
10^{12}~\mathrm{L}_{\sun}$. We note that PACS detections account for 91, 81, 
and 54 per cent at $0.7 \leq z < 1.0$, $1.0 \leq z < 1.2$, 
and $1.2 \leq z < 1.3$. Thus, we imposed $z=1.2$ as our upper limit in 
redshift, ensuring that most of our dSFGs
present a PACS detection. The $L_{\rm TIR}$ cut roughly translates to
$\log(M_\star/\mathrm{M}_{\sun})>10$ at $z\sim1$ (see discussion below). We note
that galaxies at $z \gtrsim 1$ with masses $\log(M_\star/\mathrm{M}_{\sun}) < 10$ are
scarcely detected in the PACS data of the SXDS/UDS field (see Fig. \ref{fig:mssfrse}). 
Thus, we impose this mass as lower limit for our selection. Concerning the
lower redshift limit, it is basically ruled by the low space density of ULIRGs
at $z\sim 0.6$.

\begin{figure*}
\centering
\subfloat{\includegraphics[width=0.99\textwidth]{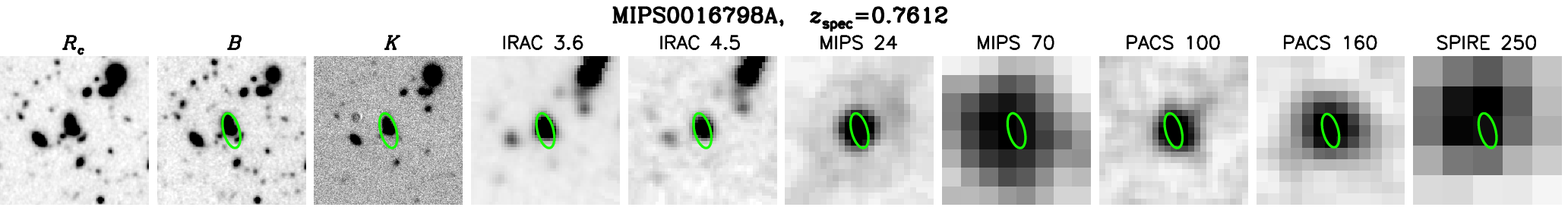}}\hfill
\subfloat{\includegraphics[width=0.99\textwidth]{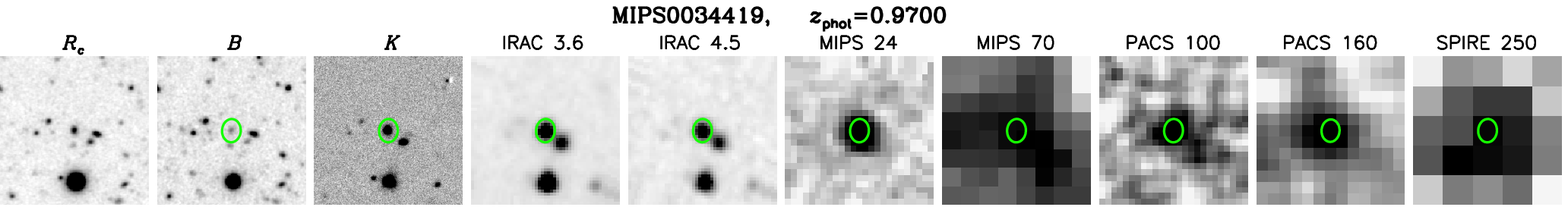}}\hfill
\subfloat{\includegraphics[width=0.99\textwidth]{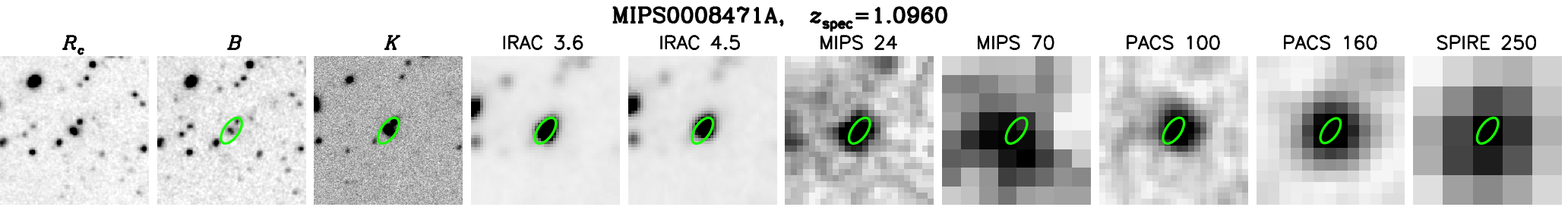}}\\
\caption{Thumbnails of 3 dusty starburst galaxies from our main sample
  ordered by increasing redshift. From left to right, we show
  30 per 30 arcsec postage stamps in the
  \textit{Subaru}/Suprime-Cam \textit{R}$_\mathrm{c}$ and \textit{B},
  and \textit{UKIRT}/WFCAM \textit{K} bands, \textit{Spitzer}/IRAC 3.6
  and $4.5~\micron$ and MIPS channels, and
  \textit{Herschel}/PACS and SPIRE 250$~\micron$ bands. The green
  ellipse shows the aperture best enclosing each one of the optical/NIR
  identifications with the less possible contamination from other objects. 
  The first and third row show examples of sources where several deblended 
  counterparts in the optical were merged based on their proximity and similar redshift. 
  In the upper part of each thumbnail we present the ID of the source, and
  the photometric or spectroscopic redshift. The thumbnails of the
  rest of the main sample and the complementary sample can be found as
  online material.\label{fig:thuna}}
\end{figure*}

We visually checked the MIPS, IRAC, Suprime-Cam, WFCAM, PACS and SPIRE
images of the 237 dSFGs satisfying the aforementioned redshift, 
$L_\mathrm{TIR}$ and stellar-mass criteria, which conform our starburst
candidates. 
As mentioned in Section \ref{sec:merca}, the MIPS-24
detections connect the wide-beam FIR channels with the
higher-resolution UV/optical/NIR bands. Thus, we used the 24-$\micron$
detections to constrain the identification at other wavelengths (to
the red and the blue). During this visual inspection
process, we removed SPIRE fluxes from the SEDs when we noticed
significant emission from near MIPS-24 neighbour(s) of the main 24-$\micron$
counterpart. The presence of such bright MIPS-24 neighbours can 
produce unphysical jumps in the FIR SED (see `clean index', \citealt{elb11}).
 A fraction of 24 per cent of the 237 dSFGs have MIPS-24 neighbours causing 
 unreliable photometry in the FIR bands. 
Given the $\sim 6$ arcsec angular resolution of MIPS-24, blending of 
several optical sources within a MIPS-24 resolution element is possible. Then,
an unambiguous optical counterpart cannot always be ascertained. Hence, we examined 
whether the position of the optical source was centred at the location of its 
MIPS-24 counterpart, and if there was any close optical neighbour possibly
emitting at $24~\micron$. If such neighbour was at similar $z_\mathrm{spec}$ 
or $z_\mathrm{phot}$ (implying that they are probably interacting), we merged the 
sources into the same object, adding up the fluxes in each band and repeating the 
$z_\mathrm{phot}$ and $M_\star$ estimations (see, e.g., \citealt{cib19}). 
Another 14 per cent of the dSFGs presented
near optical companions (at different $z_\mathrm{spec}$ or $z_\mathrm{phot}$) indicating not
being the most probable counterpart. 
In particular, around one fifth and one
twelfth of the 147 trustworthy IR-bright dSFGs could be identified with 2 and 3 optical
counterparts of very similar redshift.

After this sample purging, we repeated the $L_\mathrm{TIR}$
estimations by fitting the reliable observed fluxes at wavelengths
$\geq 70~\micron$ to the IR libraries of CE01, DH02, and R+09. We 
assumed the best-fitting result as $L_\mathrm{TIR}^\mathrm{ref}$. We did
not use the MIPS-24 flux to avoid PAH absorptions and emissions
(probed by 24-$\micron$ band in the redshift range $0.7 \leq z \leq 1.2$) 
to affect the fits, and thus obtaining a better characterization of
the FIR SED peak, which dominates the $L_\mathrm{TIR}$ estimations.
In any case, we verified that similar $L_\mathrm{TIR}$ results (with
differences $<0.1$ dex) are derived when including the 24-$\micron$
flux in the fit. The $L_\mathrm{TIR}^\mathrm{ref}$ uncertainties
(hereafter $\sigma_{L_\mathrm{TIR}}$) are calculated by adding in
quadrature the dispersion between the $L_\mathrm{TIR}$ estimates
(for $\lambda_\mathrm{obs} \ge 70\micron$) of
the 3 IR libraries, and a constant value of 0.022 dex (equivalent to
5\% of the galaxy $L_\mathrm{TIR}$). This constant value is used to
consider uncertainties in the absolute calibration and the confusion
noise of the MIPS, PACS and SPIRE instruments.

In order to select dusty starburst galaxies at intermediate redshift,
we built a \textit{SFR} vs. stellar mass plot for the dSFGs drawn from the parent sample and 
lying at $0.7 \leq z \loa 1.2$, distinguishing the 105 reliable IR-bright dSFGs resulting
with $\log(L_\mathrm{TIR}/\mathrm{L}_{\sun}) \goa 11.8$ and $\log(M_\star/\mathrm{M}_{\sun}) \ge 10$.
This is shown in Fig.~\ref{fig:mssfrse}. 
Most of these IR-bright galaxies are
located well above the so-called $z\sim1$ MS of the $SFR-M_\star$
relation (see, e.g., \citealt{elb07,sch15}).

Based on Fig.~\ref{fig:mssfrse}, we selected 93 dusty starburst galaxies     
--detected by MIPS and \textit{Herschel}--
which lie above the MS at least at the $1\sigma$ level. This is what
we call the main sample of dusty starbursts. We added to our work 18
starburst galaxies detected by MIPS at 24~$\mu$m (some also detected in 1 FIR filter), 
constituting the complementary sample. In total, our final sample is composed by
111 sources. Postage stamps of some of the galaxies in our final
sample can be found in Fig.~\ref{fig:thuna}.

\subsection{Closing remarks of the sample}

Our final sample of massive dusty starbursts lies at $0.7\le z \loa 1.2$, it consists
of 111 galaxies, 93 in the main sample, and 18 in the complementary
sample.  There are $13$ ($\sim12$ per cent) galaxies with spectroscopic
redshifts, and the normalized median absolute deviation estimated
using only these values is $\sigma_\mathrm{NMAD}=0.012$. Our massive dusty
starbursts are detected in at least 10 optical to NIR, MIPS-24, and
(only for the main sample) 2$+$ FIR bands. They have median, and 16
and 84-percentile values of $M_\star=5.3^{+7.7}_{-3.0} \times
10^{10}~\mathrm{M}_{\sun}$, $z=1.09^{+0.06}_{-0.21}$, and
$L_\mathrm{TIR}^\mathrm{ref}=1.11^{+0.48}_{-0.38}\times
10^{12}~\mathrm{L}_{\sun}$.
\begin{table*}
\begin{center}
\caption[Dusty-starburst main-sample multi-band photometry]{Dusty-starburst main-sample multi-band photometry. (1) Name of the galaxy. (2,3) Right ascension and declination (J2000) in degrees. (4) Photometric or spectroscopic redshift ($z_\mathrm{spec}$ indicated by a $\dagger$). (33-39) Flux densities in MIPS 24, 70; PACS 100, 160; SPIRE 250, 350, 500 in mJy. (5-11,19-25) Observed magnitude in FUV, NUV; \textit{B}, \textit{V}, \textit{R}$_\mathrm{c}$, \textit{i}$'$, \textit{z}$'$; \textit{J}, \textit{H}, \textit{K}; IRAC 3.6-8.0 in the AB photometric system. (40-46) Associated uncertainties to MIPS 24, 70; PACS 100, 160; SPIRE 250, 350, 500 in mJy determined as described in Sections \ref{sec:spda} and \ref{sec:heda}. (12-18,26-32) Associated uncertainties to FUV, NUV; \textit{B}, \textit{V}, \textit{R}$_\mathrm{c}$, \textit{i}$'$, \textit{z}$'$; \textit{J}, \textit{H}, \textit{K}; IRAC 3.6-8.0 as obtained from the aperture matched cataloguing (see Section \ref{sec:merca}). ``$\cdots$'' indicate bands without detections or without reliable photometric measurements.

(This Table and that for the complementary sample are available in its entirety in the online journal.)}
\label{tab:umspho}
\begin{tabular}{lcccccccccc}
\hline \hline
Galaxy & $\alpha$ & $\delta$ & $z^\dagger$ & FUV & NUV & \textit{B} & \textit{V} & \textit{R}$_\mathrm{c}$ & \textit{i}$'$ & \textit{z}$'$ \\
 & & & & $\Delta$FUV & $\Delta$NUV & $\Delta$\textit{B} & $\Delta$\textit{V} & $\Delta$\textit{R}$_\mathrm{c}$ & $\Delta$\textit{i}$'$ & $\Delta$\textit{z}$'$ \\
 & & & & \textit{J} & \textit{H} & \textit{K} & [3.6] & [4.5]  & [5.8] & [8.0] \\
 & & & & $\Delta$\textit{J} & $\Delta$\textit{H} & $\Delta$\textit{K} & $\Delta$[3.6] & $\Delta$[4.5] & $\Delta$[5.8] & $\Delta$[8.0] \\
 & & & & \textit{S}$_{24}$ & \textit{S}$_{70}$ & \textit{S}$_{100}$ & \textit{S}$_{160}$ & \textit{S}$_{250}$ & \textit{S}$_{350}$ & \textit{S}$_{500}$ \\
 & & & & $\Delta$\textit{S}$_{24}$ & $\Delta$\textit{S}$_{70}$ & $\Delta$\textit{S}$_{100}$ & $\Delta$\textit{S}$_{160}$ & $\Delta$\textit{S}$_{250}$ & $\Delta$\textit{S}$_{350}$ & $\Delta$\textit{S}$_{500}$ \\
 (1) & (2) & (3) & (4) & (5) & (6) & (7) & (8) & (9) & (10) & (11) \\
 & & & & (12) & (13) & (14) & (15) & (16) & (17) & (18) \\
 & & & & (19) & (20) & (21) & (22) & (23) & (24) & (25) \\
 & & & & (26) & (27) & (28) & (29) & (30) & (31) & (32) \\
 & & & & (33) & (34) & (35) & (36) & (37) & (38) & (39) \\
 & & & & (40) & (41) & (42) & (43) & (44) & (45) & (46) \\
\hline
\vspace{3pt} MIPS0000648A & $34.586462$ & $-5.201155$ & $1.1900$ & $ \cdots$ & $ \cdots$ & $24.48$ & $24.19$ & $23.70$ & $23.13$ & $22.33$ \\
\vspace{3pt} & & & & $ \cdots$ & $ \cdots$ & $ 0.02$ & $ 0.02$ & $ 0.01$ & $ 0.01$ & $ 0.01$ \\
\vspace{3pt} & & & & $21.80$ & $21.35$ & $20.71$ & $20.08$ & $20.05$ & $20.26$ & $20.21$ \\
\vspace{3pt} & & & & $ 0.02$ & $ 0.02$ & $ 0.01$ & $ 0.04$ & $ 0.04$ & $ 0.05$& $ 0.06$ \\
\vspace{3pt} & & & & $0.20$ & $\cdots$ & $ 8.67$ & $ 23.02$ & $26.21$ & $16.33$ & $ \cdots$ \\
\vspace{3pt} & & & & $0.02$ & $\cdots$ & $ 0.69$ & $  1.84$ & $ 2.41$ & $ 3.33$ & $ \cdots$ \\
\vspace{3pt} MIPS0000826 & $34.894621$ & $-5.300254$ & $0.7200$ & $ \cdots$ & $ \cdots$ & $24.21$ & $23.29$ & $22.33$ & $21.62$ & $21.28$ \\
\vspace{3pt} & & & & $ \cdots$ & $ \cdots$ & $ 0.02$ & $ 0.01$ & $ 0.01$ & $ 0.01$ & $ 0.01$ \\
\vspace{3pt} & & & & $20.71$ & $20.24$ & $19.70$ & $19.30$ & $19.62$ & $19.38$ & $18.86$ \\
\vspace{3pt} & & & & $ 0.01$ & $ 0.02$ & $ 0.01$ & $ 0.04$ & $ 0.04$ & $ 0.04$& $ 0.05$ \\
\vspace{3pt} & & & & $2.09$ & $66.27$ & $90.70$ & $105.05$ & $65.59$ & $ \cdots$ & $ \cdots$ \\
\vspace{3pt} & & & & $0.04$ & $ 2.92$ & $ 7.08$ & $  8.40$ & $ 4.39$ & $ \cdots$ & $ \cdots$ \\
\vspace{3pt} MIPS0002196\_1 & $34.474978$ & $-5.038923$ & $1.0980^\dagger$ & $ \cdots$ & $ \cdots$ & $26.05$ & $25.37$ & $24.65$ & $23.66$ & $22.84$ \\
\vspace{3pt} & & & & $ \cdots$ & $ \cdots$ & $ 0.05$ & $ 0.04$ & $ 0.03$ & $ 0.02$ & $ 0.01$ \\
\vspace{3pt} & & & & $21.95$ & $21.35$ & $20.58$ & $19.85$ & $19.87$ & $20.12$ & $19.64$ \\
\vspace{3pt} & & & & $ 0.02$ & $ 0.02$ & $ 0.01$ & $ 0.03$ & $ 0.03$ & $ 0.04$& $ 0.06$ \\
\vspace{3pt} & & & & $0.67$ & $ \cdots$ & $15.20$ & $ 26.28$ & $28.00$ & $18.68$ & $ \cdots$ \\
\vspace{3pt} & & & & $0.02$ & $ \cdots$ & $ 1.25$ & $  2.26$ & $ 2.60$ & $ 3.53$ & $ \cdots$ \\
\hline
\end{tabular}
\end{center}
\end{table*}

The multi-band photometry catalogue, used in the rest of the paper,
for our final sample of IR-bright starburst galaxies in SXDS/UDS field
is presented in Table~\ref{tab:umspho}.

In the next Section, we describe our method to study the stellar
population properties of our final sample of massive dusty starburst galaxies 
at the peak span end of the cosmic dusty \textit{SFR} density.
Our results are presented and compared with the literature in 
Sections \ref{sec:FIRprior} and \ref{sec:stepro}.

\section{Self-consistent SED modelling}\label{sec:scm}

In order to derive stellar population properties of $0.7\le z\loa 1.2$
starburst ($\gtrsim 1\sigma$ above the $z=1$ MS) galaxies, their
UV-to-FIR rest-frame SEDs are modelled in a self-consistent way,
accounting for their dust content and the attenuation of the stellar
light produced by that dust. In our analysis we use the
\textsc{cigale} \citep{nol09} and \textsc{synthesizer}
\citep{per03,per08} codes, which cope with the auto-consistent
modelling of the stellar emission, and dust attenuation and re-emission
using energy balance techniques.
\begin{table*}
\caption[Input parameters of the \textsc{cigale} and their explored range]{Input parameters of the \textsc{cigale} and their explored range.}
\label{tab:cinpa}
\begin{center}
\begin{tabular}{lcl}
\hline \hline
Parameter & Symbol & Range \\
\hline
\vspace{3pt} Metallicity & \textit{Z} & $\mathrm{Z}_{\sun}=0.02$ \\
\vspace{3pt} \textit{e}-folding times of the old stellar population in Gyr & $\tau_\mathrm{old}$ & 0.1, 1, 3, 10 \\
\vspace{3pt} Ages of the old stellar population in Gyr & $t_\mathrm{old}$ & 3, 5, 7 for $0.7 < z \leq 0.8$ \\
\vspace{3pt} &  & 2, 4, 6 for $0.8 < z \leq 1.0$ \\
\vspace{3pt} &  & 1, 3, 5 for $1.0 < z \loa 1.2$ \\
\vspace{3pt} \textit{e}-folding time of the young stellar population in Gyr & $\tau_\mathrm{you}$ & 20 \\
\vspace{3pt} Ages of the young stellar population in Myr & $t_\mathrm{you}$ & 1, 10, 25, 50, 75, 100, 200, 300, 400, 500, 600, \\
\vspace{3pt}  &  & 700, 800, 900, 1000 \\
\vspace{3pt} Mass fraction of young population or burst intensity & $b_\mathrm{you}$ & 0, 0.001, 0.005, 0.01, 0.05, 0.06, 0.07, 0.08\\
\vspace{3pt}  &  & 0.09, 0.1, 0.2, 0.3, 0.4, 0.5, 0.6, 0.7 \\
\vspace{3pt} \textit{V}-band attenuation for the young population in mag & $A_{V,\mathrm{you}}$ & 0.15, 0.30, 0.45, 0.6, 0.75, 0.9, 1.05, 1.2, 1.35, \\
\vspace{3pt}  &  & 1.5, 1.65, 1.8, 1.95, 2.1, 2.25, 2.4, 2.55, 2.7, \\
\vspace{3pt}  &  & 2.85, 3.0, 3.15, 3.3, 3.45, 3.6, 3.75, 3.9, 4.05, 4.2 \\
\vspace{3pt} Reduction factor of $A_{V,\mathrm{you}}$ for $A_{V,\mathrm{old}}$ & $f_\mathrm{att}$ & 0, 0.25, 0.5, 0.75, 1 \\
\hline
\end{tabular}
\end{center}
\end{table*}
\begin{table*}
\caption[Input parameters of the \textsc{synthesizer} code and their explored range]{Input parameters of the \textsc{synthesizer} code and their explored range}
\label{tab:synpa}
\begin{center}
\begin{tabular}{lcl}
\hline \hline
Parameter & Symbol & Range \\
\hline
\vspace{3pt} Metallicity & \textit{Z} & $\mathrm{Z}_{\sun}=0.02$ \\
\vspace{3pt} \textit{e}-folding times of the old stellar population in Gyr & $\tau_\mathrm{old}$ & from 0.1 to 10 in logarithmic intevals of 0.5 dex \\
\vspace{3pt} Ages of the old stellar population in Gyr & $t_\mathrm{old}$ & 3, 4, 5, 6, 7 for $0.7 < z \leq 0.8$ \\
\vspace{3pt} &  & 2, 3, 4, 5, 6 for $0.8 < z \leq 1.0$ \\
\vspace{3pt} &  & 1, 1.5, 2, 3, 4, 5 for $1.0 < z \loa 1.2$ \\
\vspace{3pt} \textit{V}-band attenuation for the old population in mag & $A_{V,\mathrm{old}}$ & from 0 to 1.5 in increments of 0.1 \\
\vspace{3pt} \textit{e}-folding time of the young stellar population in Gyr & $\tau_\mathrm{you}$ & 16, 20 \\
\vspace{3pt} Ages of the young stellar population in Myr & $t_\mathrm{you}$ & 1, 1.5, 2, 2.5, 3, 3.5, 4, 4.5, 5, 5.5, 6, 6.5, 7, 7.5, \\
\vspace{3pt}  &  & 8, 8.5, 9, 9.5, 10, 15, 20, 25, 30, 35, 40, 45, 50, \\
\vspace{3pt}  &  & 55, 60, 65, 70, 75, 80, 85, 90, 95, 100, 200, 300, \\
\vspace{3pt}  &  & 400, 500, 600, 700, 800, 900, 1000 \\
\vspace{3pt} Mass fraction of young population or burst intensity & $b_\mathrm{you}$ & from 0.1 to 0.7 in increments of 0.05\\
\vspace{3pt} \textit{V}-band attenuation for the young population in mag & $A_{V,\mathrm{you}}$ & from 0 to 4.2 in increments of 0.1 \\
\hline
\end{tabular}
\end{center}
\end{table*}

Studying the stellar population properties of dSFGs, \textit{Spitzer} 
observations have shown that SMGs house old stellar populations in place at $z\sim 2$ 
\citep{bor05}. In addition, submillimetre data have evinced that these old 
populations are accompanied of dust-enshrouded young stellar populations in 
an ongoing burst \citep{dye08}. \citet{dun11} has shown that nothing more 
complex than a SFH depicted by an old and a young population is necessary to fit 
the observed SEDs of SMGs. \citet{bua14} indicate that two stellar population
(2P) models are the best fits for the observed UV-to-FIR SEDs of PACS-detected galaxies
at $1 < z < 3$. The young population accounts for most of the UV emission, and the old 
population might significantly contribute or even dominate the optical/NIR emission. 
Such 2P SFHs have been used in many papers focusing on the study of
star-forming galaxies, whose nature is not easily explained with more
classical and simple SFHs such as decaying exponential or even delayed 
forms \citep{gaw07,lee09,cas14,gey15,cie17}.
 More importantly for our work, 2P models have been proved to recover
accurately the stellar masses of simulated SMGs
\citep{mic14}, and the photometric masses derived with these 2P
models agree well with the dynamical masses of local starburst galaxies \citep{ber16}.

In our work, the two populations are represented by a recent burst
with roughly constant SFH overlapped to and old population
with decreasing SFH. In both cases, the star formation events
are implemented with decaying exponentials. Both populations are
linked with the so-called burst intensity, defined in terms of stellar masses
for the young stars and the entire galaxy
($b_\mathrm{you}=M_{\star,\mathrm{you}}/M_\star$). The SFHs are parametrized
with $\tau_\mathrm{old}$, $t_\mathrm{old}$, $\tau_\mathrm{you}$, and
$t_\mathrm{you}$, which stand for the decay factors and ages for the
old and young populations.

Both bursts are recreated using single stellar populations of
\citet{mar05a} models, which treat minutely the
thermally pulsating asymptotic giant branch stars emission, 
necessary for the determination of rest-frame NIR colors and luminosities of
stellar populations with ages $\sim 0.1-1$~Gyr \citep{ton09,lac16}. 
However, the contribution of these stars to the NIR luminosity is still
debated \citep{zib13}. We assume a \citet{sal55} IMF.

Solar metallicity, $\mathrm{Z}_{\sun}$, is adopted for both populations,
considering that is unfeasible to break the age-metallicity degeneracy
with broadband data \citep{oco86}, and that $z \goa 1$ dSFGs typically 
present metallicities near
this value \citep{swi04,war17,boo19} for the masses we probe in our work.
The effects of dust on the UV-optical light coming from both
populations are considered using the \citet[hereafter C+00]{cal00}
attenuation law, derived for nearby starburst galaxies. \citet{sal16}
studied a galaxy sample at $z \sim 1.5-3$ from the CANDELS survey with
MIPS-24 detections, using different assumptions about the attenuation
law. They found that highly attenuated stellar populations have a
starburst-like law. Therefore, we consider the assumption of the C+00
law adequate.

In our study, the amount of attenuation in the \textit{V} band of both
populations is constrained balancing the absorbed energy with the
luminosity re-emitted by dust at IR wavelengths (the
$L_\mathrm{TIR}$).  This is what we call the FIR prior, and it
represents the main characteristic of our auto-consistent approach. 
For \textsc{cigale}, the attenuation in \textit{V} of the young 
population ($A_V^\mathrm{you}$) is estimated first, then a reduction 
factor relative to this $A_V^\mathrm{you}$ is applied to get the attenuation 
for the old stars, 
$A_V^\mathrm{old}=A_V^\mathrm{you} f_\mathrm{att}$. For the \textsc{synthesizer} 
code $A_V^\mathrm{you}$ and $A_V^\mathrm{old}$ are estimated independently.

The \textsc{cigale} code estimates $L_\mathrm{TIR}$ using the DH02
templates (the default library in \textsc{cigale fortran} 2013/11/18),
which are linked to the attenuated SPS models. The full UV-to-FIR set
of models are compared with the observed galaxy photometry using a
$\chi^2$ minimization. All available data points with
$\lambda_\mathrm{obs}<550~\micron$ are included in our fits, except
MIPS-24 data (see Section \ref{sec:sbsel}). The full set of values of
the input parameters are listed in Table \ref{tab:cinpa}. Following
\citet{gio11}, we fixed $\tau_\mathrm{you}=20$~Gyr to consider
constant SFHs for the young population. \textsc{cigale} uses
a Bayesian-like analysis to derive galaxy properties. The input
parameters, and values as $M_\star$, and \textit{SFR} are derived from
their probability distribution function. The expected value and the
standard deviation for each parameter are determined as detailed in
the `sum' method of \citet{nol09}.

In the \textsc{synthesizer} code, the dust-absorbed UV-optical stellar
energy of the models is constrained according to the MIR-FIR energy
re-radiated by dust, our $L_\mathrm{TIR}^\mathrm{ref}$, i.e., using
the FIR prior. We included in the stellar population fits all
available data points for $\lambda_\mathrm{obs}\leq4.5~\micron$
because the integrated emission in this range should be dominated by
stars for the galaxies in our sample. Uncertainties in the derived
parameters and degeneracies in the solutions are analysed using a
Monte--Carlo algorithm, varying randomly each observed flux within a
Gaussian distribution of width equal to its corresponding photometric
error, and repeating the fit with all the feasible models 3000 times.
Then, clusters of solutions are identified using a \textit{k}-means
method \citep{per13}, assigning a statistical significance to each
cluster, defined as the fraction of the 3000 solutions, which belonged
to it. So we grouped solutions providing similar results and
calculated the median value and the scatter of each group in the
multi-dimensional space formed by the fitted parameters. We considered
the most significant cluster as the best solution.

The input parameters assumed for the \textsc{synthesizer} code are
listed in Table \ref{tab:synpa}. The \textsc{synthesizer} parameters
and fitting procedure are similar to those used by \textsc{cigale}.

\section{Importance of the FIR prior in the characterization of the
  stellar populations}\label{sec:FIRprior}

In this Section, we evaluate the utility of using the FIR prior in the
SED fitting. For that purpose, we compare the solutions obtained with 2P
models including and excluding the FIR constraint. Both sets of solutions are derived
with the \textsc{synthesizer} and \textsc{cigale} codes (see Section
\ref{sec:scm}). With this exercise, we are able to test the value of
employing the FIR prior to constrain the attenuation of the stellar
emission, and to quantify the impact of this constraint in the
estimation of stellar parameters of dusty starbursts. It is important
to mention that the trustworthiness and accuracy of the resulting
stellar properties will depend on the photometric uncertainties and
the degeneracies (e.g., age-dust, age-burst strength) between such
properties.

\begin{figure*}
\centering
\subfloat{\includegraphics[width=0.8\textwidth]{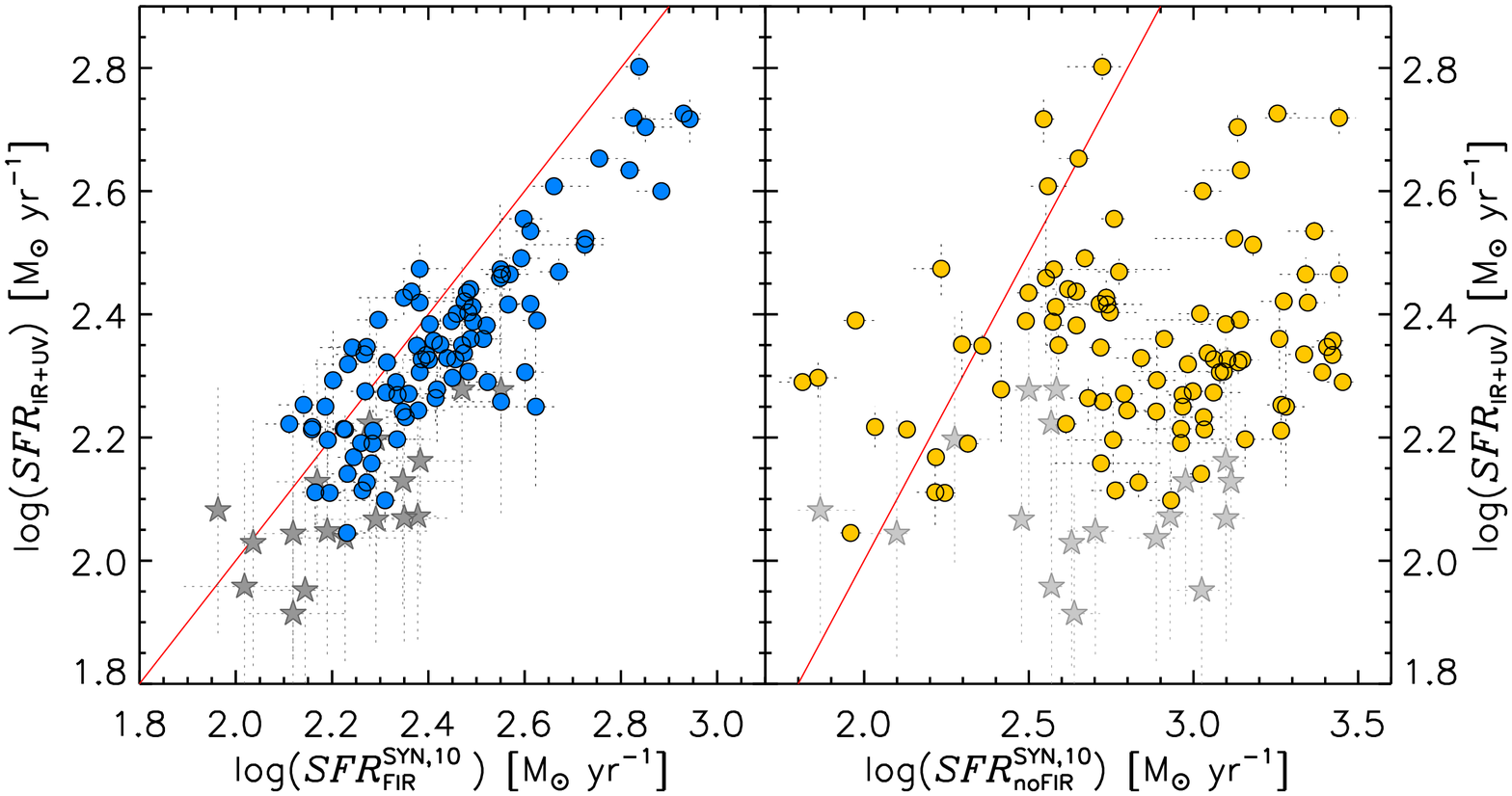}}\hfill
\subfloat{\includegraphics[width=0.8\textwidth]{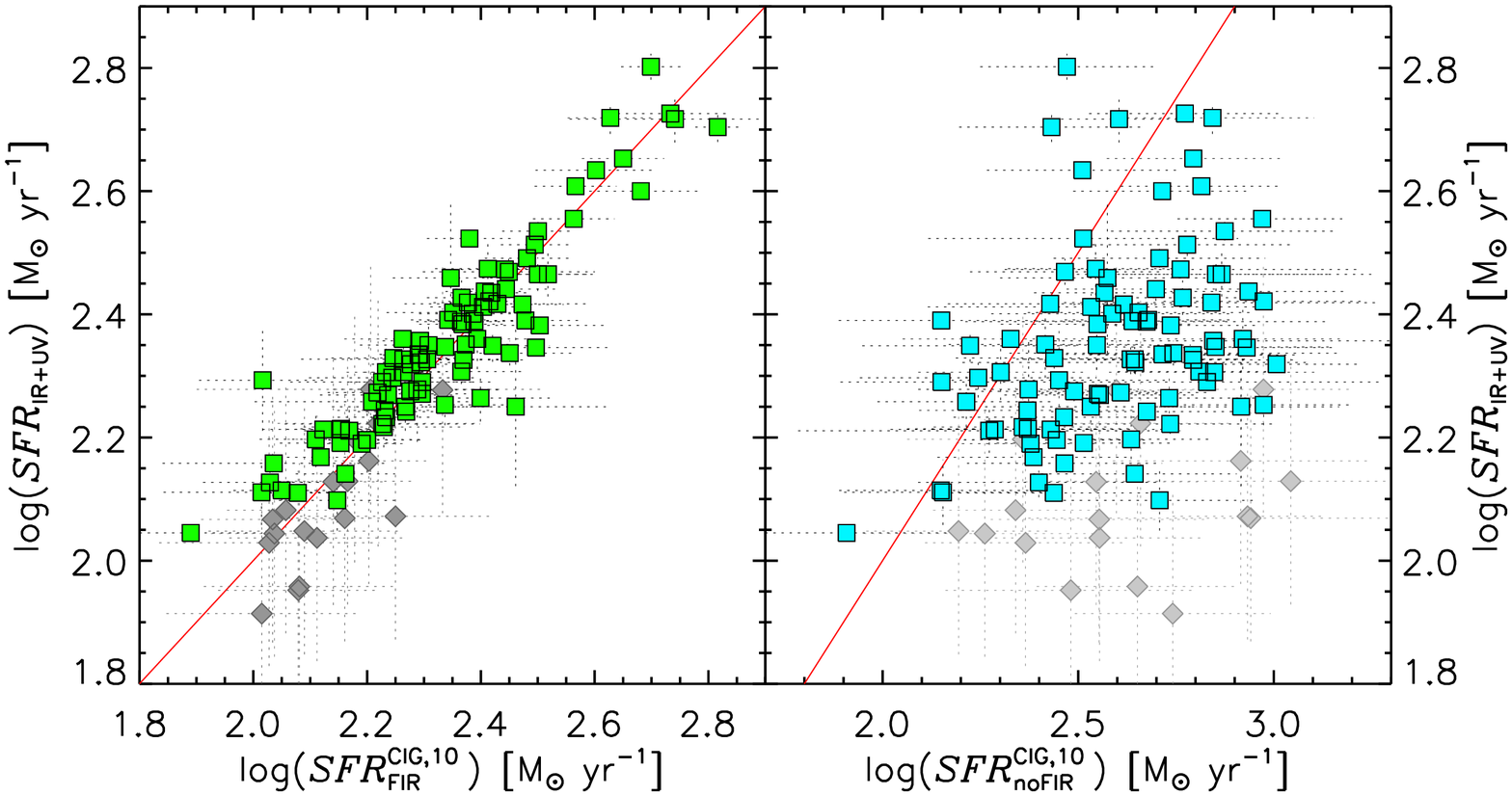}}\\
\caption{Comparison between the \textit{SFRs} determined with the
  observed FIR and UV and those derived with 2P models (averaged over
  the last 10 Myr) with \textit{(left)} and without \textit{(right)}
  the FIR prior from the \textsc{synthesizer} \textit{(top)} and
  \textsc{cigale} \textit{(bottom)} codes. The filled circles (filled
  stars) with error bars show the median values and the $1\sigma$
  uncertainties for the main (complementary) sample derived from the
  logarithmic space with the \textsc{synthesizer} code. The filled
  squares (filled diamonds) stand for the expected values
  and standard deviations determined with the \textsc{cigale}. The
  \textit{solid lines} show the one-to-one
  relation.\label{fig:sfro2pf2p}}
\end{figure*} 
\begin{figure*}
\centering
\subfloat{\includegraphics[width=0.5\textwidth]{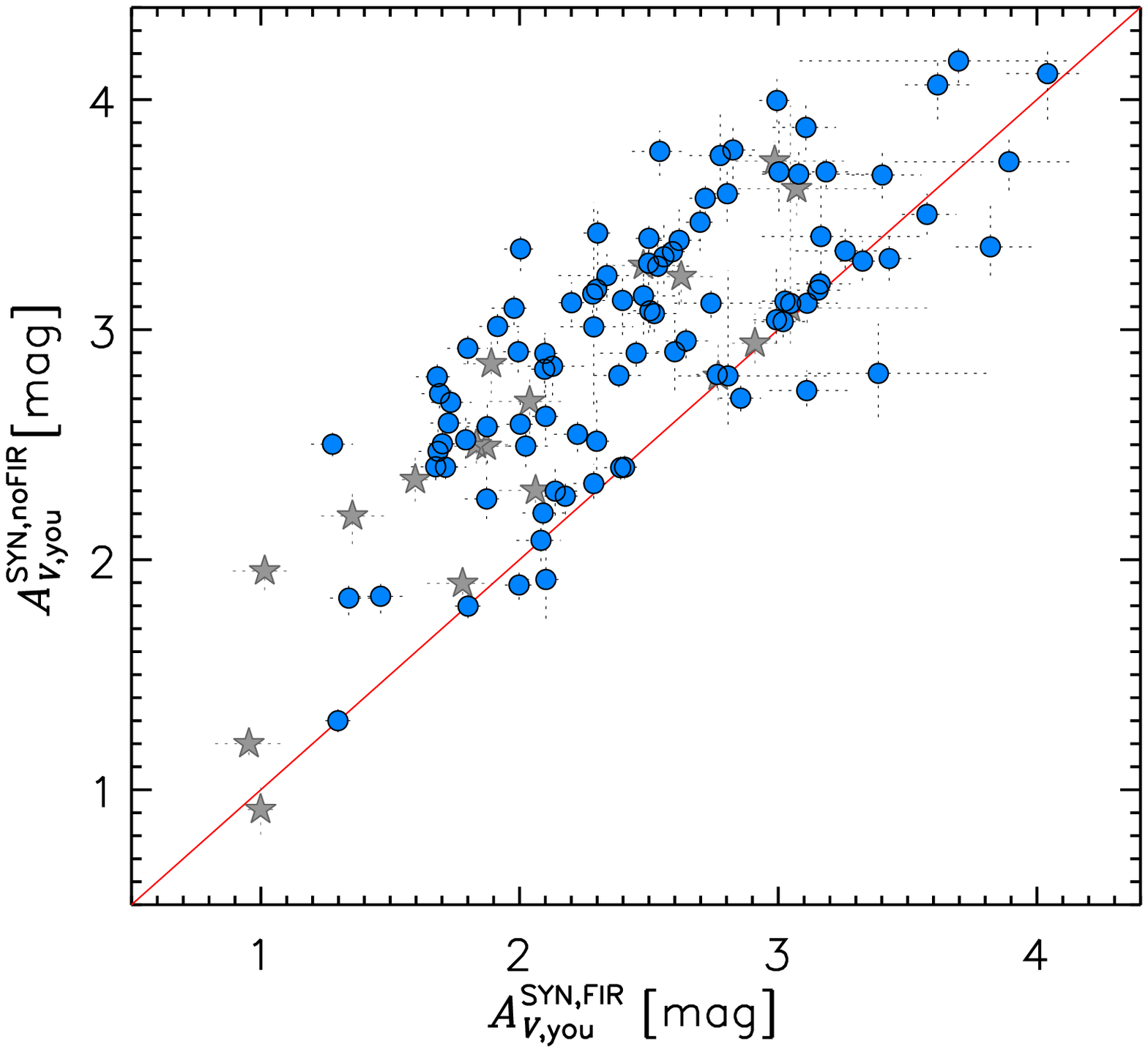}}\hfill
\subfloat{\includegraphics[width=0.5\textwidth]{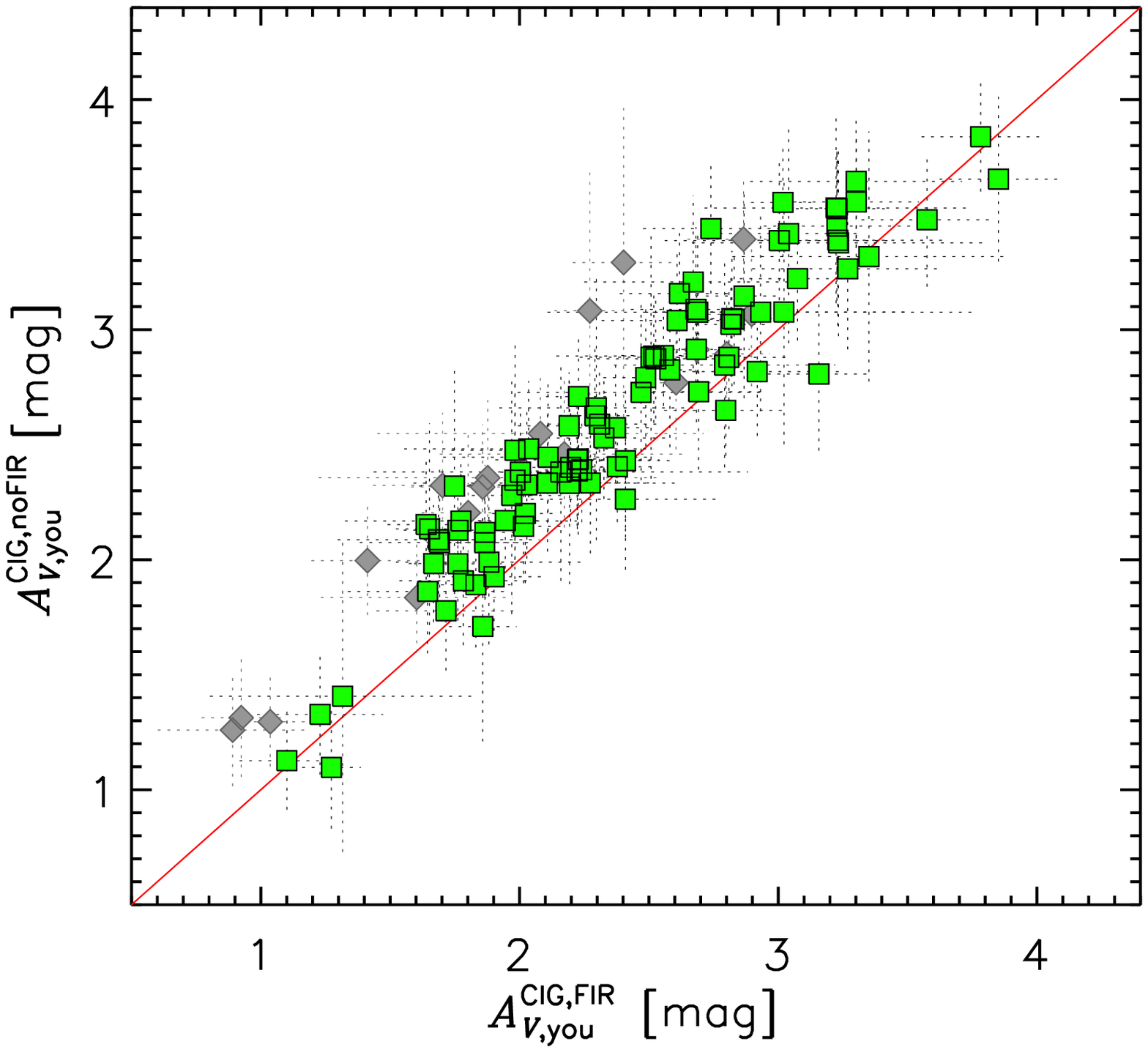}}\\
\caption{Comparison between the attenuations of the young population using the FIR prior on x-axis, and those derived without the FIR constraint on y-axis, estimated with the \textsc{synthesizer} \textit{(left panel)} and the \textsc{cigale} \textit{(right panel)} codes. The filled circles (filled stars) with error bars show the median values and $1\sigma$ uncertainties for the main (complementary) sample estimated with the \textsc{synthesizer} code. The filled squares (filled diamonds) with error bars stand for the expected values and standard deviations derived with the \textsc{cigale}. \label{fig:avy2pf2p}}
\end{figure*}

Considering that a starburst galaxy is defined by means of its stellar mass and \textit{SFR}, we compare first the values derived for these properties and then the results for attenuation and age for each modelling case and code.

\subsection{Effect of the FIR prior in the determination of stellar
  masses}\label{sec:mssfr2pc}

The \textsc{cigale} and \textsc{synthesizer} codes supply
mass-to-light ratios for all photometric bands used in the fits, which
can be compared to the measured fluxes and provide stellar mass
estimates (weighted with flux errors).

For the \textsc{synthesizer} code, the stellar masses estimated using
the FIR prior for the entire sample are in the range $5.8 \times
10^9<M_{\star}^\mathrm{SYN,FIR}/\mathrm{M}_{\sun}<1.5\times 10^{11}$
with a median of $3.8^{+3.4}_{-2.2}\times
10^{10}~\mathrm{M}_{\sun}$, and a median uncertainty of 0.07~dex. The stellar
masses obtained without the FIR data range between $6.0\times 10^9 <
M_{\star}^\mathrm{SYN,noFIR}/\mathrm{M}_{\sun}<1.6\times 10^{11}$, with
a median of $3.4^{+2.4}_{-1.3}\times 10^{10}~\mathrm{M}_{\sun}$,
and a median error of 0.06~dex. 
The median difference between the stellar masses estimated with FIR models,
and those derived with the non-FIR ones is 10 per cent. 
The stellar mass ratios range between $0.2\leq
M_{\star}^\mathrm{SYN,noFIR}/M_{\star}^\mathrm{SYN,FIR}<5.7$,
with a fraction of $\sim 14$ per cent of all objects showing mass differences
greater than a factor of 2. We find that the mass ratio increases as the 
difference in the attenuations for the old population, $A_{V,\mathrm{old}}^\mathrm{SYN,noFIR}
-A_{V,\mathrm{old}}^\mathrm{SYN,FIR}$, rises. This suggests that including
the FIR data in the fitting can help to diminish the uncertainty in 
the old-population attenuation, $A_{V,\mathrm{old}}^\mathrm{FIR}$,
which improves the determination of stellar masses of dusty starbursts (see, e.g., \citealt{dac15}).

For \textsc{cigale} and the whole sample, the stellar masses obtained
using the FIR prior are in the range $7.9 \times
10^9<M_{\star}^\mathrm{CIG,FIR}/\mathrm{M}_{\sun}<2.2\times 10^{11}$,
with a median of $5.2^{+4.8}_{-3.2}\times
10^{10}~\mathrm{M}_{\sun}$, and a median uncertainty of $0.14$~dex. The stellar
masses estimated without the FIR data range between $8.9 \times
10^9<M_{\star}^\mathrm{CIG,noFIR}/\mathrm{M}_{\sun}<1.9\times 10^{11}$,
with a median of $4.0^{+4.7}_{-2.3}\times
10^{10}~\mathrm{M}_{\sun}$, and a median error of $0.22$~dex. On median, the
stellar masses determined with FIR models are 15 per cent larger than
those derived with non-FIR models, 
which is consistent with what was found for \textsc{synthesizer}.
The stellar mass ratio values range in $0.2\leq
M_{\star}^\mathrm{CIG,noFIR}/M_{\star}^\mathrm{CIG,FIR}<1.7$,
with a fraction of $\sim 6$ per cent of all galaxies showing mass differences
greater than a factor of 2. We find that the mass ratio decreases as the 
difference in the burst intensity, $b^\mathrm{CIG,noFIR}_\mathrm{you}-
b^\mathrm{CIG,FIR}_\mathrm{you}$, increases. This indicates that using the
FIR prior aids to reduce the uncertainty in the determination of the  parameters
of the young stellar population (via better constraining the $SFR$ and 
$t^\mathrm{FIR}_\mathrm{you}$, see Sections \ref{sec:mssfr2pc2} and \ref{sec:yopo2p}),
which influences the stellar mass estimations.

It is noteworthy that both codes behave in the same way, providing
larger stellar masses on median when using the FIR prior.

Summarising the results obtained with both codes, although there
is not a significant difference between the stellar mass values derived 
with and without using the FIR prior, the FIR data seem to help to reduce the 
uncertainty in the determination of the parameters of the old and young
stellar populations. This hypothesis will be tested in the following 
Subsections.

\subsection{Effect of the FIR prior in the determination of
  \textit{SFRs}}
\label{sec:mssfr2pc2}

We compare now the \textit{SFRs} derived with the 2P models estimated
by \textsc{synthesizer} and \textsc{cigale} (with and without the FIR
prior) with \textit{SFRs} based on observables, namely the FIR and UV
emission ($SFR_\mathrm{UV+IR}$). In order to do a fair comparison
between $SFR_\mathrm{UV+IR}$ and SED-derived \textit{SFR} values, we
should consider that \citet{ken98} assumed a constant SFH with bursts
lasting $10-100$~Myr. Hence, our SED-derived \textit{SFR} values are
averaged over the last 10~Myr to put both \textit{SFR} estimates in a
similar time-scale. The comparison of \textit{SFRs} is plotted in Fig.
\ref{fig:sfro2pf2p}.

We estimated the contribution of the unobscured-UV luminosity to
$SFR_\mathrm{UV+IR}$ for our whole sample and it is less than 5 per cent on
average. Considering this minimal contribution, the main source of
uncertainty in the $SFR_\mathrm{UV+IR}$ values is due to our
$L^\mathrm{ref}_\mathrm{TIR}$ estimations, whose rms error is lower
than 0.13~dex for the main sample, and 0.2~dex for the complementary
one. For the full sample, the $SFR_\mathrm{UV+IR}$ values range in
$82 < SFR_\mathrm{UV+IR} < 635~\mathrm{M}_{\sun}\mathrm{yr}^{-1}$, 
with median of $203^{+89}_{-69}~\mathrm{M}_{\sun}\mathrm{yr}^{-1}$.

For our whole sample, the \textsc{synthesizer} 10-Myr-averaged
\textit{SFRs} derived using the FIR data range in
$91<SFR^\mathrm{SYN,10}_\mathrm{FIR}<878~\mathrm{M}_{\sun}\mathrm{yr}^{-1}$, 
with median of $241^{+155}_{-74}~\mathrm{M}_{\sun}\mathrm{yr}^{-1}$, and median
uncertainty of $\sim11$ per cent. The left panels of Fig.
\ref{fig:sfro2pf2p} show that the FIR prior succeeds to make the
SED-based \textit{SFRs} more similar to the observed \textit{SFRs},
although we detect an offset, which we attribute to the distinct
assumptions about the SFH and typical age of the recent burst. The median 
difference between the $SFR^\mathrm{SYN,10}_\mathrm{FIR}$ and 
$SFR_\mathrm{UV+IR}$ results is $\sim 25$ per cent. When no
FIR prior is used, the comparison of \textit{SFRs} is basically a
scatter plot.

The \textsc{synthesizer} 10-Myr-averaged \textit{SFRs} estimated
without the FIR prior range in $65 <
SFR^\mathrm{SYN,10}_\mathrm{noFIR} < 2838~\mathrm{M}_{\sun}\mathrm{yr}^{-1}$, 
with median of $681^{+836}_{-380}~\mathrm{M}_{\sun}\mathrm{yr}^{-1}$, and a median
error of 12 per cent. For 60 per cent of all galaxies, the \textit{SFRs} derived
without the FIR prior are 0.3~dex larger than those estimated using
the UV+FIR data. The median difference between the models without the
FIR prior and our fiducial models,
$\log(SFR^\mathrm{SYN,10}_\mathrm{noFIR})-\log(SFR^\mathrm{SYN,10}_\mathrm{FIR})$, is $0.48$~dex
(a factor $\sim3$), but differences for individual objects can
reach values larger than 1~dex.

For the whole sample, the \textsc{cigale} 10-Myr-averaged
\textit{SFRs} derived using the FIR prior range in
$77<SFR^\mathrm{CIG,10}_\mathrm{FIR}<656~\mathrm{M}_{\sun}\mathrm{yr}^{-1}$, 
with median of $196^{+105}_{-65}~\mathrm{M}_{\sun}\mathrm{yr}^{-1}$, and a median
uncertainty of $\sim26$ per cent. In the lower panels of Fig.
\ref{fig:sfro2pf2p}, it is evident that the
$SFR^\mathrm{CIG,10}_\mathrm{FIR}$ values are in better agreement
with $SFR_\mathrm{UV+IR}$ results than the
$SFR^\mathrm{CIG,10}_\mathrm{noFIR}$ values. The median difference
between the $SFR^\mathrm{CIG,10}_\mathrm{FIR}$ and
$SFR_\mathrm{UV+IR}$ results is $\sim3$ per cent. The \textsc{cigale}
10-Myr-averaged \textit{SFRs} derived without the FIR constraint range
in $80<SFR^\mathrm{CIG,10}_\mathrm{noFIR}< 1106~\mathrm{M}_{\sun}\mathrm{yr}^{-1}$, 
with median of $402^{+302}_{-170}~\mathrm{M}_{\sun}\mathrm{yr}^{-1}$, and a median
error of $\sim60$ per cent. The median difference between models without the
FIR prior and models using the FIR data,
$\log(SFR^\mathrm{CIG,10}_\mathrm{noFIR})-\log(SFR^\mathrm{CIG,10}_\mathrm{FIR})$ is $0.28$~dex, 
which is similar to the median
$SFR^\mathrm{CIG,10}_\mathrm{noFIR}$ errors.

Summarising, \textit{SFRs} estimated with observables only match
SED-based \textit{SFRs} for 2P models when using the FIR constraint. Both types of
calculations agree reasonably well for the 2 codes used in our work,
although \textsc{synthesizer} provides slightly larger values than the
UV+FIR data. We find that the \textsc{synthesizer} \textit{SFRs} become larger
than the ones based on UV+FIR data as the young-population age decreases. The galaxies
where the $SFR_\mathrm{UV+IR} > SFR^\mathrm{SYN,10}_\mathrm{FIR}$ are the ones with the 
largest young-population ages ($\goa 300$~Myr), for which the \citet{ken98} conversion
is less accurate. Thus, the explanation is tracked down in the more complex SFHs 
considered by \textsc{synthesizer} compared to the assumptions in 
\citet{ken98}, as explained in the next Subsection.

Considering that we are studying dusty starburst galaxies, we can refer to 
\textit{SFR} and $L_\mathrm{TIR}$ in an interchangeable way. Then, the scatter plots shown in
the right panels of Fig.~\ref{fig:sfro2pf2p} clearly evince that 2P models without 
the FIR information fail to reproduce the dust emission. A similar conclusion
has been found by \citet{hun19} studying a nearby galaxy sample using codes 
managing energy-balance techniques.

\subsection{Effect of the FIR prior in the determination of the
  properties of the young population}\label{sec:yopo2p}

Here we compare the young-population parameters derived with the
\textsc{cigale} and \textsc{synthesizer} codes for each 2P case, to
study what causes the difference between the SED-derived \textit{SFR}
values of both 2P models.

In Figure \ref{fig:avy2pf2p}, we compare the attenuations of the young
stellar population in our sample of dusty starbursts for the 2 types of
models and codes. The \textsc{synthesizer} young attenuations
derived using the FIR prior range in
$0.9<A_{V,\mathrm{you}}^\mathrm{SYN,FIR}<4.1$~mag, with
median of $2.4^{+0.7}_{-0.6}$~mag, and a median uncertainty
of 0.1~mag. The results computed without the FIR data are in the range
$0.9<A_{V,\mathrm{you}}^\mathrm{SYN,noFIR}<4.2$~mag, with median of
$2.9^{+0.6}_{-0.6}$~mag, and a median error of 0.1~mag. The median
difference of
$(A_{V,\mathrm{you}}^\mathrm{SYN,noFIR}-A_{V,\mathrm{you}}^\mathrm{SYN,FIR})$ is 0.6~mag, 
but individual objects can reach differences of 1.3~mag.

The \textsc{cigale} results are very similar to the \textsc{synthesizer} ones. The values determined
with the FIR constraint range in
$0.8<A_{V,\mathrm{you}}^\mathrm{CIG,FIR}<3.9$~mag, with
median of $2.3^{+0.7}_{-0.6}$~mag, and a median uncertainty of
0.2~mag. The solutions estimated omitting the FIR data are in the
range $1.0<A_{V,\mathrm{you}}^\mathrm{CIG,noFIR}<3.9$~mag, with
median of $2.5^{+0.8}_{-0.5}$~mag, and a median error of 0.3~mag. 
The median difference of 
$(A_{V,\mathrm{you}}^\mathrm{CIG,noFIR}-A_{V,\mathrm{you}}^\mathrm{CIG,FIR})$ is 0.2~mag, 
which is comparable to the median uncertainties, but individual
sources can reach differences of 0.9~mag.

\begin{figure*}
\centering
\subfloat{\includegraphics[width=0.456\textwidth]{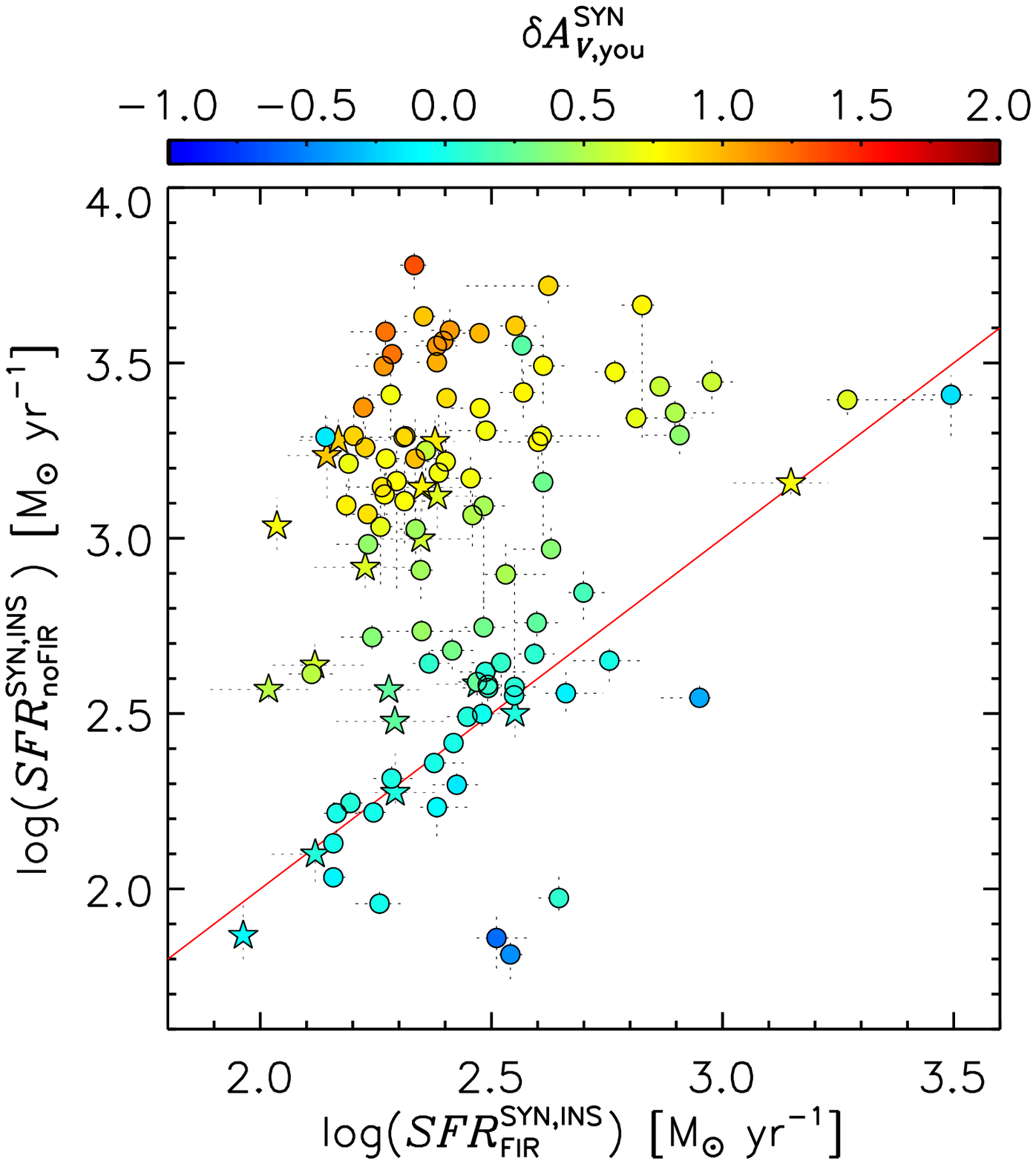}}\hfill
\subfloat{\includegraphics[width=0.456\textwidth]{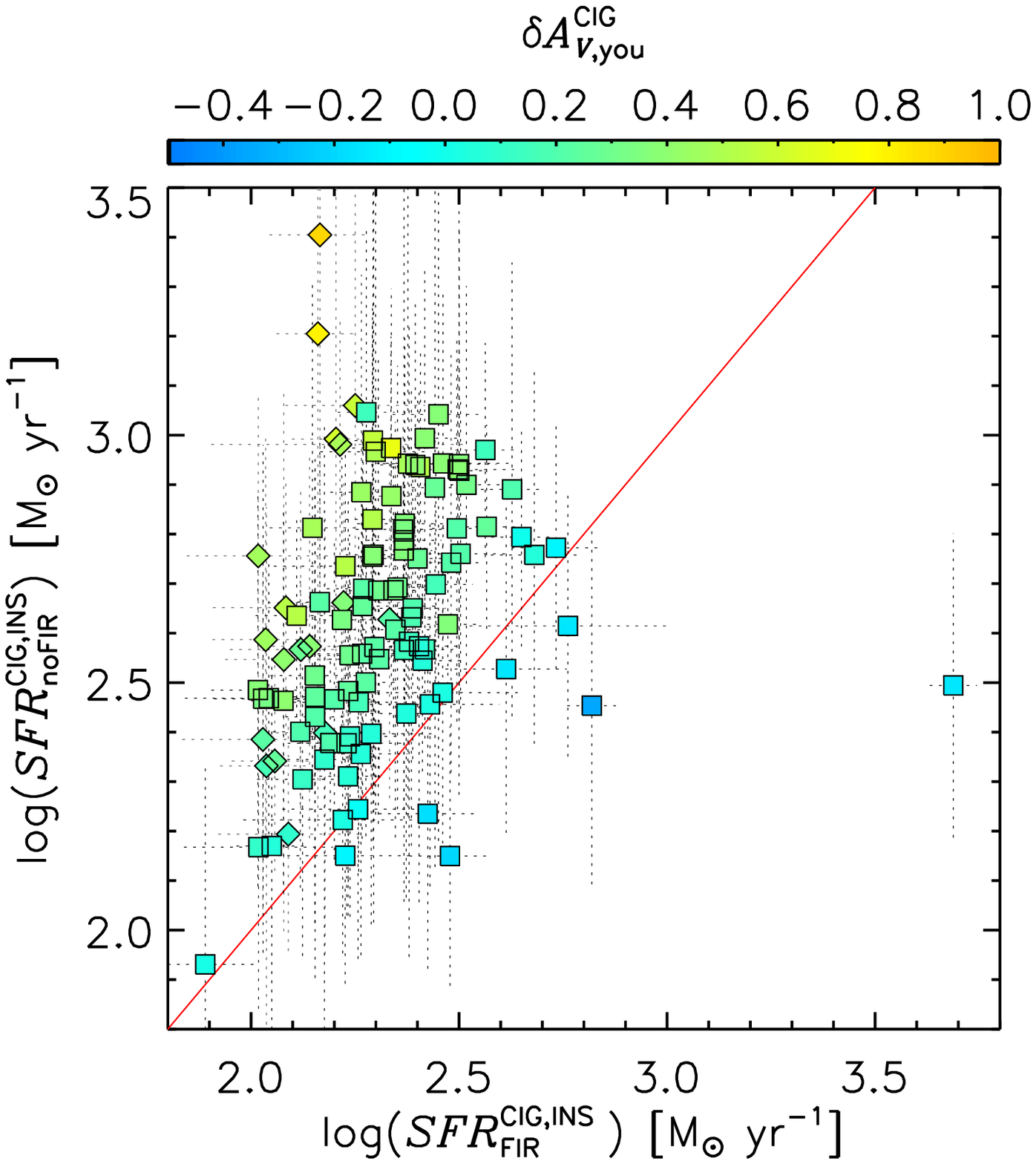}}\\
\caption{Comparison between the instantaneous \textit{SFRs} estimated with the FIR prior on x-axis, and those obtained when the FIR data are missing on y-axis derived from the 2P models with the \textsc{synthesizer} \textit{(left panel)} and the \textsc{cigale} \textit{(right panel)} codes. The filled circles (filled stars) with error bars depict the median values and $1\sigma$ uncertainties for the main (complementary) sample determined with the \textsc{synthesizer} code. The filled squares (filled diamonds) with error bars stand for the expected values and standard deviations derived with the \textsc{cigale}. The difference between $A_{V,\mathrm{you}}^\mathrm{noFIR}$ and $A_{V,\mathrm{you}}^\mathrm{FIR}$, $\delta A_{V,\mathrm{you}}$, is distinguished by colour for each code. \label{fig:sfr2pf2p}}
\end{figure*}
\begin{figure*}
\centering
\subfloat{\includegraphics[width=0.456\textwidth]{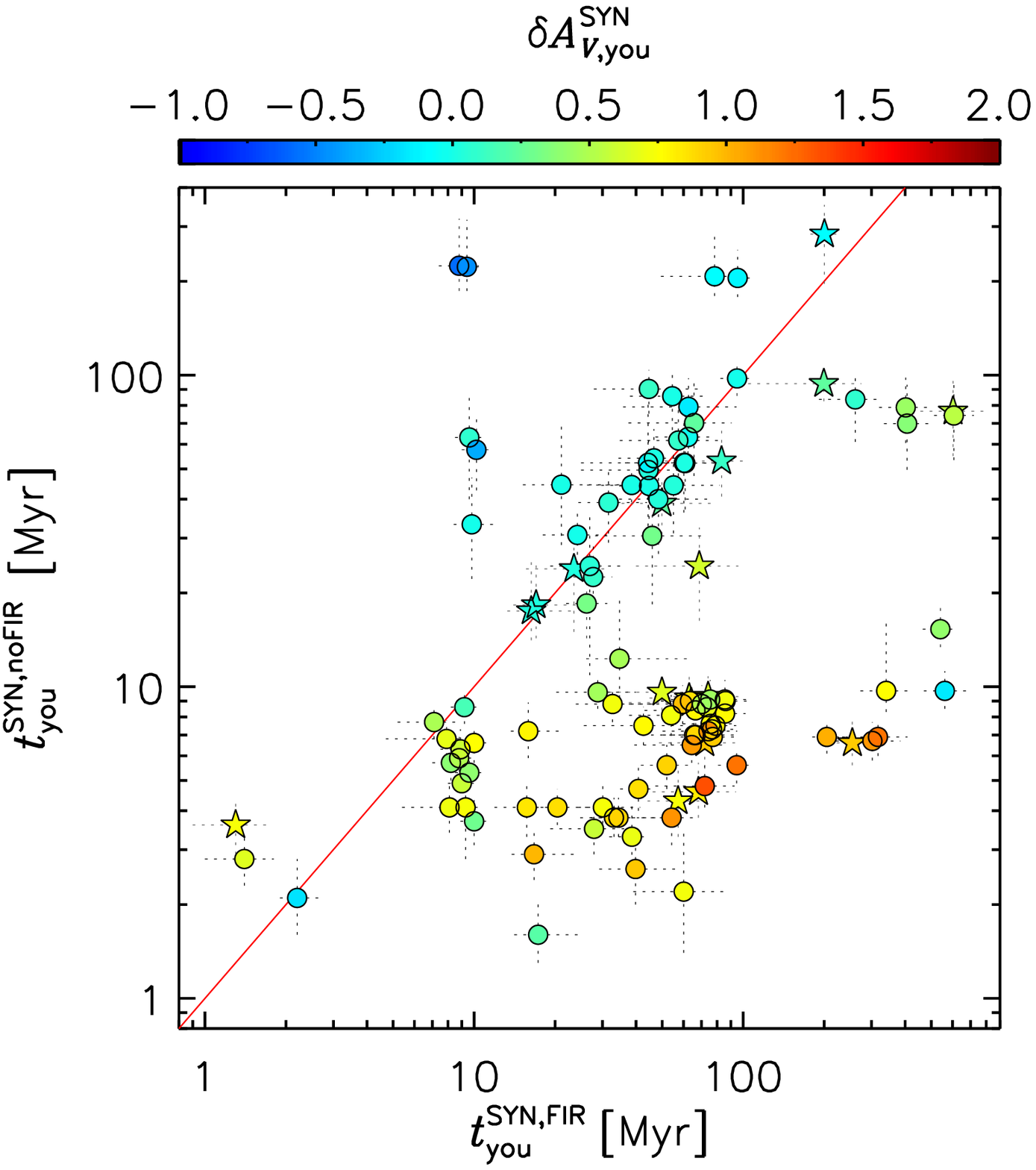}}\hfill
\subfloat{\includegraphics[width=0.456\textwidth]{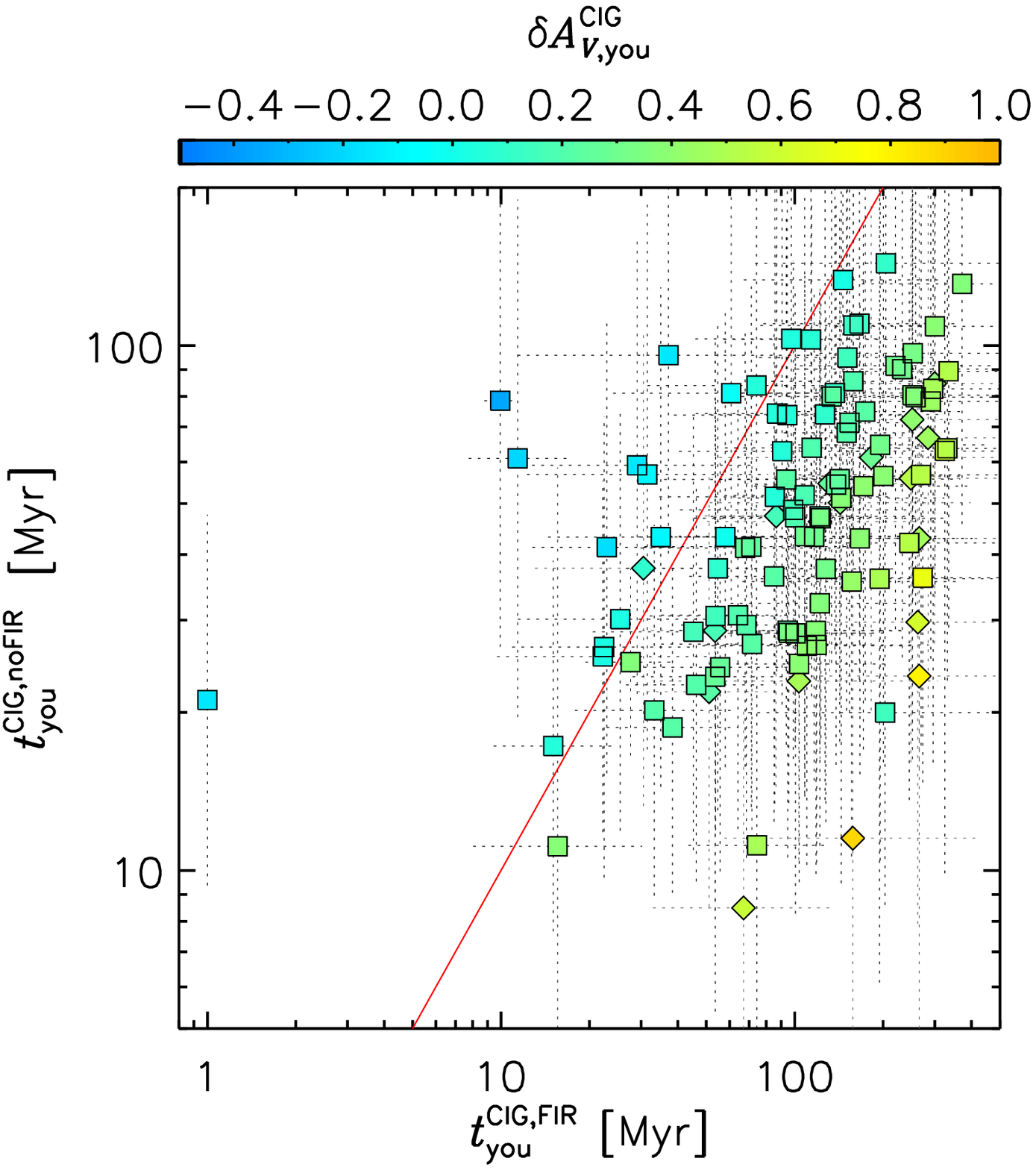}}\\
\caption{Comparison between the ages of the young population of the 2P models using the FIR prior on x-axis, and those determined when the FIR is omitted on y-axis estimated with the \textsc{synthesizer} \textit{(left panel)} and the \textsc{cigale} \textit{(right panel)} codes. The filled circles (filled stars) with error bars show the median values and $1\sigma$ uncertainties for the main (complementary) sample estimated with the \textsc{synthesizer} code. The filled squares (filled diamonds) with error bars stand for the expected values and standard deviations derived with the \textsc{cigale}. The difference between $A_{V,\mathrm{you}}^\mathrm{noFIR}$ and $A_{V,\mathrm{you}}^\mathrm{FIR}$, $\delta A_{V,\mathrm{you}}$, is distinguished by colour for each code. \label{fig:agy2pf2p}}
\end{figure*}
These comparisons evidence that the absence of FIR photometry results
mostly in an overestimation of the young-population attenuation values
when 2P models are considered. However, there is not a significant
difference on the median values obtained for the young-population attenuation
with and without the FIR data, which may suggest that these attenuation 
differences are not the main source of discrepancy between the SED-based and 
observable-based \textit{SFRs}.

As a further check of the influence of the overestimation of
$A_{V,\mathrm{you}}^\mathrm{noFIR}$ in the 2P-SED-based \textit{SFR}
determination, we show in Fig. \ref{fig:sfr2pf2p} the comparison of
the instantaneous \textit{SFRs} derived with and without the FIR prior
from the results of both codes, with the differences for the
attenuation of the young population, $\delta
A_{V,\mathrm{you}}=(A_{V,\mathrm{you}}^\mathrm{noFIR}-A_{V,\mathrm{you}}^\mathrm{FIR})$,
distinguished by colour. 
It is clear that the 2P-SED-derived \textit{SFR} differences are mainly controlled by the
young-population attenuation, with larger \textit{SFRs} obtained for objects with
larger (overestimated, as demonstrated with the FIR-prior models)
$A_{V,\mathrm{you}}$ values. The objects which do not follow the
progression of the $\delta A_{V,\mathrm{you}}^\mathrm{SYN}$ colour-bar
are galaxies obtaining $t^\mathrm{SYN}_\mathrm{you} \loa 10$~Myr in
one of the two modelling cases. This $\delta \log SFR-\delta
A_{V,\mathrm{you}}$ dependence is more accentuated for the
\textsc{synthesizer} code results, but it is also discernible in the
\textsc{cigale} solutions.

Now we study the behaviour of the ages of the young population. The
\textsc{synthesizer} values derived using the FIR prior range between
$1<t^\mathrm{SYN,FIR}_\mathrm{you}<607$~Myr, with median of
$52^{+34}_{-42}$~Myr, and a median uncertainty of $\sim27$ per cent. The
results determined without the FIR data are in the range $1<
t^\mathrm{SYN,noFIR}_\mathrm{you}<284$~Myr, with median of
$9^{+54}_{-5}$~Myr, and a median error of $\sim18$ per cent. The
\textsc{cigale} solutions derived employing the FIR constraint range
between $1\leq t^\mathrm{CIG,FIR}_\mathrm{you}<372$~Myr, with median 
of $116^{+135}_{-71}$~Myr, and a median uncertainty of $\sim80$ per cent. The
results calculated without the FIR prior are in the range $8
<t^\mathrm{CIG,noFIR}_\mathrm{you}<144$~Myr, with median of
$47^{+36}_{-22}$~Myr, and a median error of $\sim130$ per cent.

The \textsc{cigale} large errors evidence the age-dust degeneracy in
the 2P-noFIR models, i.e., without the FIR constraint there are a lot
of models, encompassing a large range in
$t^\mathrm{CIG,noFIR}_\mathrm{you}$ and
$A_{V,\mathrm{you}}^\mathrm{CIG,noFIR}$, providing similar UV-to-NIR
SEDs that compare equally well with the observed flux data points.
Even when using the FIR prior, significant differences in age and
attenuation estimations are seen for some galaxies from the 2 codes used in our work.
This means that the broad-band UV-to-FIR SEDs can be reproduced by
relatively young (50~Myr) starbursts with relatively large
attenuations and \textit{SFRs} (cf. Fig.~\ref{fig:sfro2pf2p}), or by
older stars (around 100~Myr) 
but less extincted and with more moderated 
\textit{SFRs} (comparable to those obtained from observables and the
\citealt{ken98} calibration).

In Figure \ref{fig:agy2pf2p}, we compare the ages of the young stellar
population in our sample of massive dusty starbursts for the 2 types
of models and codes, colour--coding the points by the differences in the
young-population attenuation.  Distinctly, galaxies presenting larger
attenuations have shorter young-population ages in both modelling
cases and codes, which is another manifestation of the age-attenuation
degeneracy. For objects with $\delta A_{V,\mathrm{you}}> 0.3$~mag, the
difference in young population ages of each set of models is larger
than a factor of $\sim8$ and $\sim4$ on median for
\textsc{synthesizer} and \textsc{cigale}.  There are some galaxies
with $\delta A_{V,\mathrm{you}}>0.3$~mag and young-population age
differences shorter than a factor 2. This is not surprising because
the presence of FIR data helps to break the age-dust degeneracy, but
the young-population age is also degenerated with the intensity of 
the most recent burst (i.e., \textit{SFR} and total mass created by 
the starburst).
\begin{figure*}
\centering
\subfloat{\includegraphics[width=0.5\textwidth]{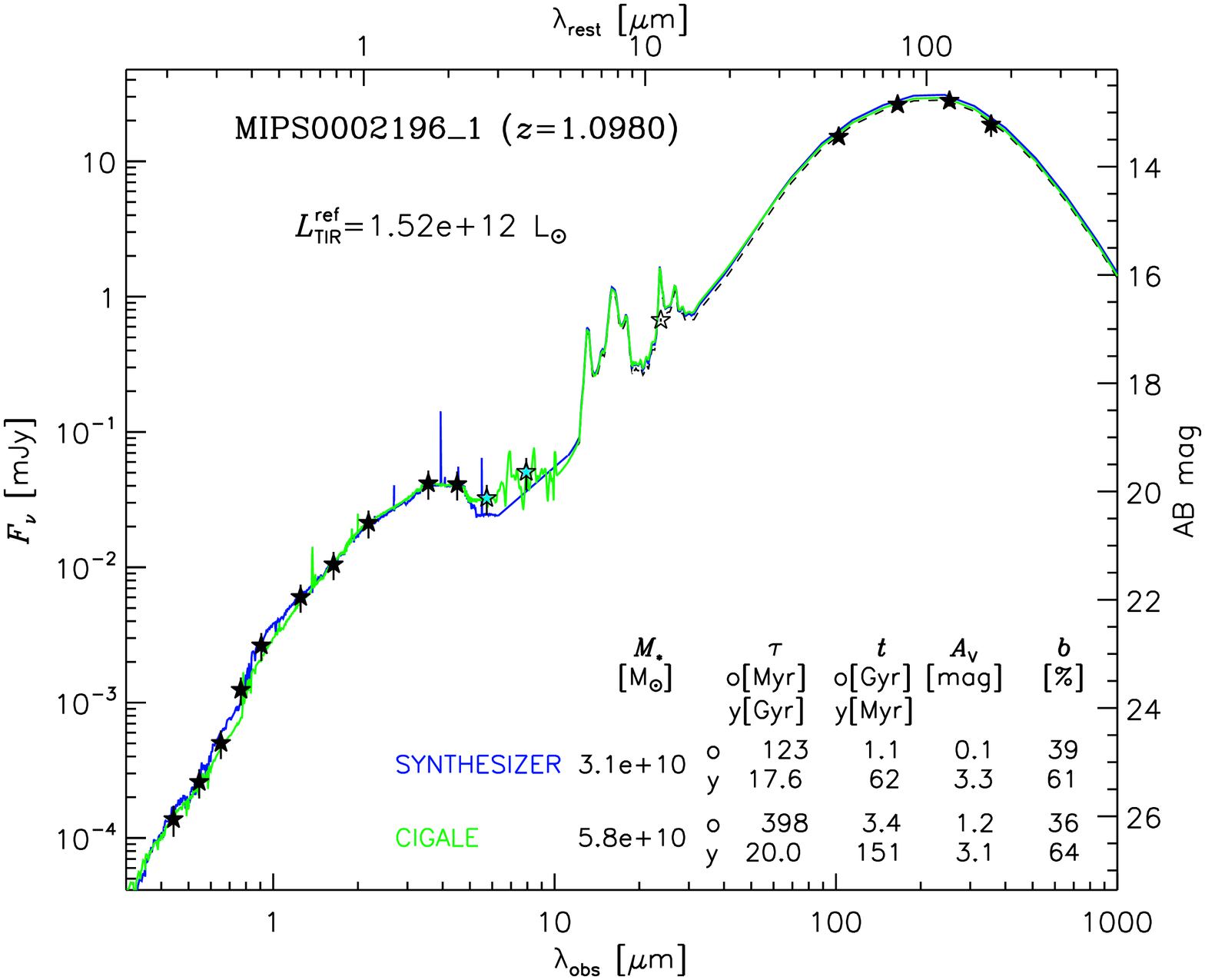}}\hfill
\subfloat{\includegraphics[width=0.5\textwidth]{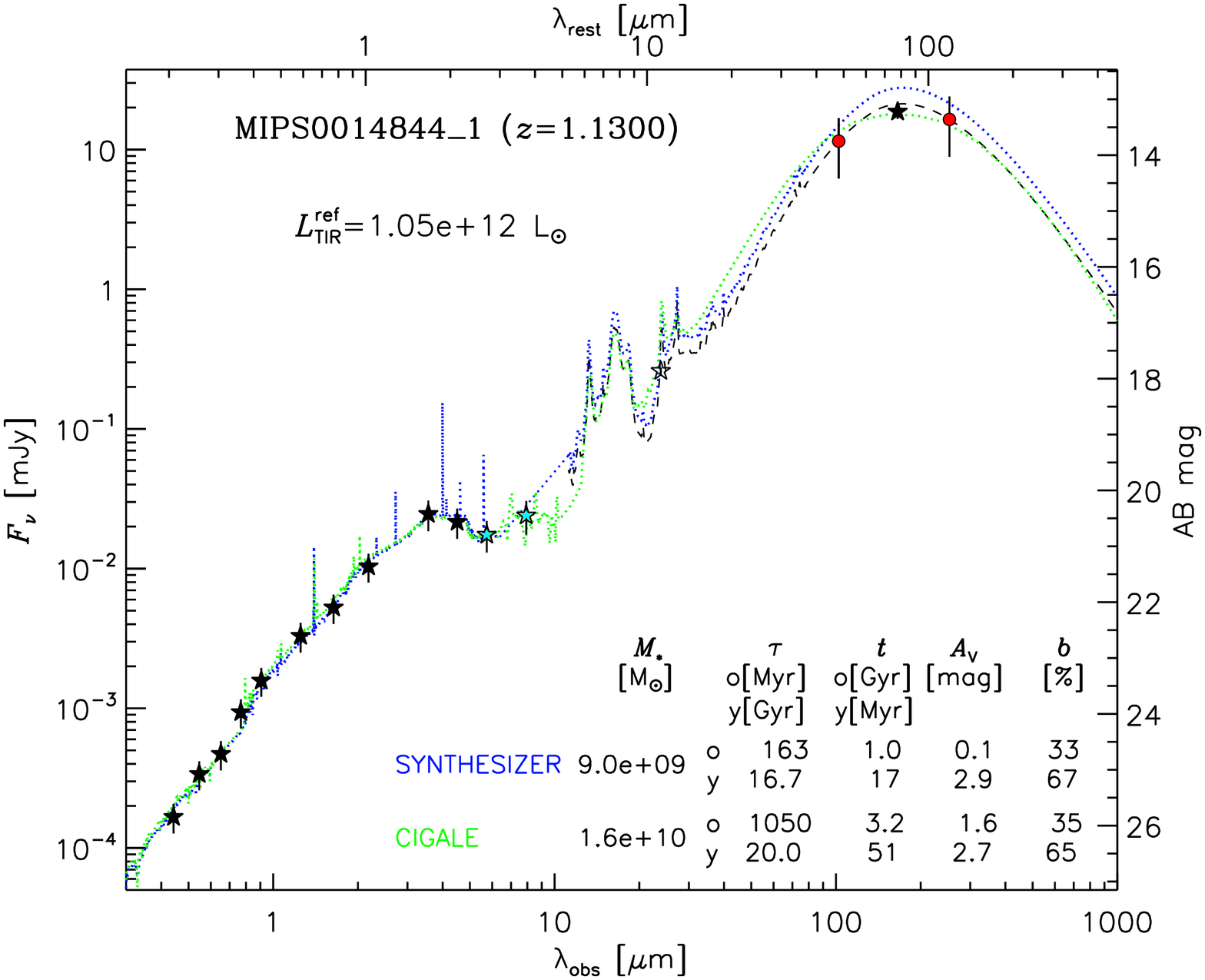}}\\
\caption{Two examples of the resulting fit to the UV-to-FIR SED for our $0.7 \leq z \lessapprox 1.2$ dusty starbursts. \textit{Left panel:} Results of the fitting process for the most significant cluster identified with the \textsc{synthesizer} code \textit{(solid blue line)}, and the fit obtained with the Bayesian analysis of the \textsc{cigale} \textit{(solid green line)} for a source in the main sample. The parameter values obtained from each code are also shown. The fit to the FIR photometry \textit{(}$\lambda_{\rm obs}\geq 70~\mu$m, \textit{dashed black line)} and its associated $L^\mathrm{ref}_\mathrm{TIR}$ are indicated. Photometric data points include uncertainties, only the \textit{filled black stars} are used in the \textsc{synthesizer} fit, while the \textit{filled black and cyan stars} are used in the \textsc{cigale} fit. \textit{Right panel:} The same as left panel for a source in the complementary sample. The fits obtained for the \textsc{synthesizer} and the \textsc{cigale} codes are shown with a \textit{dotted blue line} and a \textit{dotted green line}. The \textit{filled red circles} stand for synthetic fluxes in the FIR filters (see Section \ref{sec:zphot}), and they are also used in the fit from each code. The UV-to-FIR SEDs of all galaxies are presented as online material.\label{fig:sed2pfmc}}
\end{figure*}
To exemplify the degeneracy between the age and the stellar mass of
the young population, we return to our assumed SFHs. The instantaneous
\textit{SFR} (our $SFR^\mathrm{INS}$) is dominated by the contribution
of the young population (which is what can be expected in dSFGs). 
Hence, we can approximate
$SFR^\mathrm{INS}\simeq SFR^\mathrm{INS}_\mathrm{you}$, with
$SFR^\mathrm{INS}_\mathrm{you}$ coming from a constant recent
SFH by definition. This $SFR^\mathrm{INS}_\mathrm{you}$ can
be expressed as $M_{\star,\mathrm{you}}/t_\mathrm{you}$ or
$M_\star(b/t)_\mathrm{you}$, in terms of the stellar mass for the young
population and the total. Then, the specific SFR for a galaxy
is $sSFR^\mathrm{INS}\simeq (b/t)_\mathrm{you}$. Considering similar
stellar masses for both 2P cases, we can refer to \textit{SFR} or
\textit{sSFR} in an interchangeable way. Then, the relation of
\textit{SFR} differences and young-population attenuation variations between both
cases can be expressed in terms of $(b/t)_\mathrm{you}$ differences
and $A_{V,\mathrm{you}}$ variations. With this in mind, we can provide
a better panorama of the degeneracies presented in both 2P cases. The
$A_{V,\mathrm{you}}^\mathrm{FIR}$ values are driven by the FIR prior,
the $L_\mathrm{TIR}$, which helps to break the age-dust degeneracy.
The $b_\mathrm{you}$ and $t_\mathrm{you}$ are directly related with
the $sSFR^\mathrm{INS}$ values, where $b_\mathrm{you}$ and
$t_\mathrm{you}$ are degenerated. Thus, a high value of
$b_\mathrm{you}$ connected to a large value of $t_\mathrm{you}$ (i.e,
a long burst forming large amounts of stars) is difficult to be
disentangled from a small $b_\mathrm{you}$ associated to a small value
of $t_\mathrm{you}$ (i.e., a young burst forming less stellar mass). The
sources which do not follow the sequence
 of the $\delta A_{V,\mathrm{you}}$ colour-bar in Figs. \ref{fig:sfr2pf2p} and \ref{fig:agy2pf2p} are
mainly objects presenting extreme differences in the
$(b/t)_\mathrm{you}$ ratio between each 2P case.

\subsection{Effect of the FIR prior in the determination of the
  properties of the old population}\label{sec:fposp}

Here we compare the old-population parameters estimated with the
\textsc{cigale} and \textsc{synthesizer} codes for each 2P case.
The properties of the old population of our sample of dusty massive
starbursts are difficult to determine because the oldest stars are 
outshined by the most recent star formation event \citep{mar10}.

The \textsc{synthesizer} old attenuations
obtained using the FIR prior range in
$0.0<A_{V,\mathrm{old}}^\mathrm{SYN,FIR}<1.5$~mag, with
median of $0.1^{+0.7}_{-0.1}$~mag, and a median uncertainty
of 0.1~mag. The results estimated without the FIR data are in the range
$0.0<A_{V,\mathrm{old}}^\mathrm{SYN,noFIR}<1.5$~mag, with median of
$0.2^{+0.8}_{-0.2}$~mag, and a median error of 0.1~mag.
Significant differences in the derived $A^\mathrm{SYN}_{V,\mathrm{old}}$ values
between the 2P models with and without FIR prior translate into $M^\mathrm{SYN}_\star$
estimates that differ by a factor $>2$ (see Section
\ref{sec:mssfr2pc}).

For the \textsc{cigale}, the values derived
with the FIR constraint range in
$0.0<A_{V,\mathrm{old}}^\mathrm{CIG,FIR}<1.7$~mag, with
median of $0.9^{+0.3}_{-0.8}$~mag, and a median uncertainty of
0.6~mag. The solutions computed omitting the FIR data are in the
range $0.0<A_{V,\mathrm{old}}^\mathrm{CIG,noFIR}<1.8$~mag, with
median of $1.0^{+0.3}_{-1.0}$~mag, and a median error of 0.8~mag.

Now we investigate the ages of the old population. The
\textsc{synthesizer} values derived using the FIR prior range between
$0.9<t^\mathrm{SYN,FIR}_\mathrm{old}<6.0$~Gyr, with median of
$1.4^{+1.1}_{-0.4}$~Gyr, and a median uncertainty of $0.3$~Gyr. The
results determined without the FIR data are in the range $0.9<
t^\mathrm{SYN,noFIR}_\mathrm{old}<4.7$~Gyr, with median of
$1.4^{+0.7}_{-0.4}$~Gyr, and a median error of $0.2$~Gyr. The
\textsc{cigale} solutions obtained employing the FIR constraint range
between $1.1 < t^\mathrm{CIG,FIR}_\mathrm{old}<5.6$~Gyr, with median 
of $3.3^{+0.8}_{-0.5}$~Gyr, and a median uncertainty of $1.6$~Gyr. The
results calculated without the FIR data are in the range $1.5
<t^\mathrm{CIG,noFIR}_\mathrm{old}<5.3$~Gyr, with median of
$3.4^{+0.9}_{-0.2}$~Gyr, and a median error of $1.5$ Gyr.
 For the old-population age, the values obtained
by each code and both 2P cases are very similar for all the galaxies.
Only for a small fraction of galaxies, $21$ and $13$ per cent of all our
sample, the ages of the old stars differ significantly when comparing
\textsc{synthesizer} and \textsc{cigale} results with and without
FIR prior.  Therefore, the addition of FIR data has a minor impact in
the determination of the ages of the old population.  This is expected
because the FIR photometry traces the dust content of the galaxies,
which is related to the most recent star formation events.
 
\citet{bua14} studied a sample of $z>1$ galaxies with \textsc{cigale}
using 2P models and the FIR prior. These authors found compatible
trends in the ages of both populations with and without the
addition of FIR information. Hence, our findings are similar to their
results for the old stellar population, but not for the young
population case. We have reported a significant change in the young
population ages for more than half of the objects in our sample when
the FIR data are omitted (see Fig. \ref{fig:agy2pf2p}).

\section{A comprehensive view on the nature of
  $\lowercase{{\mathbf{0.7<}\mathbfit{z}\mathbf{<1.2}}}$ massive dusty
  starbursts}\label{sec:stepro}

In Section \ref{sec:FIRprior} we studied how the inclusion of FIR data
in the modelling procedure affects the results about the stellar
properties of massive dusty starburst galaxies at intermediate
redshift. In this Section, we discuss the models simultaneously
reproducing the stellar and dust emission (i.e., using the FIR prior)
and compare the solutions obtained by \textsc{synthesizer} and
\textsc{cigale}, putting them in the context of other works found 
in the literature.

In Table \ref{tab:parsyci} we present the results of the 2P-FIR models
applied to our sample of dusty star-forming galaxies.  We show in Fig.
\ref{fig:sed2pfmc} the complete UV-to-FIR SEDs for one source in the
main sample and one galaxy in the complementary sample. We also
provide the results of the median solution of the most significant
cluster derived from the Monte--Carlo simulation analysis with the
\textsc{synthesizer} code, and the average parameter values estimated
with the \textsc{cigale} from the Bayesian analysis of its full set of
models. We evaluated the goodness of the fits, using a reduced
$\chi^2$-estimator including the UV-to-MIR and FIR part of the
spectrum of each galaxy (see, Fig.  \ref{fig:sed2pfmc}) separately.
Although the \textsc{synthesizer} code performs better than
\textsc{cigale} in the FIR spectral range ($\chi^2_{L_\mathrm{TIR}}$ median,
 and 16 and 84-percentile of $1.5^{+2.1}_{-1.2}$ and $3.6^{+9.2}_{-3.3}$, respectively), 
we do not find any statistically relevant difference in the quality 
of the fits provided by the 2 codes.

\begin{table*}
\begin{center}
\caption[Stellar population synthesis results for main-sample starbursts.]{Stellar population synthesis results for main-sample dusty starbursts. Median and $1\sigma$ error values are shown for each parameter derived with the \textsc{synthesizer} code. The expected and standard deviation (or upper and lower limits when changing from logarithmic space to linear space) values are quoted for each parameter estimated with the \textsc{cigale}. (1) Name of the galaxy. (2) Photometric or spectroscopic redshift ($z_\mathrm{spec}$ indicated by a $\dagger$). (3) Code used to derive the parameters. (4) Stellar mass (in solar units) and its uncertainty derived from the logarithmic space. (5) Parameter values for the old or the young population. (6) \textit{e}-folding time and its uncertainty (old population in Myr and young population in Gyr). (7) Age and its uncertainty (old population in Gyr and young population in Myr). (8) Attenuation in the \textit{V}-band and its uncertainty in mag. (9) Metallicity value (fixed to the solar value) in solar units. (10) Most recent burst intensity fraction and its uncertainty in percentage.\newline(This Table and that for the complementary sample are available in its entirety in the online journal.)}
\label{tab:parsyci}
\begin{tabular}{lcllclllll}
\hline\hline
\multicolumn{1}{c}{Galaxy} & \multicolumn{1}{c}{$z^\dagger$} & \multicolumn{1}{c}{Code} & \multicolumn{1}{c}{$\log M_\star$} & \multicolumn{1}{c}{Pop.} & \multicolumn{1}{c}{$\tau$} & \multicolumn{1}{c}{Age} & \multicolumn{1}{c}{$A_V$} & \multicolumn{1}{c}{\textit{Z}} & \multicolumn{1}{c}{$b$} \\
 & & & \multicolumn{1}{c}{[$\mathrm{M}_{\sun}$]} & & \multicolumn{1}{c}{old [Myr]} & \multicolumn{1}{c}{old [Gyr]} & \multicolumn{1}{c}{[mag]} & \multicolumn{1}{c}{[$\mathrm{Z}_{\sun}$]} & \multicolumn{1}{c}{[\%]} \\
 & & & & & \multicolumn{1}{c}{you [Gyr]} & \multicolumn{1}{c}{you [Myr]} & & & \\
 \multicolumn{1}{c}{(1)} & \multicolumn{1}{c}{(2)} & \multicolumn{1}{c}{(3)} & \multicolumn{1}{c}{(4)} & \multicolumn{1}{c}{(5)} & \multicolumn{1}{c}{(6)} & \multicolumn{1}{c}{(7)} & \multicolumn{1}{c}{(8)} & \multicolumn{1}{c}{(9)} & \multicolumn{1}{c}{(10)} \\
 \hline
 \vspace{3pt} MIPS0000517A & $1.0500$ & \textsc{synthesizer} & $10.0^{+0.1}_{-0.1}$ & old & $230^{+43 }_{-125}$ & $1.0^{+0.2}_{-0.1}$ & $0.0^{+0.1}_{-0.0}$ & $1.0$ & \\
\vspace{3pt} & & & & you & $18.5^{+3.1}_{-3.2}$ & $ 39^{+6  }_{-7  }$ & $1.9^{+0.1}_{-0.1}$ & $1.0$ & $69^{+2 }_{-23}$ \\
\vspace{3pt} & & \textsc{cigale} & $10.1\pm 0.2$ & old & $1280^{+5762}_{-1047}$ & $3.2\pm 1.6$ & $1.0^{+0.7}_{-0.7}$ & $1.0$ & \\
\vspace{3pt} & & & & you & $20.0$ & $ 46^{+34 }_{-20 }$ & $1.9\pm 0.2$ & $1.0$ & $61\pm 26$ \\
\vspace{3pt} MIPS0000648A & $1.1900$ & \textsc{synthesizer} & $10.4^{+0.1}_{-0.1}$ & old & $162^{+46 }_{-45 }$ & $1.0^{+0.1}_{-0.1}$ & $0.0^{+0.1}_{-0.0}$ & $1.0$ & \\
\vspace{3pt} & & & & you & $17.7^{+3.4}_{-2.7}$ & $ 45^{+21 }_{-19 }$ & $2.3^{+0.1}_{-0.1}$ & $1.0$ & $51^{+13}_{-16}$ \\
\vspace{3pt} & & \textsc{cigale} & $10.6\pm 0.1$ & old & $1189^{+5363}_{-973 }$ & $3.2\pm 1.6$ & $1.2^{+0.7}_{-0.7}$ & $1.0$ & \\
\vspace{3pt} & & & & you & $20.0$ & $100^{+90 }_{-47 }$ & $2.2\pm 0.2$ & $1.0$ & $57\pm 27$ \\
\vspace{3pt} MIPS0000816\_1 & $0.9300$ & \textsc{synthesizer} & $10.2^{+0.2}_{-0.2}$ & old & $141^{+113}_{-45 }$ & $2.4^{+1.0}_{-0.5}$ & $0.0^{+0.3}_{-0.0}$ & $1.0$ & \\
\vspace{3pt} & & & & you & $18.9^{+2.9}_{-3.2}$ & $  8^{+1  }_{-1  }$ & $1.5^{+0.1}_{-0.1}$ & $1.0$ & $21^{+18}_{-10}$ \\
\vspace{3pt} & & \textsc{cigale} & $ 9.9\pm 0.3$ & old & $1285^{+5815}_{-1053}$ & $4.1\pm 1.6$ & $0.5^{+0.4}_{-0.4}$ & $1.0$ & \\
\vspace{3pt} & & & & you & $20.0$ & $ 22^{+29 }_{-13 }$ & $1.1\pm 0.2$ & $1.0$ & $49\pm 30$ \\
\vspace{3pt} MIPS0000826 & $0.7200$ & \textsc{synthesizer} & $10.5^{+0.1}_{-0.1}$ & old & $247^{+32 }_{-37 }$ & $3.0^{+0.4}_{-0.3}$ & $0.0^{+0.1}_{-0.0}$ & $1.0$ & \\
\vspace{3pt} & & & & you & $19.8^{+2.5}_{-2.4}$ & $  9^{+1  }_{-1  }$ & $3.2^{+0.1}_{-0.1}$ & $1.0$ & $25^{+3 }_{-5 }$ \\
\vspace{3pt} & & \textsc{cigale} & $10.7\pm 0.1$ & old & $ 314^{+686 }_{-215 }$ & $5.5\pm 1.4$ & $0.0^{+0.1}_{-0.0}$ & $1.0$ & \\
\vspace{3pt} & & & & you & $20.0$ & $ 10^{+1  }_{-1  }$ & $3.2\pm 0.2$ & $1.0$ & $11\pm  4$ \\
\vspace{3pt} MIPS0002196\_1 & $1.0980\dagger$ & \textsc{synthesizer} & $10.5^{+0.1}_{-0.1}$ & old & $123^{+50 }_{-25 }$ & $1.1^{+0.4}_{-0.2}$ & $0.1^{+0.1}_{-0.1}$ & $1.0$ & \\
\vspace{3pt} & & & & you & $17.6^{+3.4}_{-2.7}$ & $ 62^{+17 }_{-15 }$ & $3.3^{+0.1}_{-0.1}$ & $1.0$ & $61^{+6 }_{-7 }$ \\
\vspace{3pt} & & \textsc{cigale} & $10.8\pm 0.1$ & old & $ 398^{+1650}_{-320 }$ & $3.4\pm 1.6$ & $1.2^{+1.1}_{-1.1}$ & $1.0$ & \\
\vspace{3pt} & & & & you & $20.0$ & $151^{+150}_{-75 }$ & $3.1\pm 0.2$ & $1.0$ & $64\pm 25$ \\
\hline
\end{tabular}
\end{center}
\end{table*}

\begin{figure*}
\centering
\includegraphics[width=0.95\textwidth]{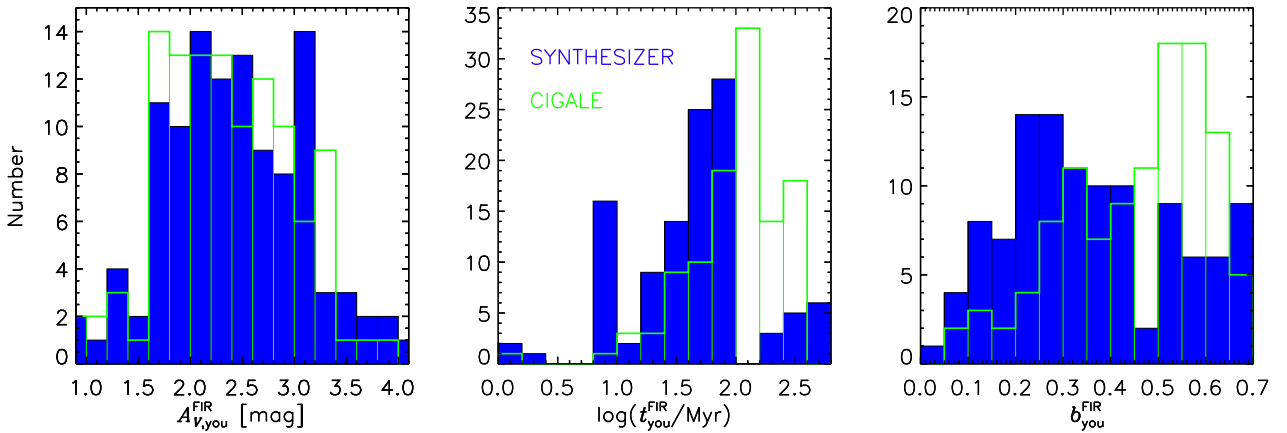}
\caption{Properties of the young stellar population of the whole sample of dusty starbursts derived using the FIR prior. From left to right, attenuation, age, and burst intensity. The \textit{fill blue histograms} stand for the results determined with the \textsc{synthesizer} code. The \textit{open green histograms} show the results obtained with the \textsc{cigale}.\label{fig:hstyp}}
\end{figure*}

\subsection{Discussion about the properties of the young stellar
  population}

Here we compare the young-population parameters derived from the
models including a FIR prior provided by the \textsc{synthesizer} and
\textsc{cigale} codes. These parameters are also contrasted with
results from the literature. Given that we fixed the value of the
metallicity for both codes and also $\tau_\mathrm{you}$ is unvaried
(assuming a constant SFH), we focus our discussion on the
attenuation $A^\mathrm{FIR}_{V,\mathrm{you}}$, the age $t^\mathrm{FIR}_\mathrm{you}$, 
and the recent burst intensity $b^\mathrm{FIR}_\mathrm{you}$. The 
distributions of values for these properties are presented  in Fig. \ref{fig:hstyp}, 
and their median solutions are shown in Table \ref{tab:parmedsyci}, 
both for the full sample of massive dusty starburst galaxies.

\subsubsection{Discussion about the attenuation of the young stellar population}

The typical attenuation of the young stellar population in our sample
of massive dusty starburst galaxies is 2.4~mag. The 2 codes used in
our paper provide similar results for
$A_{V,\mathrm{you}}^\mathrm{FIR}$ (within 0.1~mag, on median), as
expected considering the energy balance approach that the codes use.
In fact, there is a clear correlation between each set of estimated
values with Spearman correlation coefficient of $r_s=0.96$, and an
associated probability of no correlation of $p_s=7.0 \times 10^{-45}$.
The majority of our objects (70 per cent) presents
$A_{V,\mathrm{you}}^\mathrm{FIR}>2$~mag, which is not surprising given
that our sample selection is based on the FIR emission.

We have compared our results about attenuation with other studies in
the literature.  \citet{bua15} used \textsc{cigale} to study galaxies
selected at $8~\micron$ rest-frame and lying at $z\simeq 0.2-2$, with
some of these objects including PACS measurements. They found that at
$\langle z \rangle=0.9$, the average young attenuation under a
Calzetti-like curve is $\langle A_{V,\mathrm{you}}\rangle \simeq1.4$~mag. 
We have to mention that the average $\langle \log
L_\mathrm{TIR}/\mathrm{L}_{\sun}\rangle$ for these sources is 11.46, which is
smaller than the average $\langle\log
L^\mathrm{ref}_\mathrm{TIR}/\mathrm{L}_{\sun}\rangle=12.04$ of the objects in
our sample. We also note that in our study, we probed shorter young
population age values than \citeauthor{bua15} From the \textsc{cigale} results for
our sample, there are 93 galaxies (83 per cent of all sources) with
$t_\mathrm{you}^\mathrm{CIG,FIR}$ older than 50~Myr (the
\citeauthor{bua15} lowest young age value), from these 93, 64 objects
present $A_{V,\mathrm{you}}^\mathrm{CIG,FIR}>2$~mag.

We have also compared our results with sub-mm-selected samples
analysed with energy-balance techniques. \citet{dac15} examined the
ALESS sample of $z>1$ SMGs in the ECDF-S field, which
contains multi-wavelength data, including FIR detections. They used a
modified version of the \textsc{magphys} code \citep{dac08} to derive
physical properties of these objects. They found and overall
attenuation $A_V=1.9 \pm 0.2$~mag, similar to our calculations.
\citet{dan17} presented a spectroscopic catalogue of the aforenamed
ALESS sample identifying redshifts in the range
$z_\mathrm{spec}=0.7-5.0$. They also used \textsc{magphys} to fit the
SEDs for these objects, finding 
$A_V$ values spanning from $0.5$ to $7$~mag. \citet{dud20} 
analysed an 870-$\micron$ selected sample in SXDS/UDS field, including
\textit{Herschel} detections, with median redshift $z\simeq 2.6$. 
They find a median attenaution of $A_V=2.89\pm 0.04$~mag
using \textsc{magphys} too, similar to our results. \citet{cas17} 
presented NIR and optical spectra of a sub-mm-selected sample at 
$0.2<z<4$ in the COSMOS field. They determined a median attenuation 
$A_V=5.0\pm 0.4$~mag by measuring Balmer decrements. If those sources 
are similar to ours, the ionized gas attenuation would be roughly twice 
the stellar attenuation, as found by \citet{cal00}.

Considering all these findings in the literature, we conclude that the
attenuation values derived for the young population in our galaxy
sample are within the standard range of values for dSFGs 
at intermediate and high redshift, typically $A_V=2.5$~mag.

\subsubsection{Discussion about the age of the young stellar population}\label{sec:dyopo}

The recent star-forming events harboured by the dusty starburst
galaxies in our sample present ages of several tens and up to a
hundred Myr. There is a systematic difference in the derived ages for
the 2 codes, with the \textsc{synthesizer} code favouring shorter ages
than \textsc{cigale} (by around a factor of 2). To quantify the level
of correlation between the young-population age values obtained from both codes,
we estimated the Spearman correlation coefficient resulting in
$r_s=0.71$, with a probability $p_s=1.6 \times 10^{-18}$ of getting
this result as a mere coincidence. We interpret this difference in
terms of the moderately higher \textit{SFRs} and slightly smaller stellar 
masses derived by \textsc{synthesizer} (see Sections \ref{sec:mssfr2pc} and \ref{sec:mssfr2pc2}).

The SFH assumed for the young population is practically constant.
Considering that, on average, we should be observing galaxies with
ages similar to half of the lifetime that an object stays in the
starburst phase, and that our median starburst ages are $\sim 50-120$~Myr, 
we can infer a duration of the starburst phase of $\sim 100-250$~Myr. 
This value is in agreement with the results deduced for the
lifetime of the starburst phase of SMGs ($\sim 100-300$~Myr; 
\citealt{swi06,tac08,hic12,don20}). If we combine these star formation
time-scales with the median \textit{SFRs} presented in
Section~\ref{sec:mssfr2pc2} and Table \ref{tab:parmedsyci}, we obtain that 
massive dusty galaxies at
intermediate redshift typically increase their mass by 30--50 per cent in a
starburst phase.

\begin{figure*}
\centering
\includegraphics[width=0.95\textwidth]{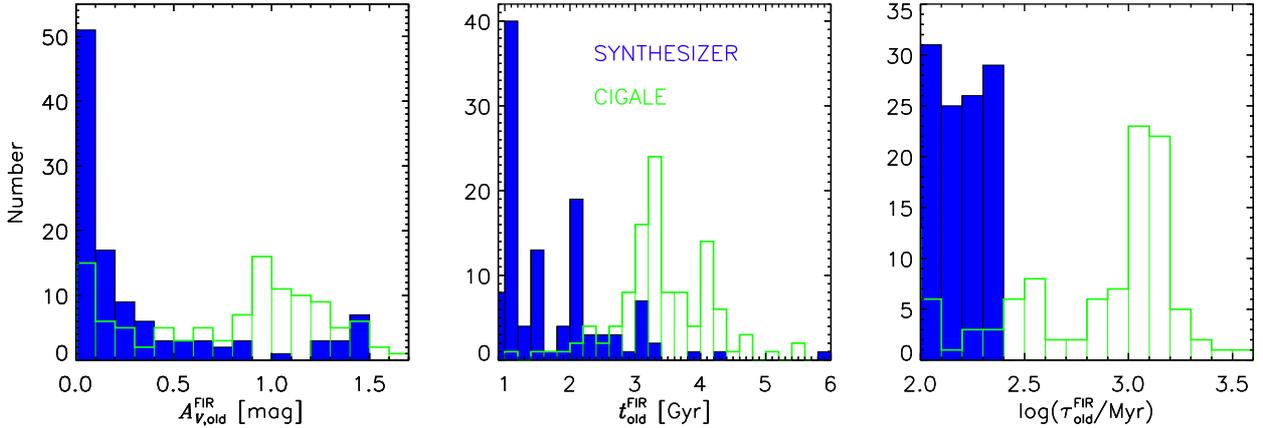}
\caption{Properties of the old stellar population of the whole sample of dusty starbursts derived using the FIR prior. From left to right, attenuation, age, and $e$-folding time. The \textit{fill blue histograms} stand for the results determined with the \textsc{synthesizer} code. The \textit{open green histograms} show the results obtained with the \textsc{cigale}.\label{fig:hstop}}
\end{figure*}

\subsubsection{Discussion about the intensity of the most recent burst}
Further discussion about how much mass the starburst phase adds can be
presented based on the burst strengths fitted by our modelling
procedure. The typical recent burst intensities ($b^\mathrm{FIR}_\mathrm{you}$) are
$0.34^{+0.26}_{-0.18}$ and $0.50^{+0.10}_{-0.21}$ (similar to the 
stellar-mass increase derived in Section \ref{sec:dyopo}), with median uncertainties 
of $\sim23$ and $\sim54$ per cent for the
\textsc{synthesizer} code and \textsc{cigale}, respectively. We do not find an
obvious correlation between the two estimations.

\citet{gio11} used \textsc{cigale} to study a 24-$\micron$ selected
sample at $z\sim 0.7$, with some galaxies including MIPS-70
detections.  They explored burst intensities (their $f_\mathrm{ySP}$) from 0 to 0.99 finding 
a broad distribution with a long tail towards high values for this 
parameter. In the work of \citet{bua14}, also carried out
with \textsc{cigale}, they explored $f_\mathrm{ySP}$ until 0.5, 
which is compatible with most of our objects
(73 and 51 per cent of all galaxies from the analysis with the
\textsc{synthesizer} code and the \textsc{cigale}).  We should
take in mind that we are selecting sources in the ULIRG regime, and these objects should
be suffering vigorous starburst events (see, e.g., \citealt{row14,lem14}).
Therefore, we consider that our $b^\mathrm{FIR}_\mathrm{you}$ results are 
within the typical values (0.2--0.6) of dSFGs.

\subsection{Discussion about the properties of the old stellar
  population}

In this Subsection, we present the main properties of the old stellar
population in our sample of dusty massive starburst galaxies at
$z\sim1$ according to \textsc{synthesizer} and \textsc{cigale}. In
particular, we discuss the attenuation
($A_{V,\mathrm{old}}^\mathrm{FIR}$), the age
($t_\mathrm{old}^\mathrm{FIR}$), and the exponential decay factor
($\tau_\mathrm{old}^\mathrm{FIR}$). The distributions of values for
these parameters are shown in Fig. \ref{fig:hstop}, and their median 
solutions are presented in Table \ref{tab:parmedsyci}, both for the
whole sample of massive dusty starburst galaxies.

\begin{table}
\begin{center}
\caption[Median, and 16 and 84-percentile values of the FIR-prior fitted parameters.]{Median, and 16 and 84-percentile values of the FIR-prior fitted parameters for the full-sample of dusty starbursts. 
From top to bottom, attenuation, age and burst intensity for the young population, attenuation, age and $e$-folding time for 
the old population, stellar mass, mass-weighted age and SED-derived 10-Myr-averaged \textit{SFR}.}
\label{tab:parmedsyci}
\begin{tabular}{lcc}
\hline\hline
\multicolumn{1}{c}{Parameter} & \multicolumn{1}{c}{\textsc{cigale}} & \multicolumn{1}{c}{\textsc{synthesizer}} \\
\hline
\vspace{3pt} $A_{V,\mathrm{you}}^\mathrm{FIR}$ [mag] & $2.3^{+0.7}_{-0.6}$ & $2.4^{+0.7}_{-0.7}$ \\
\vspace{3pt} $t_\mathrm{you}^\mathrm{FIR}$ [Myr] & $116^{+135}_{-71}$ & $51^{+34}_{-42}$ \\
\vspace{3pt} $b_\mathrm{you}^\mathrm{FIR}$ [\%] & $50^{+10}_{-21}$ & $34^{+25}_{-18}$ \\
\vspace{3pt} $A_{V,\mathrm{old}}^\mathrm{FIR}$ [mag] & $0.9^{+0.3}_{-0.8}$ & $0.1^{+0.7}_{0.1}$ \\
\vspace{3pt} $t_\mathrm{old}^\mathrm{FIR}$ [Gyr] & $3.3^{+0.8}_{-0.5}$ & $1.4^{+1.1}_{-0.4}$ \\
\vspace{3pt} $\tau_\mathrm{old}^\mathrm{FIR}$ [Myr] & $952^{+457}_{-833}$ & $158^{+63}_{-42}$ \\
\vspace{3pt} $\log(M_\star^\mathrm{FIR}/\mathrm{M}_{\sun})$ & $10.71^{+0.30}_{-0.41}$ & $10.58^{+0.28}_{-0.37}$ \\
\vspace{3pt} $t_M^\mathrm{FIR}$ [Gyr] & $0.9^{+0.9}_{-0.3}$ & $0.8^{+0.7}_{-0.3}$ \\
\vspace{3pt} $SFR_\mathrm{FIR}^{10}$ [$\mathrm{M}_{\sun}~\mathrm{yr}^{-1}$] & $196^{+105}_{-65}$ & $241^{+155}_{-74}$ \\
\hline
\end{tabular}
\end{center}
\end{table}

\subsubsection{Discussion about the attenuation of the old stellar population}

The median old-population attenuations in our sample
of massive dusty starburst are 
$0.1^{+0.7}_{-0.1}$ and $0.9^{+0.3}_{-0.8}$~mag, with median
errors of 0.1 and 0.6~mag for \textsc{synthesizer} and
\textsc{cigale}. We do not find a perceptible correlation 
between the two estimates.
The \textsc{synthesizer} median is compatible with
the SED-derived median attenuation of quiescent galaxies at $0.1 \leq
z \leq 1.1$, $A_V\sim 0.15-0.3$~mag \citep{dia19}. Although the
results of both codes range in comparable intervals
(see Section \ref{sec:fposp}), the
\textsc{cigale} yields $A_{V,\mathrm{old}}^\mathrm{CIG,FIR}>1.0$~mag
for 40 per cent of all galaxies. Such values seem to be too large for old
populations with ages $\sim3-4$~Gyr. However, \citet{mal17} found
average $A_{V,\mathrm{old}}^\mathrm{CIG,FIR}\sim0.7$~mag for a sample
of (U)LIRGs at $0.1<z<1.2$ examined with \textsc{cigale}, which is
compatible with our findings. The \textsc{synthesizer} code only
yields $A_{V,\mathrm{old}}^\mathrm{SYN,FIR}>1.0$~mag for 13 per cent of all
objects. A fraction of 50 per cent of all sources presents
$A_{V,\mathrm{old}}^\mathrm{SYN,FIR}<0.1$~mag from the
\textsc{synthesizer} code, whereas only 15 per cent fulfils this condition
from the results of \textsc{cigale}. We remark that the attenuation of
the old population is linked to the attenuation of the younger
population for \textsc{cigale}, while it is a completely independent
parameter for \textsc{synthesizer}. Considering all of the above, 
we observe that the old-population attenuation
is not constrained accurately, the 68th percentile range of the difference
between the values derived with \textsc{cigale} and \textsc{synthesizer} is
$0.8$ mag. Our analysis indicates that the old population attenuation is
around $0.1-0.9$~mag, most probably in the lowest half of that interval.

\subsubsection{Discussion about the age of the old stellar population}\label{sec:opoag}
The median old stellar ages of our dusty starburst galaxies 
are $1.4^{+1.1}_{-0.4}$ and $3.3^{+0.8}_{-0.5}$~Gyr, 
with median uncertainties of $\sim20$ and $\sim47$ per cent for \textsc{synthesizer} and
\textsc{cigale}. There is a mild correlation between each set of computed
values with Spearman correlation coefficient of $r_s=0.45$ and an associated 
probability of no correlation of $p_s=7.0\times 10^{-7}$.
The \textsc{synthesizer} code tends to obtain shorter
old-population ages for a significant part of the sample. We note
that these younger ages are accompanied with shorter $e$-folding times (cf. Fig \ref{fig:hstop}),
which are described in the Subsection \ref{sec:opotau}. Ages and $e$-folding times are,
indeed, quite degenerate. If we combine both properties in a single
parameter, a mass-weighted age, \textsc{synthesizer} and
\textsc{cigale} provide much more similar results: the medians are 
$0.8^{+0.7}_{-0.3}$ and $0.9^{+0.9}_{-0.3}$~Gyr, respectively.

\subsubsection{Discussion about the e-folding time of the old stellar population}\label{sec:opotau}
The old stellar population \textit{e}-folding times of our dusty starbursts range in
$101<\tau^\mathrm{SYN,FIR}_\mathrm{old}<249$ and
$100\le\tau^\mathrm{CIG,FIR}_\mathrm{old}<3398$~Myr, with
medians of $158^{+63}_{-42}$ and $952^{+457}_{-833}$~Myr, 
and median errors of $\sim25$ and $\sim150$ per cent for the
\textsc{synthesizer} and \textsc{cigale} codes. 
We do not find an obvious correlation between the two estimations.
Although the values of $\tau_\mathrm{old}^\mathrm{FIR}$ derived with
each code are similar for $\sim40$ per cent of all galaxies (with differences
within a factor of 2), the \textsc{cigale} systematically finds higher
values of $\tau_\mathrm{old}^\mathrm{FIR}$ for most objects, compared
to the results provided by \textsc{synthesizer}. Such differences are
compensated in \textsc{cigale} with less intense initial bursts.  This
effect, at least in part, could be caused by the few discrete values
examined in the \textsc{cigale} case (0.1, 1, 3, and 10~Gyr). We have
used in this study a similar set of
$\tau^\mathrm{CIG,FIR}_\mathrm{old}$ values to the ones explored in
several works using the \textsc{cigale} (see, e.g.,
\citealt{nol09,gio11,bua14}). In such studies, those authors found
that the $\tau^\mathrm{CIG,FIR}_\mathrm{old}$ values are
unsatisfactory estimated, and almost unconstrained.

\subsection{Comparison of stellar masses and \textit{SFRs} from 2P-FIR models}

Now we compare stellar masses and \textit{SFRs} derived using the FIR
prior with the \textsc{synthesizer} and \textsc{cigale} codes for our 
sample of massive dusty starburst galaxies. Such
parameters are contrasted with results from the literature. To do so, we 
account for different IMF and SPS models assumptions when possible\footnote{We
 transform from a \citet{cha03} IMF to a \citet{sal55} one multiplying by a 
factor of 1.6 \citep{sal09,mar09,bar11b}. Stellar masses derived with the
\citet{bru03} models have been shown to be a factor $\sim1.5$ larger than those 
determined with the \citet{mar05a} ones \citep{muz09,hai11,cas14}. Thus, 
$M_{\star}^\mathrm{M05,SAL} \simeq M_{\star}^\mathrm{BC03,CHA}$ and, consequently, 
\textit{SFRs}.}.

As mentioned in Section~\ref{sec:mssfr2pc}, the typical stellar masses
of our sample are $3.8\times 10^{10}$ and $5.2\times 10^{10}~\mathrm{M}_{\sun}$,
for \textsc{synthesizer} and \textsc{cigale}, respectively. On median, stellar
masses calculated with \textsc{cigale} are 37 per cent larger than those
obtained with \textsc{synthesizer}. We find that this mass ratio increases
as the old-population age ratio, $t^\mathrm{CIG,FIR}_\mathrm{old}/t^\mathrm{SYN,FIR}_\mathrm{old}$,
rises ($r_s=0.59$, $p_s=1.04\times 10^{-11}$). Then, this $\sim0.1$~dex difference
can be attributed to the larger old populations ages derived with \textsc{cigale} 
(see Section \ref{sec:opoag}). The larger the old-population ages, the more stellar
mass that may be enshrouded in low-mass stars \citep{hai11,mic14,dac15}.

We compared our estimates with stellar masses derived for
MIR/FIR-selected (U)LIRGs at similar redshifts.  \citet{mel08} studied
15 LIRGS at $z\sim0.8$ selected at $24~\micron$ in the GOODS-S HST
treasury field. They estimated a median stellar mass of $5.9\times
10^{10}~\mathrm{M}_{\sun}$ for such LIRGs.  \citet{gio11} examined 172 LIRGS
selected at $24~\micron$ in ECDF-S at $z\sim0.7$. They found that the
distribution of their stellar mass estimates peaks at $6.3\times
10^{10}~\mathrm{M}_{\sun}$. \citet{lem14} analysed a
\textit{Herschel}/SPIRE-selected sample of $\sim2000$ galaxies from
$0<z<4$ in the CFHTLS-D1 field. Cutting their sample to ULIRGs at
$0.7<z<1.2$, we found 109 sources with a median stellar mass of
$4.2\times 10^{10}~\mathrm{M}_{\sun}$.

We have also contrasted our stellar mass values with those of SMGs. 
\citet{row14} found 26 (U)LIRGs 
in their 250-$\micron$ selected sample with a median redshift
$z\simeq 2$. They estimated for these sources a median stellar mass
of $5.1\times 10^{10}~\mathrm{M}_{\sun}$. \citet{dac15} reported 86 (U)LIRGs in
the analysis of the ALESS sample with a median redshift of 2.7. They
derived a median stellar mass of $5.7\times 10^{10}~\mathrm{M}_{\sun}$. 
\citet{mie17} found 102 (U)LIRGs in their 1.3-mm selected sample in the COSMOS field with a median
redshift of $z\simeq 2.1$. They derived a median stellar mass of 
$1.3\times 10^{11}~\mathrm{M}_{\sun}$. When considering only the 18 (U)LIRGs 
with 68th percentile range of $z$ compatible with lying in $0.7<z<1.2$, 
the median stellar mass is $6.4\times 10^{10}~\mathrm{M}_{\sun}$.
\citet{dud20} reported 689 (U)LIRGs in their 870-$\micron$ selected sample with
a median redshift of $z\simeq 2.6$. They estimated a median stellar mass of
$1.2\times 10^{11}~\mathrm{M}_{\sun}$. When comparing only the 20 (U)LIRGs 
with $1\sigma$ range of $z$ consistent with lying in $0.7<z<1.2$, the median 
stellar mass is $7.1\times 10^{10}~\mathrm{M}_{\sun}$.

\citet{mar16} built a stellar-mass complete sample using the
UltraVISTA DR1 and 3D-HST photometric catalogues. They used optical
colours to study the fraction of dSFGs as a
function of stellar mass at $0.2<z<3.0$. They found that at $z<1.5$,
dSFGs dominate the galaxy population in the
stellar mass range $1\times 10^{10} <M_\star/\mathrm{M}_{\sun}<3.9\times 10^{10}$.
They stated that dSFGs are a factor $\sim 3-5$
more abundant at $M_\star>3.9 \times 10^{10}~\mathrm{M}_{\sun}$ than unobscured
star-forming galaxies.

Considering the stellar-mass values determined in the aforenamed
studies, we conclude that the stellar masses derived for our sample
using both codes are compatible with those of (U)LIRGs and
optical-selected dSFGs at $0.7<z<1.2$, and also
with those of SMGs at $z>1$.

\begin{figure}
\centering
\includegraphics[width=\columnwidth]{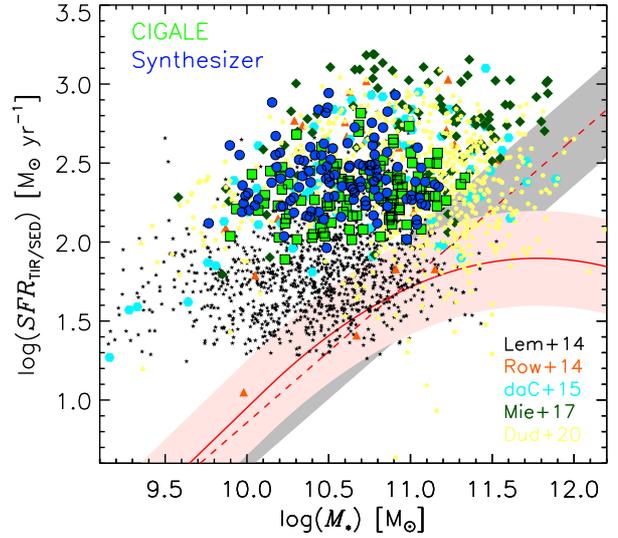}
\caption{Stellar mass, $M_\star$, vs. $SFR$ determined with the \citet{ken98} calibration, $SFR_\mathrm{TIR}$, or $SFR$ based on SED 
analysis, $SFR_\mathrm{SED}$ (see the text). The \textit{filled circles} show the median values estimated with the \textsc{synthesizer} 
code. The \textit{filled squares} show the expected values estimated with the \textsc{cigale}. The median values of (U)LIRGS from several 
FIR/sub-mm selected samples are also shown: \citet{lem14} as \textit{black stars}, \citet{row14} as \textit{orange triangles}, 
\citet{dac15} as \textit{cyan hexagons}, \citet{mie17} as \textit{dark green diamonds}, and \citet{dud20} as \textit{pale yellow 
pentagons}. Error bars are omitted for clarity. We also depict the relationships for the main sequence at $z\sim1$ of 
\citet[\textit{dashed red line}]{elb07} with its 68\ confidence level \textit{(grey shaded area)}, and of \citet[\textit{solid red line}]
{sch15} with its 0.3 dex scatter \textit{(rose shaded area)}.\label{fig:mssfr}}
\end{figure}
Concerning \textit{SFRs}, the typical SED-based value is
$200-250~M_{\sun}\mathrm{yr}^{-1}$, with very similar values provided by the 2
codes. 

\begin{figure*}
\centering
\subfloat{\includegraphics[width=0.495\textwidth]{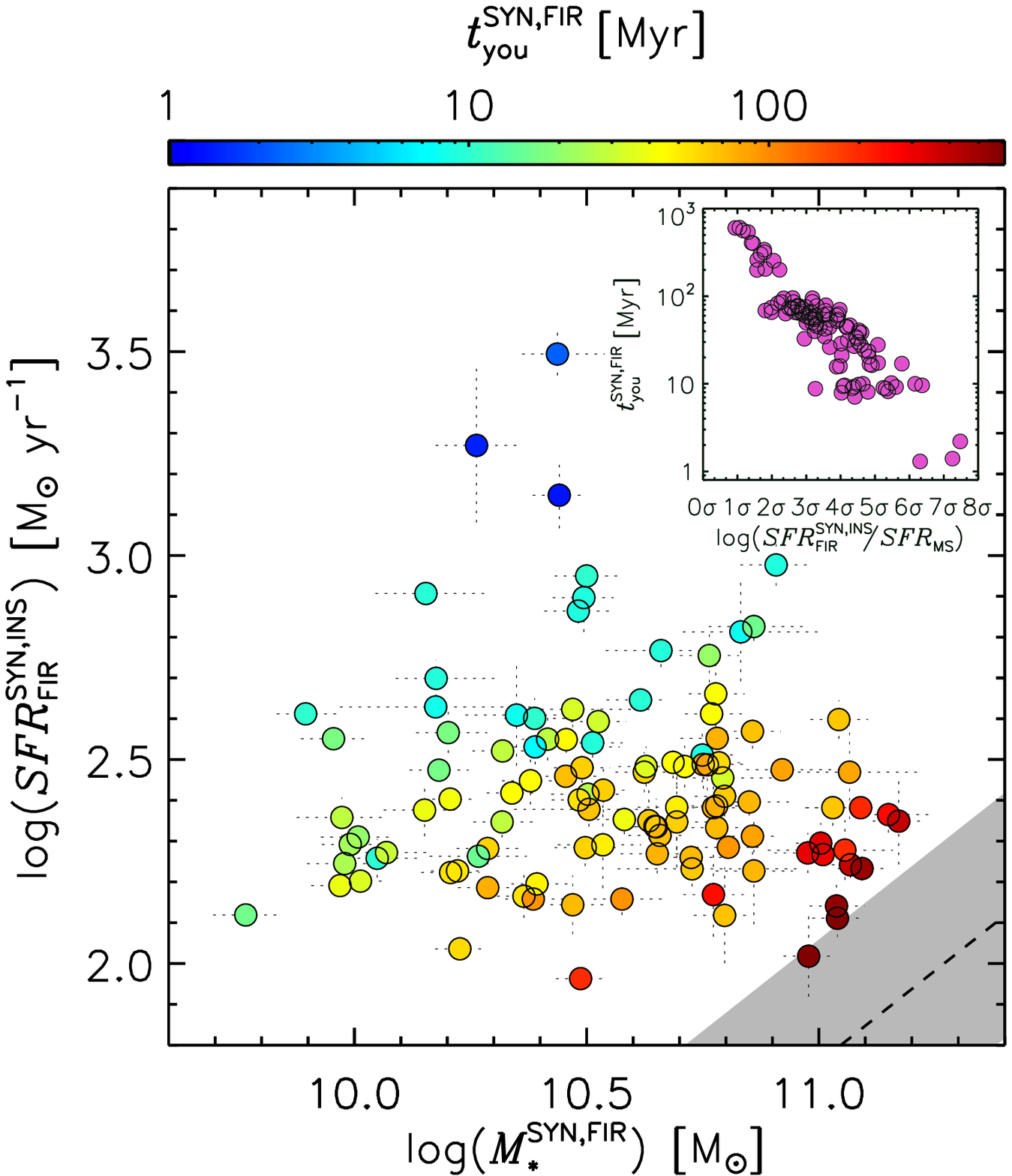}}\hfill
\subfloat{\includegraphics[width=0.495\textwidth]{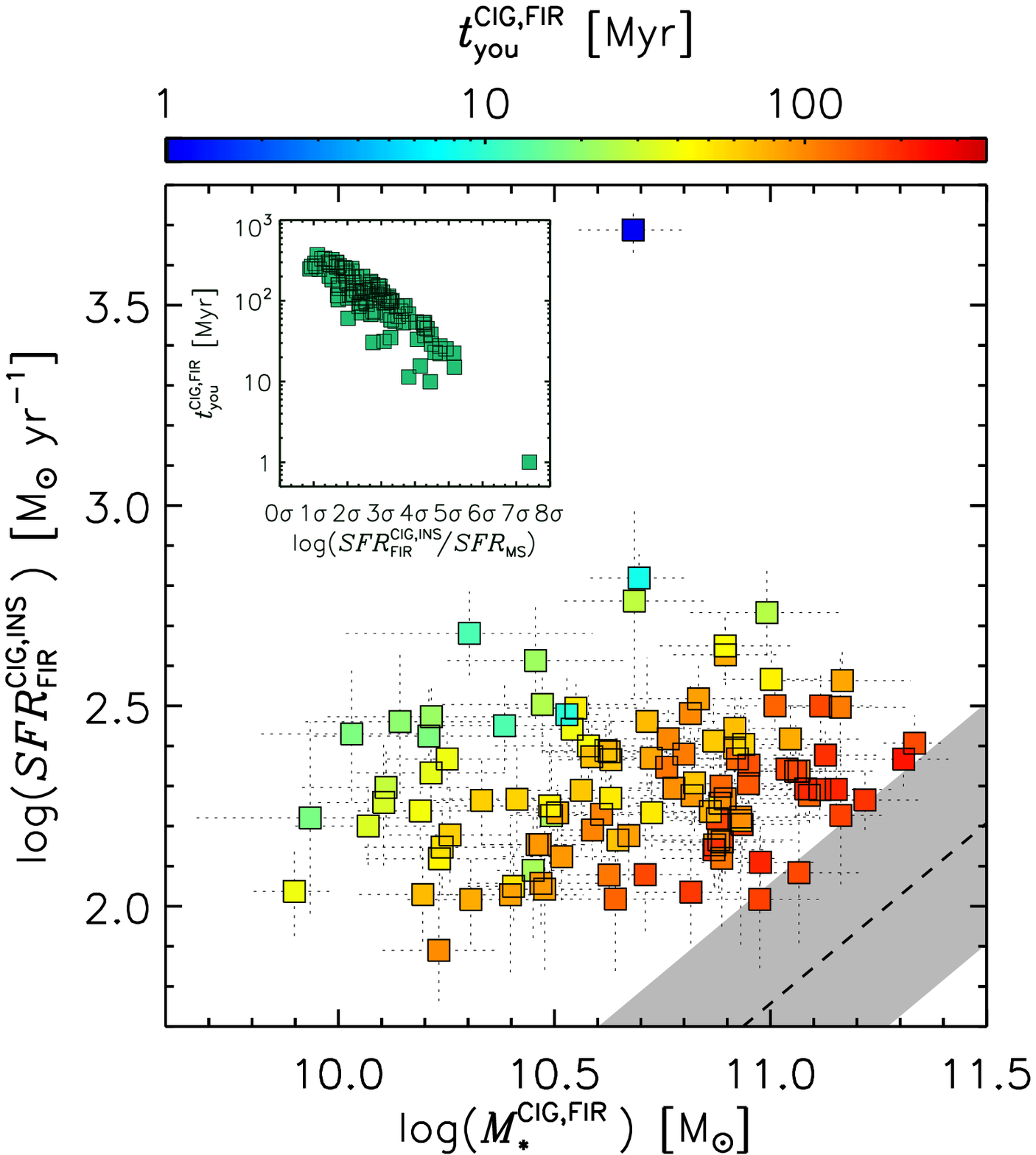}}\\
\caption{$M_\star$ vs. $SFR^\mathrm{INS}$ estimated with the FIR prior, computed with the \textsc{synthesizer} \textit{(left panel)} and the \textsc{cigale} \textit{(right panel)} codes. The filled circles with error bars depict the median values and $1\sigma$ uncertainties determined with the \textsc{synthesizer} code. The filled squares with error bars stand for the expected values and standard deviations derived with the \textsc{cigale}. The young-population age, $t_\mathrm{you}^\mathrm{FIR}$, is differentiated by colour for each code. The \textit{dashed black line} represents the relationship for the main sequence at $z\sim1$ of \citet{elb07} with its 68\% confidence level marked with the \textit{grey shaded area}. The insets show that as the objects separate from the MS they present shorter young-population ages.\label{fig:mssfra}}
\end{figure*}

Now we compare our sample results with \textit{SFR} estimations
from the literature.  \citet{lem14} determined \textit{SFRs} using the
\citet{ken98} calibration. We selected ULIRGs at $0.7<z<1.2$ from
their sample, obtaining a median of
$SFR^\mathrm{Lem}_\mathrm{TIR}=260^{+115}_{-67}$~$\mathrm{M}_{\sun}~\mathrm{yr}^{-1}$. 
\citet{row14} and \citet{dac15} presented \textit{SFRs}
derived using the \textsc{magphys} code for their SMGs, 
with medians of $SFR^\mathrm{Row}_\mathrm{SED}=407^{+168}_{-339}$ and
$SFR^\mathrm{daC}_\mathrm{SED}=251^{+304}_{-122}$~$\mathrm{M}_{\sun}~\mathrm{yr}^{-1}$ 
for the aforementioned (U)LIRGs, respectively. \citet{mie17} estimated 
\textit{SFRs} utilising the \citet{ken98} factor too. They report a median
of $SFR^\mathrm{Mie}_\mathrm{TIR}=566^{+492}_{-302}$~$\mathrm{M}_{\sun}~\mathrm{yr}^{-1}$
for (U)LIRGs. When considering only the 18 (U)LIRGs with 68th percentile range of $z$ 
compatible with lying in $0.7<z<1.2$, the median value is $270^{+186}_{-187}$~$\mathrm{M}_{\sun}~\mathrm{yr}^{-1}$.
\citet{dud20} derived \textit{SFRs} using \textsc{magphys} too, 
finding a median of $SFR^\mathrm{Dud}_\mathrm{SED}=233^{+216}_{-120}$~$\mathrm{M}_{\sun}~\mathrm{yr}^{-1}$ 
for (U)LIRGs.

Considering the \textit{SFRs} estimated in the aforementioned 
studies for our $L^\mathrm{ref}_\mathrm{TIR}$ range, we conclude that
the \textit{SFRs} determined for our sample are compatible with those
of \textit{Herschel}-selected ULIRGs at $0.7< z< 1.2$, and $1\sigma$ lower-limit values of 
$z>1$ sub-mm sources.

A natural extension after comparing stellar masses and \textit{SFRs}
is to place the galaxies of the aforenamed 6 samples in the $SFR-M_\star$
plane, that we show in Fig. \ref{fig:mssfr}. Given that our massive dusty starburst 
are at $0.7\leq z \loa 1.2$, we have also plotted in this Figure, the $SFR-M_\star$
relation of \citet{elb07} and the one of \citet{sch15}, both at $z\sim
1$. Considering our selection criteria FIR-detected dSFGs galaxies, 1$\sigma$
away from the $z\sim1$ MS, what we should recognize are sources above or to the left of such MS.
In general, samples selected by means of the dust emission (MIR/FIR or sub-mm 
selected) are expected to be found above or to the left of the MS (taking into account
 their median redshift), or lying on the MS. This is what is observed in 
Fig.~\ref{fig:mssfr}, also for the \citet{lem14} sample, and it is also found
in other works devoted to MIR/FIR selected samples (see, e.g., \citealt{gio11,lee13}).

Our dusty starbursts occupy a space in the $SFR-M_\star$ plane which is also traversed
by some galaxies of the sub-mm selected samples (see Fig. \ref{fig:mssfr}); although these sets
typically present larger stellar masses and \textit{SFRs}, reflecting their high redshift 
nature. We should also mention that experiencing short ($100-250$~Myr) bursts is more common for a 
massive galaxy at $z\sim 2$ \citep{row14,lem14,dud20}.

\subsection{Interpreting the position of our dusty starbursts in the $SFR-M_*$ plane}

Taking into account the way we selected massive dusty starbursts (see Section \ref{sec:sbsel}), 
we should consider the merger scenario of galaxy formation. Local
ULIRGs are mostly late-stage mergers, which trigger noteworthy star 
formation \citep{san96,tac02}. At $L_\mathrm{TIR} \geq 10^{11.5-12}~\mathrm{L}_{\sun}$ and 
$z\sim 1-2$, a fraction of $\sim 50-70$ per cent of sources selected with \textit{Herschel} shows 
signatures of mergers/interactions \citep{kar12,hun13}. Half of the $z\sim2$ 
\textit{Herschel}-selected (U)LIRGs defined as starbursts ($SFR/SFR_\mathrm{MS}>3$) presents clear
merger features \citep{kar12}, with the fraction of such
interacting systems increasing with the deviation above the MS \citep{hun13}.
A careful (varying the cuts) visual inspection of the \textit{Subaru} 
Suprime-Cam images of our whole sample shows sources in close pairs 
at similar redshift (see Section \ref{sec:sbsel}) or tidal (tails, shells, debris) features, 
asymmetry and multiple nuclei for $\sim 80-90$ per cent of our dusty
starbursts. These characteristics are typically originated by interactions. 

Off-main sequence galaxies are found at all redshifts, and their
position in the $SFR-M_*$ plane is used to define starburst galaxies.
The definition of starburst galaxies --based on \textit{SFR} offsets 
from the MS-- varies in the literature. 
\citet{rod11} fitted Gaussians to the logarithmic distributions of 
\textit{SFRs} at fixed stellar mass of a sample of galaxies with 
optical-to-FIR observations in the GOODS-S and COSMOS fields. Using 
these curves, they chose to define starburst galaxies as 
those sources with \textit{SFR} more than 4-times that on the MS.
According to this definition and using the $z=1$ \citet{elb07} MS, 
86 and 73 per cent of the sources in our full sample are starbursts
from the solutions of the \textsc{synthesizer} code and the \textsc{cigale}.
\citet{sch15} analysed \textit{Herschel}
images in several extragalactic fields using individual detections and
stacking. With this analysis they defined the MS from $z=4$ to 0.
Such $M_{\star}-SFR$ relation at $z=1$ is shown in Fig. \ref{fig:mssfr}. In their
study, starburst galaxies are defined as objects having 5-times larger 
\textit{SFR} than the \textit{SFR} of the MS at the redshift of each
galaxy. According to such definition 84 and 60 per cent of the galaxies in
our whole sample are starbursts from the results of \textsc{synthesizer}
 and \textsc{cigale}. Then, more than a half of the dSFGs in our sample 
satisfies both of the aforementioned starburst definitions.

\citet{cib19} analysed close pairs and morphological mergers on and above the MS 
at $0.2 \leq z \leq 2.0$ presenting UV-to-FIR data in the GOODS-N field.
They found a merger fraction $\goa70$ per cent for starburst
galaxies identified with the \citeauthor{rod11} definition. When excluding
objects in our full sample with \textit{SFR} less than 4-times that on the MS,
we observe fractions of $\sim 60$ and $\sim 70$ per cent of merger-like sources
from the \textsc{cigale} and the \textsc{synthesizer} code solutions. 
Then, our results are consistent with those on the literature \citep{kar12,cib19}.

The merger scenario as the origin of the starburst event is debated.
\citet{man16} furnished an alternative physical interpretation on the shape and
scatter of the MS of star-forming galaxies at $z\gtrsim1$.
In such explanation, young galaxies are located to the left of the MS
at a given \textit{SFR}. As time passes, they will advance at almost
constant \textit{SFR} to their MS position spending most of their life
there. Thus, off-main sequence galaxies are depicted as young sources
that have to accumulate a large fraction of their stellar mass. An
average mass-selected source lies on the MS, while FIR-selected
objects are caught on it or above. Thus, the youngest galaxies should
be located at the largest distances from the MS.

The \citeauthor{man16} interpretation is consistent with results from cosmological 
simulations, which have shown that the bulk of the
star formation in galaxies is \textit{in situ}. \citet{mar17} found 
a star-formation fraction of $\sim65$ and $\sim80$ per cent 
at $z\sim3$ and $z\sim1$ attributable to smooth gas accretion. 
For galaxies with $M_{\star} \loa 10^{11} \mathrm{M}_{\sun}$, a fraction 
of $\sim60-90$ per cent of the star formation occurs in situ \citep{lac12,pil18}.

In Figure \ref{fig:mssfra}, we show the loci of our dusty starbursts
in the $SFR-M_*$ plane, which agrees with the \citeauthor{man16}
scenario due to the SFH we assumed. 
In the aforenamed Figure (see insets), we notice that the
distance to the MS is very well correlated with $t_\mathrm{you}^\mathrm{FIR}$
($r_s=-0.88$ and $-0.86$ with $p_s=6.1\times 10^{-38}$ and $1.7\times 10^{-34}$,
for \textsc{synthesizer} and \textsc{cigale}, respectively). 
We corroborate that as galaxies are located
far above the MS, they present younger $t_\mathrm{you}^\mathrm{FIR}$
values, as expected due to our SFH formulation. We distinguish objects with
similar \textit{SFR} differing more than an order of magnitude in
stellar mass in Fig. \ref{fig:mssfra}, as these objects move to the
right approaching the MS, their young-population ages become larger.
 Thus by definiton, a larger
\textit{SFR} value can only be reached with a shorter 
$t_\mathrm{you}^\mathrm{FIR}$ value for the same stellar mass.
On the other hand, a larger stellar mass value can
only be achieved with a larger $t_\mathrm{you}^\mathrm{FIR}$
value for a constant \textit{SFR}. \citeauthor{man16} used a SFH 
defined by a slow power-law increase of
the \textit{SFR} (which can be approximated by a constant
\textit{SFR}) over a time-scale $\tau_\mathrm{B}<0.5-1$ Gyr, followed
by an exponential decline due to AGN feedback. Thus, our SFH is
similar to theirs in the almost constant part, which can explain the
similarity between both results.

Therefore, our findings and our assumed SFH show that the merger and the in situ 
star-formation scenarios are not mutually exclusive. The elevated \textit{SFRs} of our 
dusty starbursts are likely triggered by merger events. Then, these
starburst galaxies probably continue assembling stellar mass at a 
roughly constant \textit{SFR} depending on the available gas supply 
provided by smooth accretion.

\section{Summary and conclusions}\label{sec:clo}

We have studied in detail the stellar-population properties of massive ($\log(M_{\star}/\mathrm{M}_{\sun}) \ge 10$)
dusty (FIR-selected) starburst ($SFR/SFR_\mathrm{MS} > 2$) galaxies at $0.7 < z < 1.2$ 
by analysing their UV-to-FIR observed SEDs. The sample is based in at least 2 FIR flux measurements 
(above the $4\sigma$ threshold) from \textit{Spitzer}/MIPS-70 and/or 
\textit{Herschel}/PACS and SPIRE, and a MIPS-24 detection in the SXDS/UDS 
field. The multi-wavelength data has been combined by building catalogues 
in \textit{Subaru}/Suprime-Cam (\textit{B}, \textit{V}, \textit{R}$_\mathrm{c}$,
\textit{i}$'$ and \textit{z}$'$), \textit{UKIRT}/WFCAM (\textit{J}, \textit{H} and 
\textit{K}), \textit{Spitzer}/IRAC (3.6, 4.5, 5.8 and 8.0~$\micron$) and MIPS (24 and
70~$\micron$), and \textit{Herschel}/PACS (100 and 160~$\micron$) and SPIRE (250, 
350 and 500~$\micron$) bands. The full UV-to-FIR SEDs are fitted to stellar population
and dust emission models using codes (\textsc{synthesizer} and \textsc{cigale}) 
which manage the attenuation of stellar light and dust re-emission with energy 
balance techniques. We have assumed as fiducial a SFH depicted by a young stellar 
population forming stars in a roughly constant manner on top on an evolved population
each parametrized with and exponentially decaying-$\tau$ form. We have compared the 
best-fitting results obtained with and without FIR information. When using FIR data, 
their associated $L_\mathrm{TIR}$ value is used to constrain the amount of stellar-light
attenuation (what we call ``the FIR prior''). 
A summary of the main results and conclusions of our work follows:
\begin{itemize}
\item Excluding the FIR data in two population models results mainly in an overestimation
of the attenuation of the young population, $A_{V,\mathrm{you}}$, with a median value of 
0.4~mag. This overestimation translates in underestimating the young-population ages by a
factor $\sim6$ in median. Both misestimations cause overestimating the \textit{SFRs} by
0.38~dex (a factor of $\sim2.4$) in median. This evidences the efficacy of the  ``FIR 
prior'' in breaking the age-attenuation degeneracy, in providing reliable attenuation and 
age values for the young population, and in improving the determinations of the SFHs of 
massive dusty starburst galaxies.
\item The attenuation values of the young population, $A^\mathrm{FIR}_{V,\mathrm{you}}$, 
determined with both codes are compatible and fall in the range $\sim 1.5-3.5$~mag, 
with most of the galaxies (70 per cent) presenting 
 $A^\mathrm{FIR}_{V,\mathrm{you}} > 2$~mag. The median difference between codes is $<0.1$~mag.
The median value is $2.4$~mag, which is similar with median values determined for $z>1$ SMGs.
This shows that the energy-balance technique of the 2 different codes produces consistent 
results. 
\item The ages of the young population, $t^\mathrm{FIR}_\mathrm{you}$, are shorter than 
$\sim 400$ Myr, with median values of $\sim 50$ and $\sim 120$ Myr for \textsc{synthesizer}
and \textsc{cigale}. Considering these median values, our fiducial SFH, and that, on average,
we are observing our galaxies at half their lifetimes in the starburst phase, we infer a 
duration of this phase of $\sim100-250$ Myr.
\item The determination for the old-population properties are affected by large uncertainties,
linked to degenerations and the fact that new stars outshine the old population through most
of the SED. The median old-population ages are $\sim 1.5-3$ Gyr. The median $e$-folding times 
values are $\sim 200$ and $\sim 900$ Myr for \textsc{synthesizer} and \textsc{cigale}. These
$e$-folding time differences are compensated with less intense initial burst in \textsc{cigale}.
The old populations are nearly unattenuated (median $A^\mathrm{SYN,FIR}_{V,\mathrm{old}}=0.1$~mag)
as derived from \textsc{synthesizer}, and present median $A^\mathrm{CIG,FIR}_{V,\mathrm{old}}=0.8$~mag
for \textsc{cigale}.
\item Regarding the stellar masses, both codes yield similar results, with median values of 
$3.8\times 10^{10}$ and $5.2\times 10^{10}~\mathrm{M}_{\sun}$ as derived from \textsc{synthesizer}
and \textsc{cigale}. This is expected considering that the codes use the same SFH, IMF and SPS models.
These stellar mass values are compatible with those of (U)LIRGs and optical-selected dSFGs at 
intermediate redshift, and with those of SMGs at $z > 1$.
\item Concerning \textit{SFRs}, the typical SED-derived value is $\sim 200-250~\mathrm{M}_{\sun}$yr$^{-1}$, 
which is compatible with those of \textit{Herschel} selected ULIRGs at $0.7 < z < 1.2$ and $1\sigma$
lower-limit values of $z>1$ SMGs. Assuming this typical \textit{SFR} and the inferred 
duration of the starburst phase, the stellar mass added during this phase should reach $\sim 2-3\times10^{10}~\mathrm{M}_{\sun}$.
This value corresponds to a fraction of $\sim30-50$ per cent of the median stellar mass derived for
our sample.
\item The position of our dusty starburst galaxies in the $SFR-M_{\star}$ plane is mainly determined by
the young population age. The galaxies located at the largest distances of the MS present younger
$t^\mathrm{FIR}_\mathrm{you}$ values. This is in agreement with previous physical interpretations of the
outliers of the $SFR-M_{\star}$ relation.
\end{itemize}
 
\section*{Acknowledgements}
We are grateful with the anonymous referee for improving the manuscript contents.  
N.E.-B. and J.Z. acknowledge funding support from the Spanish Programa Nacional de Astronom\'ia y Astrof\'isica under grant AYA2006-02358. 
N.E.-B. acknowledges support from Fac. CC. F\'isicas, Universidad Complutense de Madrid (UCM), and from Coordinaci\'on de 
Astrof\'isica, Instituto Nacional de Astrof\'isica \'Optica y Electr\'onica (INAOE). P.G.P.-G. acknowledges funding support from the Spanish Government under grant PGC2018-093499-B-I00. L.R.-M. acknowledges support from grant PRIN MIUR 2017-20173ML3WW\_001 and funding from the Universit\`a degli studi di Padova - Dipartimento di Fisica e  Astronomia ``G.  Galilei''. This work has made use of the \textsc{rainbow} Cosmological Surveys Database, which is operated by the Centro de Astrobiolog\'ia (CAB/CSIC-INTA).

\section*{Data availability}
The data underlying this article are available in the article and in its online supplementary material.



\bibliographystyle{mnras}
\bibliography{ms} 








\bsp	
\label{lastpage}
\end{document}


\label{firstpage}
\pagerange{\pageref{firstpage}--\pageref{lastpage}}




\small
\stepcounter{table}
\LTcapwidth=\linewidth
\begin{longtable}{lcccccccccc}
\caption[Dusty-starburst main-sample multi-band photometry]{Dusty-starburst main-sample multi-band photometry. (1) Name of the galaxy. (2,3) Right ascension and declination (J2000) in degrees. (4) Photometric or spectroscopic redshift ($z_\mathrm{spec}$ indicated by a $\dagger$). (33-39) Flux densities in MIPS 24, 70; PACS 100, 160; SPIRE 250, 350, 500 in mJy. (5-11,19-25) Observed magnitude in FUV, NUV; \textit{B}, \textit{V}, \textit{R}$_\mathrm{c}$, \textit{i}$'$, \textit{z}$'$; \textit{J}, \textit{H}, \textit{K}; IRAC 3.6-8.0 in the AB photometric system. (40-46) Associated uncertainties to MIPS 24, 70; PACS 100, 160; SPIRE 250, 350, 500 in mJy determined as described in Sections 2.1 and 2.2. (12-18,26-32) Associated uncertainties to FUV, NUV; \textit{B}, \textit{V}, \textit{R}$_\mathrm{c}$, \textit{i}$'$, \textit{z}$'$; \textit{J}, \textit{H}, \textit{K}; IRAC 3.6-8.0 as obtained from the aperture matched cataloguing (see Section 3.2). ``$\cdots$'' indicate bands without detections or without reliable photometric measurements.}\\
\hline \hline
\multicolumn{1}{c}{Galaxy} & \multicolumn{1}{c}{$\alpha$} & \multicolumn{1}{c}{$\delta$} & \multicolumn{1}{c}{$z^\dagger$} & \multicolumn{1}{c}{FUV} & \multicolumn{1}{c}{NUV} & \multicolumn{1}{c}{\textit{B}} & \multicolumn{1}{c}{\textit{V}} & \multicolumn{1}{c}{\textit{R}$_\mathrm{c}$} & \multicolumn{1}{c}{\textit{i}$'$} & \multicolumn{1}{c}{\textit{z}$'$} \\
 & & & & \multicolumn{1}{c}{$\Delta$FUV} & \multicolumn{1}{c}{$\Delta$NUV} & \multicolumn{1}{c}{$\Delta$\textit{B}} & \multicolumn{1}{c}{$\Delta$\textit{V}} & \multicolumn{1}{c}{$\Delta$\textit{R}$_\mathrm{c}$} & \multicolumn{1}{c}{$\Delta$\textit{i}$'$} & \multicolumn{1}{c}{$\Delta$\textit{z}$'$} \\
 & & & & \multicolumn{1}{c}{\textit{J}} & \multicolumn{1}{c}{\textit{H}} & \multicolumn{1}{c}{\textit{K}} & \multicolumn{1}{c}{[3.6]} & \multicolumn{1}{c}{[4.5]}  & \multicolumn{1}{c}{[5.8]} & \multicolumn{1}{c}{[8.0]} \\
 & & & & \multicolumn{1}{c}{$\Delta$\textit{J}} & \multicolumn{1}{c}{$\Delta$\textit{H}} & \multicolumn{1}{c}{$\Delta$\textit{K}} & \multicolumn{1}{c}{$\Delta$[3.6]} & \multicolumn{1}{c}{$\Delta$[4.5]} & \multicolumn{1}{c}{$\Delta$[5.8]} & \multicolumn{1}{c}{$\Delta$[8.0]} \\
 & & & & \multicolumn{1}{c}{\textit{S}$_{24}$} & \multicolumn{1}{c}{\textit{S}$_{70}$} & \multicolumn{1}{c}{\textit{S}$_{100}$} & \multicolumn{1}{c}{\textit{S}$_{160}$} & \multicolumn{1}{c}{\textit{S}$_{250}$} & \multicolumn{1}{c}{\textit{S}$_{350}$} & \multicolumn{1}{c}{\textit{S}$_{500}$} \\
 & & & & \multicolumn{1}{c}{$\Delta$\textit{S}$_{24}$} & \multicolumn{1}{c}{$\Delta$\textit{S}$_{70}$} & \multicolumn{1}{c}{$\Delta$\textit{S}$_{100}$} & \multicolumn{1}{c}{$\Delta$\textit{S}$_{160}$} & \multicolumn{1}{c}{$\Delta$\textit{S}$_{250}$} & \multicolumn{1}{c}{$\Delta$\textit{S}$_{350}$} & \multicolumn{1}{c}{$\Delta$\textit{S}$_{500}$} \\
\multicolumn{1}{c}{(1)} & \multicolumn{1}{c}{(2)} & \multicolumn{1}{c}{(3)} & \multicolumn{1}{c}{(4)} & \multicolumn{1}{c}{(5)} & \multicolumn{1}{c}{(6)} & \multicolumn{1}{c}{(7)} & \multicolumn{1}{c}{(8)} & \multicolumn{1}{c}{(9)} & \multicolumn{1}{c}{(10)} & \multicolumn{1}{c}{(11)} \\
 & & & & \multicolumn{1}{c}{(12)} & \multicolumn{1}{c}{(13)} & \multicolumn{1}{c}{(14)} & \multicolumn{1}{c}{(15)} & \multicolumn{1}{c}{(16)} & \multicolumn{1}{c}{(17)} & \multicolumn{1}{c}{(18)} \\
 & & & & \multicolumn{1}{c}{(19)} & \multicolumn{1}{c}{(20)} & \multicolumn{1}{c}{(21)} & \multicolumn{1}{c}{(22)} & \multicolumn{1}{c}{(23)} & \multicolumn{1}{c}{(24)} & \multicolumn{1}{c}{(25)} \\
 & & & & \multicolumn{1}{c}{(26)} & \multicolumn{1}{c}{(27)} & \multicolumn{1}{c}{(28)} & \multicolumn{1}{c}{(29)} & \multicolumn{1}{c}{(30)} & \multicolumn{1}{c}{(31)} & \multicolumn{1}{c}{(32)} \\
 & & & & \multicolumn{1}{c}{(33)} & \multicolumn{1}{c}{(34)} & \multicolumn{1}{c}{(35)} & \multicolumn{1}{c}{(36)} & \multicolumn{1}{c}{(37)} & \multicolumn{1}{c}{(38)} & \multicolumn{1}{c}{(39)} \\
 & & & & \multicolumn{1}{c}{(40)} & \multicolumn{1}{c}{(41)} & \multicolumn{1}{c}{(42)} & \multicolumn{1}{c}{(43)} & \multicolumn{1}{c}{(44)} & \multicolumn{1}{c}{(45)} & \multicolumn{1}{c}{(46)} \\ \hline
\endfirsthead

\multicolumn{11}{c}{\tablename \thetable{} -- continued from previous page}\\
\hline
\multicolumn{1}{c}{Galaxy} & \multicolumn{1}{c}{$\alpha$} & \multicolumn{1}{c}{$\delta$} & \multicolumn{1}{c}{$z^\dagger$} & \multicolumn{1}{c}{FUV} & \multicolumn{1}{c}{NUV} & \multicolumn{1}{c}{\textit{B}} & \multicolumn{1}{c}{\textit{V}} & \multicolumn{1}{c}{\textit{R}$_\mathrm{c}$} & \multicolumn{1}{c}{\textit{i}$'$} & \multicolumn{1}{c}{\textit{z}$'$} \\
 & & & & \multicolumn{1}{c}{$\Delta$FUV} & \multicolumn{1}{c}{$\Delta$NUV} & \multicolumn{1}{c}{$\Delta$\textit{B}} & \multicolumn{1}{c}{$\Delta$\textit{V}} & \multicolumn{1}{c}{$\Delta$\textit{R}$_\mathrm{c}$} & \multicolumn{1}{c}{$\Delta$\textit{i}$'$} & \multicolumn{1}{c}{$\Delta$\textit{z}$'$} \\
 & & & & \multicolumn{1}{c}{\textit{J}} & \multicolumn{1}{c}{\textit{H}} & \multicolumn{1}{c}{\textit{K}} & \multicolumn{1}{c}{[3.6]} & \multicolumn{1}{c}{[4.5]}  & \multicolumn{1}{c}{[5.8]} & \multicolumn{1}{c}{[8.0]} \\
 & & & & \multicolumn{1}{c}{$\Delta$\textit{J}} & \multicolumn{1}{c}{$\Delta$\textit{H}} & \multicolumn{1}{c}{$\Delta$\textit{K}} & \multicolumn{1}{c}{$\Delta$[3.6]} & \multicolumn{1}{c}{$\Delta$[4.5]} & \multicolumn{1}{c}{$\Delta$[5.8]} & \multicolumn{1}{c}{$\Delta$[8.0]} \\
 & & & & \multicolumn{1}{c}{\textit{S}$_{24}$} & \multicolumn{1}{c}{\textit{S}$_{70}$} & \multicolumn{1}{c}{\textit{S}$_{100}$} & \multicolumn{1}{c}{\textit{S}$_{160}$} & \multicolumn{1}{c}{\textit{S}$_{250}$} & \multicolumn{1}{c}{\textit{S}$_{350}$} & \multicolumn{1}{c}{\textit{S}$_{500}$} \\
\end{longtable}

\newpage
\addtocounter{table}{-1}
\begin{longtable}{lcccccccccc}
\caption[Dusty-starburst complementary-sample multi-band photometry]{Dusty-starburst complementary-sample multi-band photometry. (1) Name of the galaxy. (2,3) Right ascension and declination (J2000) in degrees. (4) Photometric or spectroscopic redshift ($z_\mathrm{spec}$ indicated by a $\dagger$). (33-39) Flux densities in MIPS 24, 70; PACS 100, 160; SPIRE 250, 350, 500 in mJy (syntethic fluxes indicated by $s$). (5-11,19-25) Observed magnitude in FUV, NUV; \textit{B}, \textit{V}, \textit{R}$_\mathrm{c}$, \textit{i}$'$, \textit{z}$'$; \textit{J}, \textit{H}, \textit{K}; IRAC 3.6-8.0 in the AB photometric system. (40-46) Associated uncertainties to MIPS 24, 70; PACS 100, 160; SPIRE 250, 350, 500 in mJy determined as described in Sections 2.1 and 2.2. (12-18,26-32) Associated uncertainties to FUV, NUV; \textit{B}, \textit{V}, \textit{R}$_\mathrm{c}$, \textit{i}$'$, \textit{z}$'$; \textit{J}, \textit{H}, \textit{K}; IRAC 3.6-8.0 as obtained from the aperture matched cataloguing (see Section 3.2). ``$\cdots$'' indicate bands without detections or without reliable photometric measurements.}\\
\hline \hline
\multicolumn{1}{c}{Galaxy} & \multicolumn{1}{c}{$\alpha$} & \multicolumn{1}{c}{$\delta$} & \multicolumn{1}{c}{$z^\dagger$} & \multicolumn{1}{c}{FUV} & \multicolumn{1}{c}{NUV} & \multicolumn{1}{c}{\textit{B}} & \multicolumn{1}{c}{\textit{V}} & \multicolumn{1}{c}{\textit{R}$_\mathrm{c}$} & \multicolumn{1}{c}{\textit{i}$'$} & \multicolumn{1}{c}{\textit{z}$'$} \\
 & & & & \multicolumn{1}{c}{$\Delta$FUV} & \multicolumn{1}{c}{$\Delta$NUV} & \multicolumn{1}{c}{$\Delta$\textit{B}} & \multicolumn{1}{c}{$\Delta$\textit{V}} & \multicolumn{1}{c}{$\Delta$\textit{R}$_\mathrm{c}$} & \multicolumn{1}{c}{$\Delta$\textit{i}$'$} & \multicolumn{1}{c}{$\Delta$\textit{z}$'$} \\
 & & & & \multicolumn{1}{c}{\textit{J}} & \multicolumn{1}{c}{\textit{H}} & \multicolumn{1}{c}{\textit{K}} & \multicolumn{1}{c}{[3.6]} & \multicolumn{1}{c}{[4.5]}  & \multicolumn{1}{c}{[5.8]} & \multicolumn{1}{c}{[8.0]} \\
 & & & & \multicolumn{1}{c}{$\Delta$\textit{J}} & \multicolumn{1}{c}{$\Delta$\textit{H}} & \multicolumn{1}{c}{$\Delta$\textit{K}} & \multicolumn{1}{c}{$\Delta$[3.6]} & \multicolumn{1}{c}{$\Delta$[4.5]} & \multicolumn{1}{c}{$\Delta$[5.8]} & \multicolumn{1}{c}{$\Delta$[8.0]} \\
 & & & & \multicolumn{1}{c}{\textit{S}$_{24}$} & \multicolumn{1}{c}{\textit{S}$_{70}$} & \multicolumn{1}{c}{\textit{S}$_{100}$} & \multicolumn{1}{c}{\textit{S}$_{160}$} & \multicolumn{1}{c}{\textit{S}$_{250}$} & \multicolumn{1}{c}{\textit{S}$_{350}$} & \multicolumn{1}{c}{\textit{S}$_{500}$} \\
 & & & & \multicolumn{1}{c}{$\Delta$\textit{S}$_{24}$} & \multicolumn{1}{c}{$\Delta$\textit{S}$_{70}$} & \multicolumn{1}{c}{$\Delta$\textit{S}$_{100}$} & \multicolumn{1}{c}{$\Delta$\textit{S}$_{160}$} & \multicolumn{1}{c}{$\Delta$\textit{S}$_{250}$} & \multicolumn{1}{c}{$\Delta$\textit{S}$_{350}$} & \multicolumn{1}{c}{$\Delta$\textit{S}$_{500}$} \\
\multicolumn{1}{c}{(1)} & \multicolumn{1}{c}{(2)} & \multicolumn{1}{c}{(3)} & \multicolumn{1}{c}{(4)} & \multicolumn{1}{c}{(5)} & \multicolumn{1}{c}{(6)} & \multicolumn{1}{c}{(7)} & \multicolumn{1}{c}{(8)} & \multicolumn{1}{c}{(9)} & \multicolumn{1}{c}{(10)} & \multicolumn{1}{c}{(11)} \\
 & & & & \multicolumn{1}{c}{(12)} & \multicolumn{1}{c}{(13)} & \multicolumn{1}{c}{(14)} & \multicolumn{1}{c}{(15)} & \multicolumn{1}{c}{(16)} & \multicolumn{1}{c}{(17)} & \multicolumn{1}{c}{(18)} \\
 & & & & \multicolumn{1}{c}{(19)} & \multicolumn{1}{c}{(20)} & \multicolumn{1}{c}{(21)} & \multicolumn{1}{c}{(22)} & \multicolumn{1}{c}{(23)} & \multicolumn{1}{c}{(24)} & \multicolumn{1}{c}{(25)} \\
 & & & & \multicolumn{1}{c}{(26)} & \multicolumn{1}{c}{(27)} & \multicolumn{1}{c}{(28)} & \multicolumn{1}{c}{(29)} & \multicolumn{1}{c}{(30)} & \multicolumn{1}{c}{(31)} & \multicolumn{1}{c}{(32)} \\
 & & & & \multicolumn{1}{c}{(33)} & \multicolumn{1}{c}{(34)} & \multicolumn{1}{c}{(35)} & \multicolumn{1}{c}{(36)} & \multicolumn{1}{c}{(37)} & \multicolumn{1}{c}{(38)} & \multicolumn{1}{c}{(39)} \\
 & & & & \multicolumn{1}{c}{(40)} & \multicolumn{1}{c}{(41)} & \multicolumn{1}{c}{(42)} & \multicolumn{1}{c}{(43)} & \multicolumn{1}{c}{(44)} & \multicolumn{1}{c}{(45)} & \multicolumn{1}{c}{(46)} \\ \hline
\endfirsthead

\multicolumn{11}{c}{\tablename \thetable{ } -- continued from previous page}\\
\hline
\multicolumn{1}{c}{Galaxy} & \multicolumn{1}{c}{$\alpha$} & \multicolumn{1}{c}{$\delta$} & \multicolumn{1}{c}{$z^\dagger$} & \multicolumn{1}{c}{FUV} & \multicolumn{1}{c}{NUV} & \multicolumn{1}{c}{\textit{B}} & \multicolumn{1}{c}{\textit{V}} & \multicolumn{1}{c}{\textit{R}$_\mathrm{c}$} & \multicolumn{1}{c}{\textit{i}$'$} & \multicolumn{1}{c}{\textit{z}$'$} \\
 & & & & \multicolumn{1}{c}{$\Delta$FUV} & \multicolumn{1}{c}{$\Delta$NUV} & \multicolumn{1}{c}{$\Delta$\textit{B}} & \multicolumn{1}{c}{$\Delta$\textit{V}} & \multicolumn{1}{c}{$\Delta$\textit{R}$_\mathrm{c}$} & \multicolumn{1}{c}{$\Delta$\textit{i}$'$} & \multicolumn{1}{c}{$\Delta$\textit{z}$'$} \\
 & & & & \multicolumn{1}{c}{\textit{J}} & \multicolumn{1}{c}{\textit{H}} & \multicolumn{1}{c}{\textit{K}} & \multicolumn{1}{c}{[3.6]} & \multicolumn{1}{c}{[4.5]}  & \multicolumn{1}{c}{[5.8]} & \multicolumn{1}{c}{[8.0]} \\
 & & & & \multicolumn{1}{c}{$\Delta$\textit{J}} & \multicolumn{1}{c}{$\Delta$\textit{H}} & \multicolumn{1}{c}{$\Delta$\textit{K}} & \multicolumn{1}{c}{$\Delta$[3.6]} & \multicolumn{1}{c}{$\Delta$[4.5]} & \multicolumn{1}{c}{$\Delta$[5.8]} & \multicolumn{1}{c}{$\Delta$[8.0]} \\
 & & & & \multicolumn{1}{c}{\textit{S}$_{24}$} & \multicolumn{1}{c}{\textit{S}$_{70}$} & \multicolumn{1}{c}{\textit{S}$_{100}$} & \multicolumn{1}{c}{\textit{S}$_{160}$} & \multicolumn{1}{c}{\textit{S}$_{250}$} & \multicolumn{1}{c}{\textit{S}$_{350}$} & \multicolumn{1}{c}{\textit{S}$_{500}$} \\
\hline
\end{longtable}

\newpage
\addtocounter{table}{2}
\begin{longtable}{lcllclllll}
\caption[Stellar population synthesis results for main-sample starbursts.]{Stellar population synthesis results for main-sample dusty starbursts. Median and $1\sigma$ error values are shown for each parameter derived with the \textsc{synthesizer} code. The expected and standard deviation (or upper and lower limits when changing from logarithmic space to linear space) values are quoted for each parameter estimated with the \textsc{cigale}. (1) Name of the galaxy. (2) Photometric or spectroscopic redshift ($z_\mathrm{spec}$ indicated by a $\dagger$). (3) Code used to derive the parameters. (4) Stellar mass (in solar units) and its uncertainty derived from the logarithmic space. (5) Parameter values for the old or the young population. (6) \textit{e}-folding time and its uncertainty (old population in Myr and young population in Gyr). (7) Age and its uncertainty (old population in Gyr and young population in Myr). (8) Attenuation in the \textit{V}-band and its uncertainty in mag. (9) Metallicity value (fixed to the solar value) in solar units. (10) Most recent burst intensity fraction and its uncertainty in percentage.}\\
\hline\hline
\multicolumn{1}{c}{Galaxy} & \multicolumn{1}{c}{$z^\dagger$} & \multicolumn{1}{c}{Code} & \multicolumn{1}{c}{$\log M_\star$} & \multicolumn{1}{c}{Pop.} & \multicolumn{1}{c}{$\tau$} & \multicolumn{1}{c}{Age} & \multicolumn{1}{c}{$A_V$} & \multicolumn{1}{c}{\textit{Z}} & \multicolumn{1}{c}{$b$} \\
 & & & \multicolumn{1}{c}{[$\mathrm{M}_{\sun}$]} & & \multicolumn{1}{c}{old [Myr]} & \multicolumn{1}{c}{old [Gyr]} & \multicolumn{1}{c}{[mag]} & \multicolumn{1}{c}{[$\mathrm{Z}_{\sun}$]} & \multicolumn{1}{c}{[\%]} \\
 & & & & & \multicolumn{1}{c}{you [Gyr]} & \multicolumn{1}{c}{you [Myr]} & & & \\
 \multicolumn{1}{c}{(1)} & \multicolumn{1}{c}{(2)} & \multicolumn{1}{c}{(3)} & \multicolumn{1}{c}{(4)} & \multicolumn{1}{c}{(5)} & \multicolumn{1}{c}{(6)} & \multicolumn{1}{c}{(7)} & \multicolumn{1}{c}{(8)} & \multicolumn{1}{c}{(9)} & \multicolumn{1}{c}{(10)} \\
\hline
\endfirsthead

\multicolumn{10}{c}{\tablename \thetable{ } -- continued from previous page}\\
\hline
\multicolumn{1}{c}{Galaxy} & \multicolumn{1}{c}{$z^\dagger$} & \multicolumn{1}{c}{Code} & \multicolumn{1}{c}{$\log M_\star$} & \multicolumn{1}{c}{Pop.} & \multicolumn{1}{c}{$\tau$} & \multicolumn{1}{c}{Age} & \multicolumn{1}{c}{$A_V$} & \multicolumn{1}{c}{\textit{Z}} & \multicolumn{1}{c}{$b$} \\
 & & & \multicolumn{1}{c}{[$\mathrm{M}_{\sun}$]} & & \multicolumn{1}{c}{old [Myr]} & \multicolumn{1}{c}{old [Gyr]} & \multicolumn{1}{c}{[mag]} & \multicolumn{1}{c}{[$\mathrm{Z}_{\sun}$]} & \multicolumn{1}{c}{[\%]} \\
 & & & & & \multicolumn{1}{c}{you [Gyr]} & \multicolumn{1}{c}{you [Myr]} & & & \\
 \multicolumn{1}{c}{(1)} & \multicolumn{1}{c}{(2)} & \multicolumn{1}{c}{(3)} & \multicolumn{1}{c}{(4)} & \multicolumn{1}{c}{(5)} & \multicolumn{1}{c}{(6)} & \multicolumn{1}{c}{(7)} & \multicolumn{1}{c}{(8)} & \multicolumn{1}{c}{(9)} & \multicolumn{1}{c}{(10)} \\
\hline
\endhead
\hline \multicolumn{10}{r}{Continued in the next page}\\
\endfoot
\endlastfoot
\vspace{3pt} MIPS0000517A & $1.0500$ & Synthesizer & $10.0^{+0.1}_{-0.1}$ & old & $230^{+43 }_{-125}$ & $1.0^{+0.2}_{-0.1}$ & $0.0^{+0.1}_{-0.0}$ & $1.0$ & \\
\vspace{3pt} & & & & you & $18.5^{+3.1}_{-3.2}$ & $ 39^{+6  }_{-7  }$ & $1.9^{+0.1}_{-0.1}$ & $1.0$ & $69^{+2 }_{-23}$ \\
\vspace{3pt} & & CIGALE & $10.1\pm 0.2$ & old & $1280^{+5762}_{-1047}$ & $3.2\pm 1.6$ & $1.0^{+0.7}_{-0.7}$ & $1.0$ & \\
\vspace{3pt} & & & & you & $20.0$ & $ 46^{+34 }_{-20 }$ & $1.9\pm 0.2$ & $1.0$ & $61\pm 26$ \\
\vspace{3pt} MIPS0000648A & $1.1900$ & Synthesizer & $10.4^{+0.1}_{-0.1}$ & old & $162^{+46 }_{-45 }$ & $1.0^{+0.1}_{-0.1}$ & $0.0^{+0.1}_{-0.0}$ & $1.0$ & \\
\vspace{3pt} & & & & you & $17.7^{+3.4}_{-2.7}$ & $ 45^{+21 }_{-19 }$ & $2.3^{+0.1}_{-0.1}$ & $1.0$ & $51^{+13}_{-16}$ \\
\vspace{3pt} & & CIGALE & $10.6\pm 0.1$ & old & $1189^{+5363}_{-973 }$ & $3.2\pm 1.6$ & $1.2^{+0.7}_{-0.7}$ & $1.0$ & \\
\vspace{3pt} & & & & you & $20.0$ & $100^{+90 }_{-47 }$ & $2.2\pm 0.2$ & $1.0$ & $57\pm 27$ \\
\hline
\end{longtable}

\newpage
\addtocounter{table}{-1}
\begin{longtable}{lcllclllll}
\caption[Stellar population synthesis results for complementary-sample starbursts.]{Stellar population synthesis results for complementary-sample dusty starbursts. Median and $1\sigma$ error values are shown for each parameter derived with the \textsc{synthesizer} code. The expected and standard deviation (or upper and lower limits when changing from logarithmic space to linear space) values are quoted for each parameter estimated with the \textsc{cigale}. (1) Name of the galaxy. (2) Photometric or spectroscopic redshift ($z_\mathrm{spec}$ indicated by a $\dagger$). (3) Code used to derive the parameters. (4) Stellar mass (in solar units) and its uncertainty derived from the logarithmic space. (5) Parameter values for the old or the young population. (6) \textit{e}-folding time and its uncertainty (old population in Myr and young population in Gyr). (7) Age and its uncertainty (old population in Gyr and young population in Myr). (8) Attenuation in the \textit{V}-band and its uncertainty in mag. (9) Metallicity value (fixed to the solar value) in solar units. (10) Most recent burst intensity fraction and its uncertainty in percentage.}\\
\hline\hline
\multicolumn{1}{c}{Galaxy} & \multicolumn{1}{c}{$z^\dagger$} & \multicolumn{1}{c}{Code} & \multicolumn{1}{c}{$\log M_\star$} & \multicolumn{1}{c}{Pop.} & \multicolumn{1}{c}{$\tau$} & \multicolumn{1}{c}{Age} & \multicolumn{1}{c}{$A_V$} & \multicolumn{1}{c}{\textit{Z}} & \multicolumn{1}{c}{$b$} \\
 & & & \multicolumn{1}{c}{[$\mathrm{M}_{\sun}$]} & & \multicolumn{1}{c}{old [Myr]} & \multicolumn{1}{c}{old [Gyr]} & \multicolumn{1}{c}{[mag]} & \multicolumn{1}{c}{[$\mathrm{Z}_{\sun}$]} & \multicolumn{1}{c}{[\%]} \\
 & & & & & \multicolumn{1}{c}{you [Gyr]} & \multicolumn{1}{c}{you [Myr]} & & & \\
 \multicolumn{1}{c}{(1)} & \multicolumn{1}{c}{(2)} & \multicolumn{1}{c}{(3)} & \multicolumn{1}{c}{(4)} & \multicolumn{1}{c}{(5)} & \multicolumn{1}{c}{(6)} & \multicolumn{1}{c}{(7)} & \multicolumn{1}{c}{(8)} & \multicolumn{1}{c}{(9)} & \multicolumn{1}{c}{(10)} \\
\hline
\endfirsthead

\multicolumn{10}{c}{\tablename \thetable{ } -- continued from previous page}\\
\hline
\multicolumn{1}{c}{Galaxy} & \multicolumn{1}{c}{$z^\dagger$} & \multicolumn{1}{c}{Code} & \multicolumn{1}{c}{$\log M_\star$} & \multicolumn{1}{c}{Pop.} & \multicolumn{1}{c}{$\tau$} & \multicolumn{1}{c}{Age} & \multicolumn{1}{c}{$A_V$} & \multicolumn{1}{c}{\textit{Z}} & \multicolumn{1}{c}{$b$} \\
 & & & \multicolumn{1}{c}{[$\mathrm{M}_{\sun}$]} & & \multicolumn{1}{c}{old [Myr]} & \multicolumn{1}{c}{old [Gyr]} & \multicolumn{1}{c}{[mag]} & \multicolumn{1}{c}{[$\mathrm{Z}_{\sun}$]} & \multicolumn{1}{c}{[\%]} \\
 & & & & & \multicolumn{1}{c}{you [Gyr]} & \multicolumn{1}{c}{you [Myr]} & & & \\
 \multicolumn{1}{c}{(1)} & \multicolumn{1}{c}{(2)} & \multicolumn{1}{c}{(3)} & \multicolumn{1}{c}{(4)} & \multicolumn{1}{c}{(5)} & \multicolumn{1}{c}{(6)} & \multicolumn{1}{c}{(7)} & \multicolumn{1}{c}{(8)} & \multicolumn{1}{c}{(9)} & \multicolumn{1}{c}{(10)} \\
\hline
\endhead
\hline \multicolumn{10}{r}{Continued in the next page}\\
\endfoot
\endlastfoot
\vspace{3pt} MIPS0000373\_1 & $1.1100$ & Synthesizer & $10.5^{+0.1}_{-0.1}$ & old & $121^{+52 }_{-25 }$ & $1.0^{+0.1}_{-0.1}$ & $0.6^{+0.2}_{-0.3}$ & $1.0$ & \\
\vspace{3pt} & & & & you & $18.5^{+2.9}_{-3.1}$ & $ 50^{+30 }_{-17 }$ & $2.1^{+0.1}_{-0.1}$ & $1.0$ & $30^{+12}_{-10}$ \\
\vspace{3pt} & & CIGALE & $10.8\pm 0.1$ & old & $1255^{+5646}_{-1027}$ & $3.7\pm 1.5$ & $1.1^{+0.6}_{-0.6}$ & $1.0$ & \\
\vspace{3pt} & & & & you & $20.0$ & $285^{+402}_{-167}$ & $1.9\pm 0.3$ & $1.0$ & $49\pm 31$ \\
\vspace{3pt} MIPS0000814\_1 & $1.1200$ & Synthesizer & $10.7^{+0.1}_{-0.1}$ & old & $115^{+42 }_{-20 }$ & $1.0^{+0.1}_{-0.1}$ & $0.4^{+0.2}_{-0.2}$ & $1.0$ & \\
\vspace{3pt} & & & & you & $18.6^{+2.9}_{-3.1}$ & $ 50^{+26 }_{-18 }$ & $1.8^{+0.1}_{-0.1}$ & $1.0$ & $25^{+11}_{-9 }$ \\
\vspace{3pt} & & CIGALE & $10.9\pm 0.1$ & old & $1309^{+5897}_{-1071}$ & $3.5\pm 1.5$ & $1.1^{+0.6}_{-0.6}$ & $1.0$ & \\
\vspace{3pt} & & & & you & $20.0$ & $263^{+356}_{-151}$ & $1.7\pm 0.3$ & $1.0$ & $50\pm 31$ \\
\hline
\end{longtable}

\bsp	
\label{lastpage}